\documentclass[aps,prb,twocolumn,amsmath,amssymb,nofootinbib,superscriptaddress,floatfix,scrartcl]{revtex4-1}

\usepackage{amssymb}
\usepackage{hyperref}
\usepackage{color}
\usepackage{graphicx}
\usepackage{subfigure}
\usepackage{srcltx}
\usepackage{mathrsfs}
\usepackage{multirow}
\usepackage{ulem}
\usepackage{mathtools}
\usepackage{umoline}
\usepackage{lipsum}
\usepackage{cleveref}

\def\t#1{\textrm{#1}}
\def\n{\nonumber \\}

\newcommand{\verteq}{\rotatebox{90}{$\,=$}}
\newcommand{\equalto}[2]{\underset{\scriptstyle\overset{\mkern4mu\verteq}{#2}}{#1}}
\newcommand{\vertoplus}{\rotatebox{90}{$\,\oplus$}}
\newcommand{\oplusto}[2]{\underset{\scriptstyle\overset{\mkern4mu\vertoplus}{#2}}{#1}}

\begin{document}

\title{
CPT theorem and classification of topological insulators
and superconductors
}

   \author{Chang-Tse Hsieh} 
 \affiliation{
 Department of Physics, University of Illinois
 at Urbana-Champaign, 
 1110 West Green St, Urbana IL 61801
             }

\author{Takahiro Morimoto}
 \affiliation{
Condensed Matter Theory Laboratory, 
RIKEN, Wako, Saitama, 351-0198, Japan
             }

   \author{Shinsei Ryu}
 \affiliation{
 Department of Physics, University of Illinois
 at Urbana-Champaign, 
 1110 West Green St, Urbana IL 61801
             }

\date{\today}

\begin{abstract}
We present a systematic topological classification of fermionic and 
bosonic topological phases protected by
time-reversal, particle-hole, parity, and combination of these 
symmetries. 
We use two complementary approaches: one in terms of 
K-theory classification of gapped quadratic fermion theories
with symmetries, 
and the other in terms of the K-matrix theory description of
the edge theory of (2+1)-dimensional bulk theories. 
The first approach is specific to free fermion theories in general spatial dimensions
while the second approach is limited to two spatial dimensions but incorporates effects of interactions.  
We also clarify the role of CPT theorem in classification of
symmetry-protected topological phases, and show, 
in particular, topological superconductors discussed before
are related by CPT theorem. 
\end{abstract}


\maketitle

\tableofcontents

\section{Introduction}

Topological insulators (TIs) and topological superconductors (TSCs) are 
states of matter that are not adiabatically connected to,
in the presence of a set of symmetry conditions, 
topologically trivial states of matter.
\cite{HasanKane2010,QiZhang2011} 
TIs and TSCs 
characterized by Altland-Zirnbauer symmetries, 
\cite{Altland1997}
time-reversal symmetry (TRS), particle-hole symmetry (PHS), 
and combinations thereof, 
have been theoretically predicted 
\cite{Kane2005a, Kane2005b, Bernevig2006, Kane2007, Moore2007, Roy2009, Schnyder2008, Kitaev2009, Volovik} 
and experimentally discovered.  
\cite{Konig2007, Hsieh2008, Chen2009, Hsieh2009, Xia2009, Wada2008, Murakawa2009, Murakawa2011}

In more recent years,
interplay between 
on-site symmetries 
(such as TRS and PHS listed in Altland-Zirnbauer symmetry classes)
and non-on-site symmetries 
(such as space group symmetries) 
has enriched the topological phases of matters. 
A novel class of topological matter characterized (additionally) by non-on-site symmetries, 
such as topological crystalline insulators (TCIs)
\cite{Fu2013}
and topological crystalline superconductors (TSCSs), 
have been discovered.  
\cite{Teo2008, Fang2012, Hsieh2012, Xu2012, Tanaka2012, Dziawa2012, Zhang2013, Ueno2013} 
The topological classification, 
originally studied 
in the presence/absence of various on-site symmetries in 
Altland-Zirnbauer classes, 
\cite{Schnyder2008, Ryu2010, Kitaev2009} are also extended to include the non-on-site symmetries 
such as reflection symmetry, 
\cite{Chiu2013, Morimoto2013} 
inversion symmetry,
\cite{Turner2012, Lu2014} 
and (crystal) point group symmetries, 
\cite{Slager2013, Fang2013a, Teo2013, Benalcazar2013} 
recently. 

Motivated by these recent works,
in this paper, 
we further study TIs and TSCs protected by a wider set of 
symmetries than symmetries in Altland-Zirnbauer classes 
by including, in particular, a parity symmetry (PS), 
which is a symmetry under the reflection of an odd number of spatial coordinates. 
One of our focuses is, 
in addition to the cases where parity is conserved, 
on situations where a combination of parity with some other symmetries, 
such as CP (product of PHS and PS) or PT (product of PS and TRS), 
are preserved. 
For earlier related works, see, for example,  
Refs.\ \onlinecite{
 Ryu2010, Hu2011, Mizushima2013, Fang2013b}.

Another issue we will discuss in this paper
is the effect of interactions 
on the classification of those topological phases
protected by parity and other symmetries (such as 
combination of parity and other symmetries).  
It has been demonstrated, in various examples,
that there are phases that appear to be topologically distinct 
from trivial phases at non-interacting level,
which, in fact, can adiabatically be deformable into
a trivial state of matter in the presence of interactions. 
\cite{Fidkowski2010, Fidkowski2011, Turner2011,Qi2013, Ryu2012, Yao2013}
For example, Ref.\ \onlinecite{Yao2013} discusses 
(2+1)-dimensional [or 2 spatial-dimensional (2D)] superconducting systems 
in the presence of parity and time-reversal symmetries, 
which are classified, at the quadratic level, by an integer topological invariant,
while once inter-particle interactions are included, 
states with an integer multiple 
of eight units of the non-interacting topological invariant
are shown to be unstable.
Focusing on 2D bulk topological states that support 
an edge state described by a K-matrix theory, i.e.,
Abelian states,
we will study 
the stability of the edge state (and hence the bulk state) in
the presence of 
parity symmetry or parity symmetry combined with other symmetries.  

\begin{figure}[tbp]
\centering

\subfigure[]{
   \scalebox{0.4}{\includegraphics{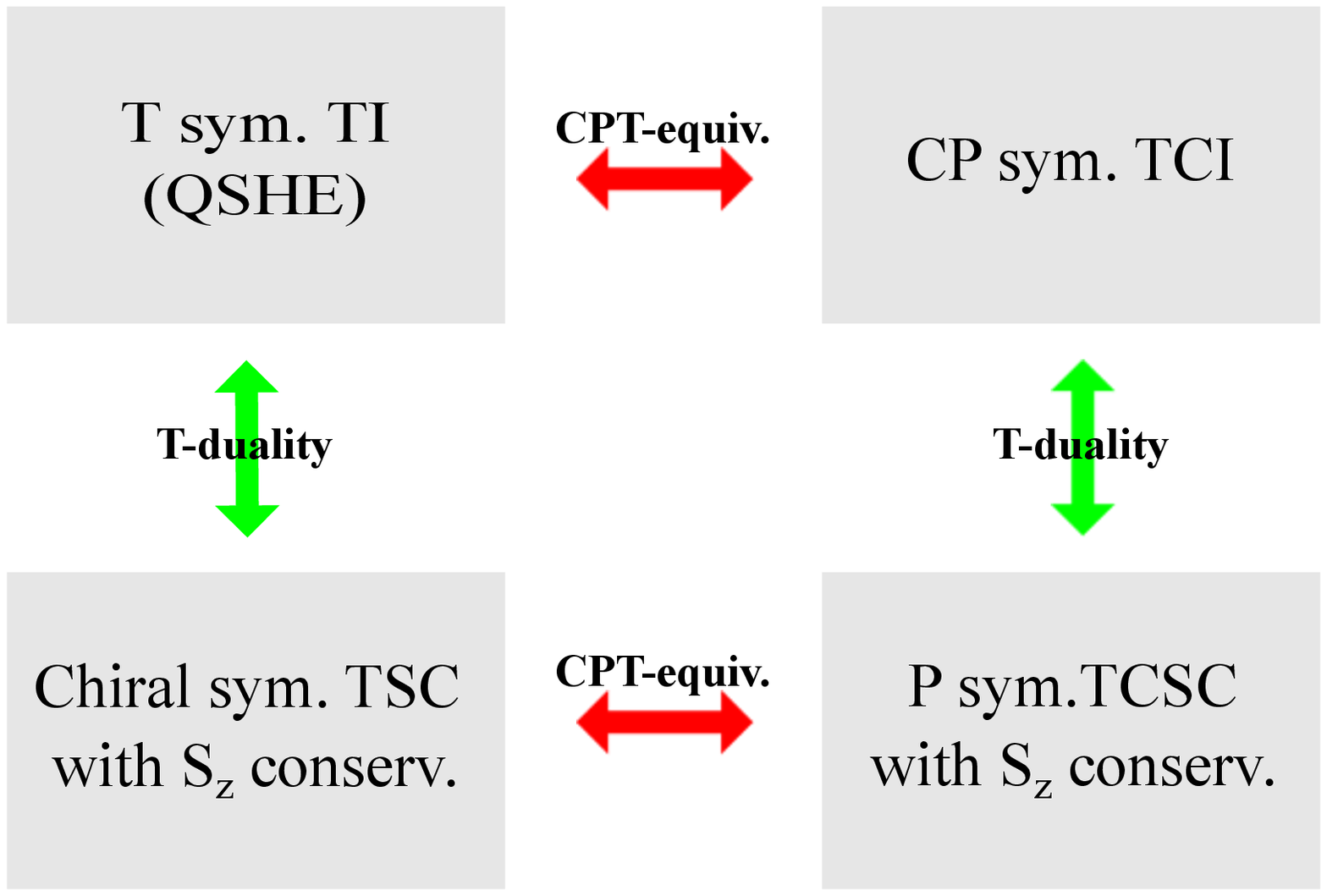}}
   \label{CPT-equiv_T-dual_TI&TSC_a}
 }

 \subfigure[]{
   \scalebox{0.4}{\includegraphics{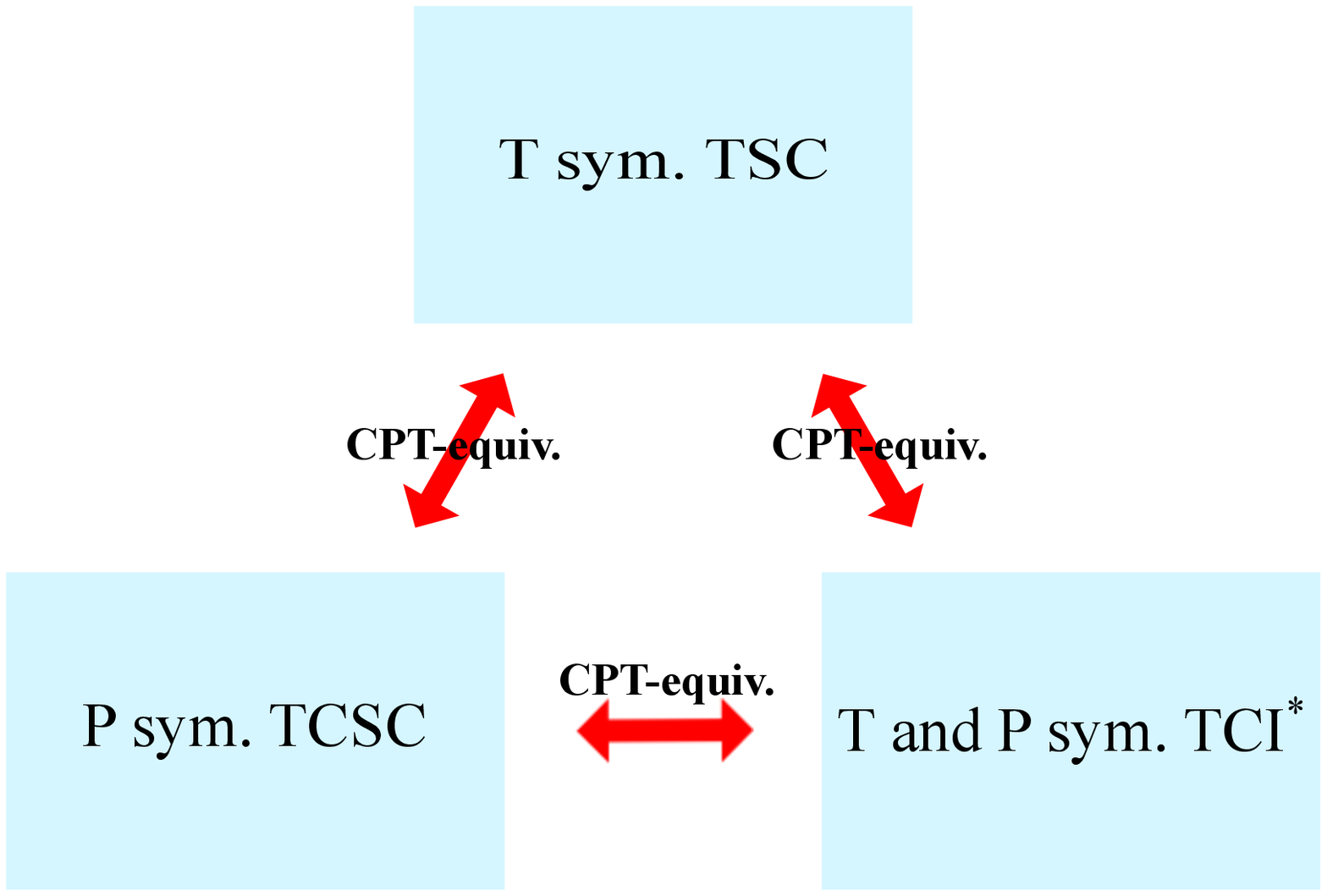}}
   \label{CPT-equiv_T-dual_TI&TSC_b}
 }

\label{CPT-equiv_T-dual_TI&TSC}
\caption{Two sets of topological (crystalline) insulators and superconductors related by CPT-equivalence and/or T-duality: (a) T symmetric TI (QSHE), CP symmetric TCI, chiral (TC) symmetric TSC with $S_z$ conservation, and P symmetric TCSC with $S_z$ conservation;
(b) T symmetric TSC, P symmetric TCSC, and T and P symmetric TCI that can support gapless edge states even in the absence of charge U(1) symmetry. 
}

\end{figure}

We will also show that, once 
parity symmetry or parity symmetry combined with other discrete symmetries
is included into our consideration, 
CPT theorem plays an important role
in classifying topological states of matter.
CPT theorem holds in Lorentz invariant quantum field theories, 
which says, C, P, T, when combined into CPT, is always 
conserved, i.e., $CPT=1$, schematically. 
For example, a Lorentz invariant CP symmetric field theory 
also possesses TRS, and vice versa.

In condensed matter systems, however, such relations between these discrete symmetries (T, C, and P) do not arise 
since we are not to be restricted to relativistic systems; 
symmetries can be imposed independently.  
Nevertheless, some physical properties of these non-relativistic systems at long wavelength limit, 
such as the band topology or the electromagnetic response, can be encoded in the so-called topological field theory, 
which respects the Lorentz symmetry. 
When these topological properties are protected (or determined) 
by some symmetry, TRS, say,  
they can also be protected solely by 
CP symmetry, which is a ``CPT-equivalent'' partner to TRS.  
For example, the magnetoelectric effect in 
3D time-reversal symmetric TIs\cite{Qi2008}
is also expected to be observed in a CP symmetric TI, because they are both described by the axion term (effective action for electromagnetic response) 
with the same nontrivial (quantized) value of $\theta$ angle. 

In addition, 
from the prospect of topological classification, 
classifying symmetry-protected topological (SPT) phases
of a free fermion systems
[characterized either by a gapped Hamiltonian of the $(d+1)$-dimensional bulk or by a gapless Hamiltonian of $d$-dimensional boundary with symmetries] 
is equivalent to classifying the corresponding Dirac operators 
with symmetry restrictions\cite{Kitaev2009}. 
It is thus natural to associate TIs protected by TRS with TIs protected by CP symmetry, as the Dirac Hamiltonian has a CPT invariant form.

In this paper, 
by going through classification problems of non-interacting fermions
in the presence of various symmetry conditions,
and also microscopic stability analysis of interacting edge theories, 
we will demonstrate explicitly such CPT theorem holds at the level 
of topological classification for all cases that we studied. 
Through this analysis, we can see, for example, 
that 
2D TSCs protected by spin parity conservation,
\cite{Qi2013, Ryu2012}
and 
2D TSCs by parity and time-reversal,
\cite{Yao2013}
both of which are classified in terms of $\mathbb{Z}_8$, 
are related by CPT theorem.

As mentioned above, CPT theorem (i.e., 
topological classification problems with different set of symmetries 
related by CPT relations) may largely be expected,
for example, once we anticipate description of SPT phases
by an underlying topological field theories. 
However, perhaps more fundamentally, we will also discuss
that while physical Hamiltonians may not obey CPT theorem,
their entanglement Hamiltonians obey a form of CPT theorem.  
\cite{Turner2012, Hughes2011, Chang2014}

We will also discuss yet another duality relation, ``T-duality'', 
for a wide range of topological insulators and superconductors.
T-duality is a duality that exchanges a phase field ($\phi \sim \varphi_L + \varphi_R$)
and its dual ($\theta \sim \varphi_L -\varphi_R$)
in the (1+1)-dimensional boson theory or in string theory.
[Here, $\phi$ and $\theta$ are the compact boson fields in the (1+1)-dimensional boson theory
and $\varphi_{L/R}$ are their left/right-moving parts.]  
Similarly to CPT theorem, this duality relation (and its proper generalization to 
K-matrix theory with multi-component boson fields) relates topological classification 
of (2+1)-dimensional fermionic systems 
with CP symmetry and charge U(1) symmetry
to topological classification with
parity symmetry and spin U(1) symmetry. 
The latter system is a Bogoliubov-de Genne (BdG) system with
conserved parity and spin U(1) symmetry. 
Therefore,
topological classification of 
(i) time-reversal symmetric insulators conserving charge U(1) (the quantum spin Hall effect),
(ii) CP-symmetric insulators with conserving charge U(1),
(iii) parity-symmetric BdG systems with conserved spin U(1),
and
(iv) TC-symmetric BdG system with conserved spin U(1), 
are all related (equivalent); all these systems are classified by 
a $\mathbb{Z}_2$ topological number. 
Such relation is shown pictorially in FIG. \ref{CPT-equiv_T-dual_TI&TSC_a}. FIG. \ref{CPT-equiv_T-dual_TI&TSC_b} shows another example for CPT-equivalent insulators and BdG systems with related symmetries.

While we were preparing the draft, a preprint that is related to this paper 
appeared on arXiv.
\cite{Shiozaki2014} 
While our analysis in terms the K-theory largely overlaps with this preprint,
our analysis in terms of K-matrix theories and our discussion in terms of 
CPT-theorem and T-duality were not discussed therein.

\subsection{Outline and main results}

The structure of the paper and the main results are summarized as follows.

In Sec.\ \ref{2D fermionic topological phases protected by symmetries}, 
we start our discussion by considering 
2D fermionic topological phases protected by CP symmetry. 
As inferred
from 
the CPT theorem
and
the existence of 
time-reversal symmetric topological insulators in 
two dimensions, 
we will show that there are two topologically distinct classes of 
insulators with CP symmetry in two dimensions,
i.e., $\mathbb{Z}_2$ classification.  
We will present
in Sec.\ \ref{CP symmetric insulators}
a simple example (tight-binding model) 
of CP symmetric band insulators,
which is constructed from 
two copies of two-band Chern insulators 
with opposite chiralities.
While we defer a systematic classification of CP symmetric insulators 
until
Sec.\ \ref{Classification of CP symmetric TI in arbitrary dimension},  
we discuss the edge theory of the CP symmetric topological insulators,
and perturbations to it such as mass terms.

We will also show,
in Sec.\ \ref{BdG systems with spin U(1) and P symmetries}, 
that  
systems with CP symmetry and charge conservation [charge U(1) symmetry] 
can also be 
interpreted as BdG systems 
that preserve parity and one component of SU(2) spin, $S_z$, say 
[spin U(1) symmetry].  
That is, 
CP symmetric topological insulators can be realized as 
topological crystalline superconductors with $S_z$ conservation. 
In terms of edge theories, 
the relation between CP symmetric insulators
and P symmetric BdG systems 
is nothing but T-duality or the Kramers-Wannier duality. 
To the best of our knowledge, this topological superconductor protected by 
spin U(1) and parity symmetry has not been discussed in the literature.

In Sec.\ \ref{Connection to T-symmetric insulators},
following Ref.\ \onlinecite{Turner2012}, 
we make a further connection between CP symmetric insulators
and time-reversal (T) symmetric insulators by considering entanglement Hamiltonians
and effective symmetries thereof. 
Then we introduce the ideas of CPT-equivalence and T-duality of topological phases in Sec. \ref{CPT-equivalence and T-duality of topological phases: an example}, taking the CP symmetric TI and its related systems [shown in FIG.\ \ref{CPT-equiv_T-dual_TI&TSC_a}] as an example that shows such equivalence.

In Sec.\ \ref{Classification of CP symmetric TI in arbitrary dimension}, 
we use K-theory to classify noninteracting CP symmetric TI in arbitrary dimensions. 
We found the topological classification of this symmetry class is exactly the same as that of the symmetry class possessing TRS. 
An explicit construction of the "effective" TRS operator from the Dirac Hamiltonian with CP symmetry is given. 
Then the topological invariants of CP symmetric TI are also constructed. 
On the other hand, using the extension problem of the Clifford algebra, 
the BdG systems with spin U(1) and P symmetries 
are also shown to fall in the classification equivalent to the symmetry class possessing TRS in any dimensions. 
We thus extend CPT-equivalence and T-duality that we observe  
in Sec.\ \ref{2D fermionic topological phases protected by symmetries}
in terms of 2D fermionic systems to all dimensions.

With the same idea, 
in Sec.\ \ref{Topological classification of other symmetries}, 
we also study topological phases protected by PT symmetry, 
which are CPT-equivalent partners of topological phases protected by PHS. While the latter is usually implemented by a BdG Hamiltonian that breaks charge U(1) symmetry (superconducting system), it is interesting to find a system of insulator with nontrivial topology protected by PT that manifests the same topological features as TSCs with PHS. 
While there is no nontrivial topological phase protected by PT 
in 3D and 2D (here we are not interested in the chiral topological phases in 2D), 
a nontrivial PT symmetric TI, which is characterized by a $\mathbb{Z}_2$ class, exists in a 1D (and 0D) system.

In Sec. \ref{Topological CPT theorem and topological CPT-equivalent symmetry classes} 
we discuss CPT-equivalence for more general symmetry classes. It can be stated as 
"topological CPT theorem" for non-interacting fermionic systems 
in arbitrary dimensions. Furthermore, the complete classification of TIs and TSCs 
(and TCIs and TCSCs if spatial symmetries are present) 
for non-interacting fermionic systems with T, C , P, 
and/or their combinations is obtained by considering symmetry classes "AZ+CPT":
(a) CPT-equivalent symmetry classes "generated" from AZ classes by a trivial CPT symmetry;
(b) Other symmetry classes "generated" from AZ classes by nontrivial CPT symmetries.
The result is summarized in TABLE \ref{tab_classification_C,P,T_free_fermion}.

In Sec. \ref{Classification of 2D interacting SPT phases: K-matrix formulation} 
we use (Abelian) K-matrix theory to classify 2D interacting topological phases protected by T, C, P
the combined symmetries, and/or U(1) symmetries, for either bosonic or fermionic systems. 
The results are summarized in TABLE \ref{tab_boson_nonchiral} and \ref{tab_fermion_nonchiral}. 
Comparing with the case of non-interacting fermions, 
we also give an interacting version of "topological CPT theorem" 
for 2D interacting bosonic and fermionic topological phases.
The key point is that any perturbations
(not necessary Lorentz scalars) that can gap the edges of a 2D bulk must be invariant under a "trivial" CPT symmetry. 
We also discuss T-duality in the K-matrix formalism.
Both CPT-equivalence and T-duality for 2D interacting topological phases can be seen manifestly in 
Table \ref{tab_boson_nonchiral} and \ref{tab_fermion_nonchiral}.

\section{2D fermionic topological phases protected by symmetries}
\label{2D fermionic topological phases protected by symmetries}

In this section, we start our discussion
by considering a simple fermionic tight-binding model 
which is invariant under CP symmetry. 
We will also note that the fermionic system can also be interpreted as
a topological superconductor (BdG system) that conserves 
the $z$-component of spin. Later, we will comment on the connection between
CP symmetric TIs and T symmetric TIs (QSHE), and introduce the ideas of CPT-equivalence and T-duality for topological phases.

\subsection{CP symmetric insulators}
\label{CP symmetric insulators}

\subsubsection{tight-binding model}

Let us consider  
the following tight-binding Hamiltonian: 
\begin{align}
T
&=
\sum_{r}
\psi^{\dag}(r)
\left(
\begin{array}{cc}
t & {i}\Delta\\
{i}\Delta  & -t
\end{array}
\right)
\psi(r+\hat{x})
+\mathrm{h.c.}
\nonumber \\
&
\qquad
+
\psi^{\dag}(r)
\left(
\begin{array}{cc}
t & \Delta\\
-\Delta  & -t
\end{array}
\right)
\psi(r+\hat{y})
+\mathrm{h.c.}
\nonumber \\
&
\qquad
+
\psi^{\dag}(r)
\left(
\begin{array}{cc}
\mu & 0\\
0  & -\mu
\end{array}
\right)
\psi(r),
\label{lattice p-wave}
\end{align}
where 
the two-component fermion annihilation 
operator at site $r$ on the two-dimensional square lattice,
$\psi(r)$, 
is given in terms of 
the electron annihilation operators
with spin up and down,
$c_{r,1/2}^{\ }$,
as
$\psi^{T}({r})=
(c_{r,1}, c_{r,2})$,
and we take
$t=\Delta=1$. 
There are four phases separated by three quantum critical points
at $\mu=0,\pm 4$, which are labeled by
the Chern number as
$\mathrm{Ch}=0$ $( |\mu| > 4)$,
$\mathrm{Ch}=-1$ $(-4 < \mu < 0)$, and
$\mathrm{Ch}=+1$ $( 0 < \mu < +4)$.
In the following, we are interested in the phase with
$\mathrm{Ch}=\pm 1$.
In momentum space, 
$T 
= 
\sum_{k \in \mathrm{BZ}}
\psi^{\dag}(k)
\left[
\vec{n}(k) \cdot\vec{\sigma}
\right]
\psi(k),
$ 
\begin{align}
\vec{n}(k)
&=
\left(
\begin{array}{c}
-2 \Delta \sin k_x \\
-2 \Delta \sin k_y \\
2t (\cos k_x + \cos k_y) + \mu
\end{array}
\right).
\label{n vector}
\end{align}
We will mostly focus on the case of $\mathrm{Ch}=\pm 1$.  

A lattice model of the topological insulator 
with CP symmetry can be constructed by 
taking two copies of the above two-band Chern insulator 
with opposite chiralities.
Consider the Hamiltonian in momentum space, 
\begin{align}
H 
&= 
\sum_{k \in \mathrm{BZ}}
\sum_{s=\uparrow,\downarrow}
\psi^{\dag}_{s}(k)
\left[
\vec{n}_{s}(k) \cdot\vec{\sigma}
\right]
\psi_{s}(k)
\nonumber \\
&=
\sum_{k\in \mathrm{BZ}}
\Psi^{\dag}(k)
\mathcal{H}(k)
\Psi(k), 
\label{chiral+R}
\end{align}
where 
$s=\uparrow,\downarrow$
represent ``pseudo spin'' degrees of freedom, 
$\Psi(k)$ is a four-component fermion field,
and
$\vec{n}_{s}(k)$ is given, in terms of $\vec{n}$ as
$\vec{n}_{\uparrow}(k)
=
\vec{n}(k) 
$,
$\vec{n}_{\downarrow}(k)
=
\vec{n}_{\uparrow}(\tilde{k})
=
\vec{n}(\tilde{k})
$,
where $\tilde{k} = (-k_1, k_2)$.
I.e.,
the single particle Hamiltonian in momentum 
space is given in terms of the $4\times 4$ matrix, 
\begin{align}
\mathcal{H}
(k)
=
n_x(k) \tau_z \sigma_x
+
n_y(k) \tau_0 \sigma_y
+
n_z(k) \tau_0 \sigma_z,  
\end{align}
where $\tau_{0, z}$ is the Pauli matrix acting
on the pseudo spin index. 

The Hamiltonian is invariant under 
the following CP transformations:
\begin{align}
 \mathcal{U}
\Psi(x)
\mathcal{U}^{-1}=
U_{\mathrm{CP}} \Psi^{\dag}(\tilde{x}), 
\end{align}
where $\tilde{x}:=(-x_1, x_2)$ and
$U_{\mathrm{CP}}$ is given either by 
\begin{align}
 U_{\mathrm{CP}}&=
\tau_x \sigma_x
\quad
U_{\mathrm{CP}}^T
=
+U_{\mathrm{CP}},  
\quad 
(\eta=+1),
\nonumber \\
U_{\mathrm{CP}}&=
\tau_y \sigma_x, 
\quad 
U_{\mathrm{CP}}^T
= 
-U_{\mathrm{CP}},  
\quad
(\eta=-1). 
\label{choice_U_CP}
\end{align}
To ditinguish these two cases,
we introduced an index $\eta$;
$\eta=\pm 1$ refers to the first/second case. 
We will also use notation
$\eta = e^{2\pi i \epsilon}$
where 
$\epsilon=0, 1/2$
for 
$\eta = 1, -1$, 
respectively.

It 
turns out 
imposing 
$U_{\mathrm{CP}}=\tau_x\sigma_x$
leads to
the CP symmetric topological
insulator.
This can be seen
by
looking at the stability of 
the edge mode
that can appear
when we terminate the system
in $y$-direction
(i.e., the edge is along the $x$-direction.)
One can check, numerically, and 
also in terms of the continuum edge theory
(see below), 
$U_{\mathrm{CP}}=\tau_x\sigma_x$
protects the edge state
while 
$U_{\mathrm{CP}}=\tau_y\sigma_x$ 
does not.
In the following, these CP transformations
will be combined with charge U(1) gauge transformation,
and the corresponding transformation will be denoted by
$\mathcal{U}$ (See below).

\subsubsection{Edge theory}

We now develop
a continuum theory for the edge state
that exists when the system is
terminated in $y$ direction.
I.e., the edge is along $x$ direction. 
Let us consider the free fermion:
\begin{align}
  H &=
  \frac{v}{2\pi}
  \int dx
  \big(
    \psi^{\dag}_L {i}\partial_x \psi^{\ }_L
     -\psi^{\dag}_R {i}\partial_x \psi^{\ }_R 
    \big).
    \nonumber \\
    &=
    \frac{v}{2\pi}   \int dx
    \Psi^{\dag} {i}\partial_x \sigma_z \Psi^{\ },
    \quad
    \Psi =
    \left(
    \begin{array}{c}
     \psi_L \\
     \psi_R 
    \end{array}
    \right). 
\end{align}
We consider two types of CP symmetry operation
\begin{align}
 \mathcal{U}
 \psi^{\ }_L(x)
 \mathcal{U}^{-1}
&=
e^{i\alpha} \psi^{\dag}_{R}(-x), 
\nonumber \\
 \mathcal{U}
 \psi^{\ }_R(x)
\mathcal{U}^{-1}
&=
\eta e^{i\alpha} \psi^{\dag}_{L}(-x), 
\end{align}
where CP transformation is combined with
the EM U(1) charge twist with
an arbitray phase factor 
$\alpha$.
The sign $\eta=\pm$ is 
$+/-$ for topological/non-topological cases:
\begin{align}
\eta &
= e^{2\pi i \epsilon}
=
\left\{
 \begin{array}{ll}
  +1& \mbox{topological} \\
  -1& \mbox{trivial} \\
 \end{array}
 \right.
\end{align}
This is CP symmetry with twisting by the charge operator,
\begin{align}
 \mathcal{U}
 =
 e^{ -i \alpha F_V}
 \mathcal{CP}, 
\end{align}
where $F_V$ is the total charge operator,
\begin{align}
 F_V &:= 
 F_R + F_L
 =
 \int dx 
 \left(\psi^{\dag}_R \psi^{\ }_R
 + \psi^{\dag}_L \psi^{\ }_L
\right). 
\end{align}

There are two fermion mass bilinears 
that are consistent with 
the charge U(1) symmetry:
These masses are odd under CP 
when $\eta=+1$. 
We thus conclude that the edge theory is
at least at the quadratic level stable (ingappable). 

%
%
%
%
%
%
%
%
%
%
%
%
%

\subsection{BdG systems with spin U(1) and P symmetries}
\label{BdG systems with spin U(1) and P symmetries}

In this section, 
we show that systems with charge U(1) and CP symmetry
can  be derived from BdG systems with conserved one component of spin 
($S_z$, say)  and parity symmetry. 

The system of our interest preserves spin U(1) but not charge U(1). 
At the quadratic level, this situation is described by 
the BdG Hamiltonian. 
Following Altland and Zirnbauer,\cite{Altland1997}
we consider the following general form of a BdG Hamiltonian
for the dynamics of quasiparticles 
\begin{align}
H
&=
\frac{1}{2}
\left(
\begin{array}{cc}
\boldsymbol{c}^{\dag}, & \boldsymbol{c}
\end{array}
\right)
\mathcal{H}_4
\left(
\begin{array}{c}
\boldsymbol{c}^{\ } \\
\boldsymbol{c}^{\dag}
\end{array}
\right),
\nonumber \\
\mathcal{H}_4
&=
\left(
\begin{array}{cc}
\Xi & \Delta \\
 -\Delta^{*} & -\Xi^{{T}}
\end{array}
\right),
\label{eq: BdG hamiltonian}
\end{align}
where $\mathcal{H}_4$ is a $4N \times 4N$ matrix for a system with $N$
orbitals (lattice sites),
and $\boldsymbol{c}=
\left(\boldsymbol{c}_{\uparrow},\boldsymbol{c}_{\downarrow}\right)$.
[$\boldsymbol{c}$ and $\boldsymbol{c}^{\dag}$ 
can be either column or row vector
depending on the context.]
The matrix elements obey 
$\Xi=\Xi^{\dag}$ (hermiticity)
and
$\Delta=-\Delta^{{T}}$ (Fermi statistics). 
The presence of $\mathrm{SU}(2)$ spin rotation symmetry is represented by
\begin{eqnarray}
&&
\big[\mathcal{H}_4,J_{a}\big]=0,
\quad
J_{a}:=
\left(
\begin{array}{cc}
s_{a} & 0 \\
0 & -s_{a}^{{T}}
\end{array}
\right),
\end{eqnarray}
where $a=x,y,z$.

With conservation of one component of spin, 
say, $z$-component, we have a U(1) symmetry
associated with rotation around $z$-axis. 
With this $S_z$ conservation, 
one can reduce this BdG Hamiltonian into the following form:
\begin{align}
H &=
\left(
\begin{array}{cc}
 \boldsymbol{c}^{\dag}_{\uparrow},  &
 \boldsymbol{c}^{\ }_{\downarrow}
\end{array}
\right)
\mathcal{H}_2
\left(
\begin{array}{c}
 \boldsymbol{c}^{\ }_{\uparrow}
 \\
 \boldsymbol{c}^{\dag}_{\downarrow}
\end{array}
\right), \ 
\mathcal{H}_2&=
\left(
\begin{array}{cc}
 \xi_{\uparrow} & \delta \\
 \delta^{\dag} & -\xi^T_{\downarrow}
\end{array}
\label{BdG_Sz_conserv}
\right),
\end{align}
up to a term which is proportional to the
identity matrix.
At the quadratic level, this Hamiltonian is a member of symmetry class A 
(unitary symmetry class in AZ classes). 
[with $\boldsymbol{c}_{\downarrow}\to \boldsymbol{c}_{\downarrow}^{\dagger}$, 
one can ``convert'' spin U(1) to fictitious charge U(1)]. 
This can be seen as follows. 
Let us consider BdG Hamiltonians
which are invariant under rotations about the $z$- (or any fixed)
axis in spin space,
yielding to the condition
$\left[ \mathcal{H}_4, J_{z} \right]= 0$,
which implies that the Hamiltonian can be brought into the form
\begin{eqnarray}
\mathcal{H}_4=
\left(
\begin{array}{cccc}
a & 0 & 0& b \\
 0 & a' & -b^T & 0\\
0  & -b^{*} & -a^T & 0\\
b^{\dag} & 0&0 & -a^{\prime T}
\end{array}
\right)\!,
\,\,
a^{\dag}=a,
\,\,
a^{\prime \dag}=a^{\prime}.
\end{eqnarray}
Due to the sparse structure of $\mathcal{H}_4$,
we can rearrange the elements of this
$4N \times 4N$ matrix into the form of a $2N\times 2N$ matrix $\mathcal{H}_2$
above.
%

Let us now consider parity symmetry.
For simplicity, we assume orbitals transform
trivially under parity, and hence assume the following
form:
\begin{align}
 \mathcal{P}\left(
 \begin{array}{c}
  \boldsymbol{c}_{\uparrow}(r) \\
  \boldsymbol{c}_{\downarrow}(r)
 \end{array}
 \right)
 \mathcal{P}^{-1}
 =
\left(
 \begin{array}{c}
  e^{-i\alpha} \boldsymbol{c}_{\downarrow}(\tilde{r}) \\
  \eta e^{i\alpha}
  \boldsymbol{c}_{\uparrow}(\tilde{r})
 \end{array}
 \right). 
\end{align}
Within the reduced $2N\times 2N$ basis, 
parity symmetry looks like
CP symmetry.
To see this, let us write out 
the Hamiltonian in the following form:
\begin{align}
 H&=
\left(
\begin{array}{cc}
 \boldsymbol{c}^{\dag}_{\uparrow}, &
 \boldsymbol{c}^{\ }_{\downarrow}
\end{array}
\right)
\left(
\begin{array}{cc}
 \xi_{\uparrow} & \delta \\
 \delta^{\dag} & -\xi^T_{\downarrow}
\end{array}
\right)
\left(
\begin{array}{c}
 \boldsymbol{c}^{\ }_{\uparrow}
 \\
 \boldsymbol{c}^{\dag}_{\downarrow}
\end{array}
\right)
\\
 &=
 c^{\dag}_{\uparrow a}(r)
 \xi_{\uparrow ab} (r,r')
 c^{\ }_{\uparrow b}(r')
 -
 c^{\ }_{\downarrow a}(r)
 \xi^T_{\downarrow ab} (r,r')
 c^{\dag}_{\downarrow b}(r')
 \nonumber \\
 &\quad
 +
c^{\dag}_{\uparrow a}(r)\delta_{ab}(r,r') 
c^{\dag}_{\downarrow b}(r')
 +
 c^{\ }_{\downarrow a}(r)\delta^{\dag}_{ab}(r,r') 
c^{\ }_{\uparrow b}(r')
\nonumber 
\end{align}
(summation over repeated indices are implicit).
Then,
\begin{align}
 \mathcal{P} H
 \mathcal{P}^{-1}
 &=
 \left(
 \begin{array}{cc}
  c^{\dag}_{\uparrow}, &
  c^{\ }_{\downarrow}
 \end{array}
 \right)_{b \tilde{r}'}
 \left(
\begin{array}{cc}
 \xi_{\downarrow} & -\eta \delta^T 
 \\
 -\eta \delta^{*} &
-\xi^T_{\uparrow}
\end{array}
\right)_{br',ar}
\left(
\begin{array}{c}
 c^{\ }_{\uparrow} \\
 c^{\dag}_{\downarrow}
\end{array}
\right)_{a\tilde{r}}
\end{align}
(The transpose $T$ here acts
both $a$ and $r$).
Thus, the invariance under P 
implies
\begin{align}
 \left(
\begin{array}{cc}
 \xi_{\downarrow} & -\eta \delta^T 
 \\
 -\eta \delta^{*} &
-\xi^T_{\uparrow}
\end{array}
\right)_{a\tilde{r},b\tilde{r}'}
&=
 \left(
\begin{array}{cc}
 \xi_{\uparrow} &  \delta 
 \\
\delta^{\dag} &
-\xi^T_{\downarrow}
\end{array}
\right)_{a {r},b{r}'}
\end{align}

With the transformation
or relabeling
\begin{align}
 c^{\ }_{\uparrow}=:\Psi^{\ }_{\uparrow},
 \quad
 c^{\ }_{\downarrow}=: \Psi^{\dag}_{\downarrow}, 
\end{align}
we can write the Hamiltonian as
\begin{align}
H 
&=
\sum_{r,r'} 
\Psi^{\dag}(r)\ 
\mathcal{H}_2(r,r')\, 
\Psi(r'). 
\end{align}
Provided the system has 
translational symmetry,
$\mathcal{H}_2(r,r') = 
\mathcal{H}_2(r-r')$,
with periodic boundary conditions in each spatial direction
(i.e., the system is defined on a torus $T^d$),
we can perform the Fourier transformation and obtain 
in momentum space
\begin{align}
H 
=
\sum_{k \in \mathrm{Bz}} 
\Psi^{\dag}(k) \,
\mathcal{H}_2(k)\, 
\Psi (k) ,
\end{align}
where the crystal momentum $k$
runs over the first Brillouin zone (Bz), 
and the Fourier component of the fermion operator
and the Hamiltonian are given by 
$\Psi(r)
=
V^{-1/2}
\sum_{k\in \mathrm{Bz}}
e^{{i} k \cdot r}
\Psi(k)
$
and
$
\mathcal{H}_2(k)
=
\sum_{r} 
e^{- {i} {k}\cdot {r}}
\mathcal{H}_2({r})
$,
respectively.

Then the P invariance demands
\begin{align}
\left(
\begin{array}{cc}
 \xi_{\downarrow} & -\eta \delta^T 
 \\
 -\eta \delta^{*} &
-\xi^T_{\uparrow}
\end{array}
\right)_{-\tilde{k}} =
\left(
\begin{array}{cc}
 \xi_{\uparrow} & \delta 
 \\
 \delta^{\dag} &
-\xi^T_{\downarrow}
\end{array}
\right)_{k}. 
\end{align}
Observing that 
\begin{align}
\left(
\begin{array}{cc}
 0 & 1 \\
 \eta & 0
\end{array}
\right)
\left(
\begin{array}{cc}
 -\xi_{\downarrow} & \eta \delta^T 
 \\
 \eta \delta^{*} &
\xi^T_{\uparrow}
\end{array}
\right)^T
\left(
\begin{array}{cc}
 0 & \eta \\
 1 & 0
\end{array}
\right)
=
\eta 
\left(
\begin{array}{cc}
\xi_{\uparrow}
& \delta
\\
\delta^{\dag}
&
-\xi^T_{\downarrow}
\end{array}
\right)
\end{align}
we then conclude that, 
when $\eta=1$, 
\begin{align}
 \tau_x\mathcal{H}_2(-\tilde{k})^T \tau_x 
 =
 -\mathcal{H}_2(k),
 \quad
 \tau^T_x = +\tau_x
\end{align}
whereas when $\eta=-1$, 
\begin{align}
 \tau_y\mathcal{H}_2(-\tilde{k})^T \tau_y 
 =
 -\mathcal{H}_2(k),
 \quad
 \tau_y^T= -\tau_y. 
\end{align}
I.e., the single-particle Hamiltonian 
$\mathcal{H}_2$ is CP symmetric.

It should be noted that 
P symmetry with $\eta=+1$ is somewhat unusual. 
When acting twice on spinors, $\mathcal{P}^2=+1$,
whereas we usually expect $\mathcal{P}^2=-1$.
This is so since parity should reverse the sign of 
angular momentum, either of orbital or spin origin. 
The P symmetry with $\eta=1$ can be considered 
as a composition of a P symmetry with $\eta=-1$
and spin parity $(-1)^{N_{\uparrow}}$
where $N_{\uparrow}$ is the number operator
associated to up spins. 

\paragraph{T-duality (Kramers-Wannier duality)}

By taking T-dual or Kramers-Wannier-dual of the above setting,
\begin{align}
 \psi^{\dag}_L (x) \to \psi^{\ }_L(x), 
\end{align}
we obtain the P symmetric system. 
In the bosonized language, this amounts to exchange phase field $\phi$
and its dual $\theta$.
Also, if we decompose the complex fermion $\psi_L$ in terms of 
two real (Majorana) fermions $\chi^{1,2}_L$, $\psi_L = \chi^1_L + i \chi^2_L$,  
the above transformation amounts to $\chi^2_L \to -\chi^2_L$ while keeping 
the right moving intact. This is nothing but the Kramers-Wannier duality in the Ising model. 

The P symmetry, dualized from CP symmetry above, is given by
\begin{align}
 \mathcal{U}
 \psi^{\ }_L(x)
 \mathcal{U}^{-1}
&=
e^{-i\alpha} \psi^{\ }_{R}(-x)
\nonumber \\
 \mathcal{U}
 \psi^{\ }_R(x)
\mathcal{U}^{-1}
&=
\eta e^{i\alpha} \psi^{\ }_{L}(-x). 
\end{align}
This is P symmetry with $S_z$ twisting,
\begin{align}
 \mathcal{U}
 =
 e^{ -i \alpha F_A}
 \mathcal{P}, 
\end{align}
where $F_A$ is the axial charge,
\begin{align}
 F_A &:= 
 F_R - F_L
 =
 \int dx 
 \left(\psi^{\dag}_R \psi^{\ }_R
 - \psi^{\dag}_L \psi^{\ }_L
\right). 
\end{align}

\subsection{Connection to T symmetric insulators}
\label{Connection to T-symmetric insulators}

The CP symmetric model introduced above is in fact
also time-reversal invariant in the absence of perturbations. 
If there is Lorentz invariance, 
because of CPT theorem,  
any perturbation to the model that is CP symmetric 
is also T symmetric. 
Hence, 
within Lorentz invariant theories,
the same set of perturbations is prohibited by CP and T symmetries. 
The topological phase protected by CP symmetry can thus be also viewed as
a T symmetric topological phase. 

However, the above argument based on CPT theorem 
of course raises a question as 
we do not want to be confined to relativistic systems,
and Lorentz invariance is absent in the lattice model. 
Note, however, the following: 
(i)
CPT theorem tells us the presence of antiparticles.
This seems a necessary ingredient to have a topological phase
(topologically non-trivial ``vacua''). 
(ii)
Topological phases that are characterized by
a term of topological origin in the response theory,
such as the Chern-Simons term or the axion term
for the external (background) U(1) gauge field, 
are Lorentz invariant. 
This in particular means 
CP symmetry dictates the theta angle to be 0 or $\pi$
(mod $2\pi$), just as TRS does. 

Finally, 
while Hamiltonians may violate Lorentz invariance,
and hence CPT theorem, 
a version of CPT like theorem applies to
wavefunctions (= projection operators),
or the ``entanglement Hamiltonian''. 
In other words,
wavefunctions or the entanglement Hamiltonian
have more symmetries than the physical Hamiltonian. 
Due to this, for any CP symmetric system,
one can define ``effective'' time-reversal symmetry for
the projector or the entanglement Hamiltonian. 
See Appendix \ref{Entanglement spectrum and effective symmetries}. 

\subsection{CPT-equivalence and T-duality of topological phases: an example}
\label{CPT-equivalence and T-duality of topological phases: an example}

The above discussion reveals a "CPT-equivalence" between CP and T symmetric topological phases. Furthermore, from the fact that, both CP symmetric TI with $(\mathcal{CP})^2=1$ [$(\mathcal{CP})^2=(-1)^{N_f}$] and T symmetric TI with $\mathcal{T}^2=(-1)^{N_f}$ ($\mathcal{T}^2=1$), where $N_f$ is the total fermion number operator, possess the same nontrivial $\mathbb{Z}_2$ (trivial) classification in two dimensions, we expect a specific correspondence between these two "CPT-equivalent" topological phases. In general, such correspondence can be observed among topological phases protected by discrete symmetries T, C, P, and/or their combinations. We will discuss it in the following sections.

On the other hand, there is a duality -- which we call "T-duality" in this paper -- between topological phases of insulating and superconducting systems with corresponding symmetries. Imposing a symmetry $g$ on a BdG system with $S_z$ conservation (\ref{BdG_Sz_conserv}) will result a constraint on the reduced  BdG Hamiltonian $\mathcal{H}_2$ by the dual symmetry $\widetilde{g}$, which is exactly in the same symmetry class as a tight-binding Hamiltonian $\mathcal{H}$ constrained by the symmetry $\widetilde{g}$ in a insulating system (with charge conservation implicitly). For example, as we discussed in Sec. \ref{BdG systems with spin U(1) and P symmetries}, the CP symmetric topological phase can also be realized in  a BdG system with P symmetry and $S_z$ conservation. Another known example is that a chiral symmetric topological phases (class AIII in AZ class) can also be interpreted as a BdG Hamiltonian possessing TRS and $S_z$ conservation. \cite{Schnyder2008, Hosur2010} Interestingly (and expectedly), a 2D T symmetric TI, i.e., the QSHE, also has a dual realization in a superconducting system -- a BdG system with chiral symmetry and $S_z$ conservation. This can be seen by a similar discussion from Sec. \ref{BdG systems with spin U(1) and P symmetries}. For a reduced BdG Hamiltonian (by $S_z$ conservation) (\ref{BdG_Sz_conserv}), if we impose a "chiral" symmetry $\mathcal{S}$ (which is defined as a combination of T and C symmetries) as
\begin{align}
 \mathcal{S}\left(
 \begin{array}{c}
  \boldsymbol{c}_{\uparrow} \\
  \boldsymbol{c}_{\downarrow}
 \end{array}
 \right)
 \mathcal{S}^{-1}
 &=
\left(
 \begin{array}{c}
  e^{-i\alpha} \boldsymbol{c}_{\downarrow}^{\dagger} \\
  \eta e^{i\alpha}
  \boldsymbol{c}_{\uparrow}^{\dagger}
 \end{array}
 \right)
=
e^{-i\alpha\tau_z}U\left(
 \begin{array}{c}
  \boldsymbol{c}_{\uparrow}^{\dagger} \\
  \boldsymbol{c}_{\downarrow}^{\dagger}
 \end{array}
 \right), \n
U=
&\left\{
 \begin{array}{ll}
  \tau_x&  \mbox{for $\eta=1$} \\
  i\tau_y& \mbox{for $\eta=-1$}, \\
 \end{array}
 \right.
\end{align}
then
\begin{align}
 H=
\mathcal{S}H\mathcal{S}^{-1}
&=
\mathcal{S}
\left(
\begin{array}{cc}
 \boldsymbol{c}^{\dag}_{\uparrow},  &
 \boldsymbol{c}^{\ }_{\downarrow}
\end{array}
\right)
\mathcal{H}_2
\left(
\begin{array}{c}
 \boldsymbol{c}^{\ }_{\uparrow}
 \\
 \boldsymbol{c}^{\dag}_{\downarrow}
\end{array}
\right) 
\mathcal{S}^{-1}\n
&=
\left(
\begin{array}{cc}
 \boldsymbol{c}^{\dag}_{\uparrow},  &
 \boldsymbol{c}^{\ }_{\downarrow}
\end{array}
\right)
\left(U^{\dagger}\mathcal{H}_2^*U\right)
\left(
\begin{array}{c}
 \boldsymbol{c}^{\ }_{\uparrow}
 \\
 \boldsymbol{c}^{\dag}_{\downarrow}
\end{array} 
\right) 
\end{align}
(note that $\mathcal{S}=\mathcal{TC}$ is antiunitary)
implies $ U^{\dagger}\mathcal{H}_2^TU=\mathcal{H}_2$, i.e., the single-particle Hamiltonian $\mathcal{H}_2$ is TRS. In conclusion, dual symmetries between the tight-binding Hamiltonian (with charge conservation) and the BdG reduced Hamiltonian (with $S_z$ conservation) have the following correspondences:
\begin{align}
&\mathcal{T} \leftrightarrow \mathcal{TC}, \quad\mathcal{P} \leftrightarrow \mathcal{CP}.
\label{TCP_duality_noninteracting}
 \end{align}

FIG.\ \ref{CPT-equiv_T-dual_TI&TSC} shows some examples about CPT-equivalence and T-duality among topological (crystalline) insulators and superconductors. Especially, FIG. \ref{CPT-equiv_T-dual_TI&TSC_a} shows the connection between T symmetric TIs, CP symmetric TIs, and their dual realizations in BdG systems with $S_z$ conservations introduced in this section. Another example, as shown in FIG. \ref{CPT-equiv_T-dual_TI&TSC_b}, is the CPT-equivalence between T symmetric TSCs, P symmetric TCSCs, and T and P symmetric TCIs that can support gapless edge states even in the absence of charge U(1) symmetry. 

 In the following section, we make a more precise discussion for the idea of CPT-equivalence and T-duality introduced here, focusing on non-interacting fermionic CP symmetric TIs in arbitrary dimensions


\section{Classification of CP symmetric TIs in arbitrary dimension}
\label{Classification of CP symmetric TI in arbitrary dimension}

In this section we consider systems of non-interacting fermions
with CP symmetry and
classify CP symmetric TIs in arbitrary dimensions using K-theory.

Relevant symmetries are written as constraints on the Hamiltonian 
matrix $\mathcal{H}$ as follows.
The particle-hole symmetry (PHS) is an anti-unitary operator $C$
that anti-commutes with the Hamiltonian as
$\{C,\mathcal{H}\}=0$,
which is equivalently written using an unitary operator $U_{\mathrm{C}}$ as
\begin{align}
 U_{\mathrm{C}} \mathcal{H}^*(-k_1,\ldots,-k_d) U_{\mathrm{C}}^{-1}= -\mathcal{H}(k_1,\ldots,k_d).
\end{align}
The parity symmetry $P$ is a symmetry that swaps left-handed and right-handed coordinates,
which can be implemented as a mirror symmetry with respect to a particular direction (here we take $k_1$) as
\begin{align}
 P \mathcal{H}(-k_1,k_2,\ldots,k_d) P^{-1}= \mathcal{H}(k_1,k_2,\ldots,k_d),
\label{eq:P-symmetry}
\end{align}
with a unitary operator $P$.
Combining these two symmetries $C$ and $P$,
we define CP symmetry by an unitary operator $U_{\mathrm{CP}}$ satisfying
\begin{align}
 U_{\mathrm{CP}} \mathcal{H}^*(k_1,-k_2,\ldots,-k_d) 
 U_{\mathrm{CP}}^{-1}= -\mathcal{H}(k_1,k_2,\ldots,k_d).
\label{eq:CP-symmetry}
\end{align}
A CP symmetric TI is a topological insulator that
does not possess C nor P symmetry 
but is characterized with a combined CP symmetry.

\subsection{Classification by K-theory}

In non-interacting fermion systems, 
CP symmetric TIs are classified using K-theory in a way similar to 
the classification of topological defects discussed by Teo and Kane.\cite{teo-kane10}

A TI with CP symmetry (\ref{eq:CP-symmetry}) is regarded as
a TSC with PHS $\tilde C=CP$ 
in the $\tilde d=d-1$ dimensions with momenta $k_2,\ldots,k_d$ 
(that are flipped by an action of $\tilde C$),
containing a defect with a co-dimension $D=1$ parameterized with $k_1$
(that is not flipped by $\tilde C$).
When we have PHS $\tilde C$ with $\tilde C^2=+1$ or $\tilde C^2=-1$,
the symmetry class is class D or class C
and the associated classifying spaces $R_q$ are given as 
\cite{Kitaev2009,Morimoto2013}
\begin{align}
\t{class D}&\quad \tilde C^2=1 &: \quad R_2 &\quad (q=2), \n
\t{class C}&\quad \tilde C^2=-1 &: \quad R_6 &\quad (q=6).
\label{eq:Rq-PHS}
\end{align}
Then the classification for CP symmetric TI is given by a homotopy group\cite{teo-kane10}
\begin{align}
\pi_D(R_{q-\tilde d}) \simeq \pi_0(R_{q-\tilde d +D})=\pi_0(R_{q+2-d}).
\label{eq:CP-classification}
\end{align}
This can be interpreted that the relevant classifying space changes from $R_q$
 to $R_{q+2}$,
which corresponds to the symmetry class AII ($R_4$) or class AI ($R_8 \simeq R_0$),
both possessing TRS.
Thus CP symmetric TI behaves similar to TR symmetric TI in terms of topological classification and corresponding edge states.
This is consistent with the CPT theorem for Lorentz-invariant systems where CP symmetry can be effectively converted into time reversal symmetry.

While we adopted the parity symmetry (\ref{eq:P-symmetry}) that 
flips only one momentum $k_1$,
we can generally reverse $2n+1$ coordinates for the parity as
\begin{align}
 &P \mathcal{H}(-k_1,\ldots,-k_{2n+1},k_{2n+2},\ldots,k_d) P^{-1} \n
 &= \mathcal{H}(k_1,\ldots,k_{2n+1},k_{2n+2},\ldots,k_d).
\end{align}
The TI with CP symmetry constructed from above $P$ can be 
regarded as a TSC with PHS in the $\tilde d=d-2n-1$ dimensions with momenta $k_{2n+2},\ldots,k_d$,
containing a defect with a co-dimension $D=2n+1$ parameterized with $k_1,\ldots,k_{2n+1}$.
Then the classification is given by a homotopy group
\begin{align}
\pi_D(R_{q-\tilde d})=\pi_0(R_{q+4n+2-d}),
\end{align}
where the relevant classifying space looks like $R_{q+4n+2}$.
Since we have $q=2$ or $q=6$, 
$R_{q+4n+2}$ becomes $R_0$ or $R_4$, i.e,
the classifying space associated with the symmetry class with TRS,
which is consistent with the CPT theorem.

\subsection{Dirac models and CPT theorem}
\label{Dirac model and CPT theorem}
While we cannot explicitly construct an anti-unitary operator $T$ for TRS
from CP symmetry in general cases,
we can construct $T$ operator from CP symmetry in a Dirac Hamiltonian
\begin{align}
 \mathcal{H}(k)=m\gamma_0 + \sum_{i=1}^d k_i \gamma_i, 
\label{Dirac_Hamiltonian}
\end{align}
where $\gamma_i$'s are anti-commuting gamma matrices, $m$ is a mass,
and $k_i$'s are momenta.
The CP symmetry (\ref{eq:CP-symmetry}) then leads to relations
\begin{align}
 \{ U_{\mathrm{CP}}K,\gamma_i\}&=0, \qquad   i=0,1, \nonumber \\
 [ U_{\mathrm{CP}}K,\gamma_i]&=0, \qquad   2\le i \le d, 
\end{align}
with a complex conjugation $K$.
Now we can construct an effective TRS from CP symmetry as
$T=\gamma_1 U_{\mathrm{CP}}K$,
satisfying
\begin{align}
 \gamma_1 U_{\mathrm{CP}} 
 \mathcal{H}^*(-k_1,\ldots,-k_d) (\gamma_1 U_{\mathrm{CP}})^{-1} =
 \mathcal{H}(k_1,\ldots,k_d).
\label{eq:effecitve-TRS}
\end{align}

The existence of $\gamma_1$ in the Dirac model enables us to convert the CP symmetry into a TRS,
which is not the case for a general lattice model
where a kinetic term along reflected coordinate is not necessarily written by a gamma matrix $\gamma_1$.

\subsection{Topological invariants}
Topological invariants of CP symmetric TIs 
are constructed in the same way as those for topological defects. \cite{teo-kane10}
For $q+2-d=0,4$ in Eq.\ (\ref{eq:CP-classification}),
we have topological invariants $\mathbb{Z}$.
Due to $q=2,6$ [Eq.\ (\ref{eq:Rq-PHS})],
the topological invariants $\mathbb{Z}$ are realized in even dimensions $d$,
where we can define the Chern number over the Brillouin zone.
The Chern number gives the topological invariants,
which is written, by putting $d=2n$, as
\begin{align}
Ch_n
&= \frac{1}{n!}\int d^{2n}k\, \t{tr}\left(\frac{iF}{2\pi}\right)^n,\n
F&=dA+ A\wedge A, \qquad
A=\langle u_k| d |u_k\rangle,
\end{align}
with valence bands $|u_k\rangle$ and a derivative $d$ with respect to momenta $k$.

Next, first descendant $\mathbb{Z}_2$ is given by a Chern-Simons form,
which takes place for $q+2-d=1$ in Eq.~(\ref{eq:CP-classification}),
so that the dimension $d$ is odd. 
When we have the first descendant $\mathbb{Z}_2$,
we can choose a continuous gauge $A$ over the entire Brillouin zone,
and an integration of the Chern-Simons form,
which is defined for odd dimensions,
 gives topological invariant $\mathbb{Z}_2$.

Second descendant $\mathbb{Z}_2$ is given by a dimension reduction of the above $\mathbb{Z}_2$.
We consider a one parameter family of the Hamiltonian $\tilde{\mathcal{H}}(\theta, k)$
connecting the original Hamiltonian $\tilde{\mathcal{H}}(0, k)=\mathcal{H}(k)$ 
and a reference CP symmetric Hamiltonian $\tilde{\mathcal{H}}(\pi, k)=\mathcal{H}_0$ with a parameter $0 \le \theta \le \pi$.
If we extend a range of $\theta$ into  $-\pi \le \theta \le \pi$ by a relation
\begin{align}
 &\tilde{\mathcal{H}}(\theta, k_1,k_2,\ldots,k_d) \n
&=-U_{\mathrm{CP}}
\tilde{\mathcal{H}}^*(-\theta, k_1,-k_2,\ldots,-k_d) U_{\mathrm{CP}}^{-1},
\end{align}
we can define a CP symmetric Hamiltonian $\tilde{\mathcal{H}}(\theta, k)$ over    $-\pi \le \theta \le \pi$ and $k$.
Then the second descendant $\mathbb{Z}_2$ characterizing the Hamiltonian $\mathcal{H}(k)$
is given by an integration of Chern-Simons form for $\tilde{\mathcal{H}}(\theta, k)$.

\subsection{BdG systems with spin U(1) and P symmetries}
As we have seen in Sec.~\ref{2D fermionic topological phases protected by symmetries}B,
CP symmetric TI can be realized by a BdG system with a reflection symmetry 
and spin U(1) symmetry ($S_z$ conservation).
Here we interpret their equivalence to class AII TIs
in terms of  K-theory and Clifford algebras.
When we have a unitary operator commuting with the Hamiltonian,
we should block-diagonalize the Hamiltonian when we consider a topological classification.
When the $S_z$ anti-commutes with the PHS $C$ and the reflection symmetry $P$,
the block-diagonalized Hamiltonian does not possess $C$ nor $P$ any further,
while the combined $CP$ still remains as a symmetry of the block Hamiltonian.
The situation is summarized as follows,
\begin{align}
 \{C,\mathcal{H}\}&=0, & [S_z,\mathcal{H}]&=0, & \{C,S_z\}&=0, & \{P,S_z\}&=0, \n
&&S_z^2&=1, & P^2&=1, &&
\label{eq:Sz-and-P}
\end{align}
along with a parity symmetry (\ref{eq:P-symmetry}).
[Note: we can choose $P^2=1$ by appropriately multiplying ``$i$'',
which may change a commutation/anticommutation relation with $C$.]

Now let us look at a topologically non-trivial example of this construction.
We start with a BdG Hamiltonian in class D and two dimensions as
\begin{align}
 \mathcal{H}(k)&=n_x(k) \sigma_x + n_y(k) \sigma_y + n_z(k) \sigma_z, 
\end{align}
where $\vec{n}(k)$  
is defined in
Eq.\ (\ref{n vector})
and 
we have PHS of $C=\sigma_x K$. 
A parity symmetry (\ref{eq:P-symmetry}) is implemented by taking two copies of the above BdG Hamiltonian (denoted by $\tau$) and a spin U(1) symmetry is implemented by taking two copies representing spin degrees of freedom (denoted by $s$),
which yields 
\begin{align}
 \mathcal{H}(k) &=n_x(k) \sigma_x \tau_z s_x + n_y(k) \sigma_y s_x + n_z(k) \sigma_z, 
\end{align}
where we have 
 the spin U(1) symmetry $S_z=\sigma_z s_z$.
We have two parity symmetries $P$
(a reflection symmetry with respect to $x$-direction)
written as 
\begin{align}
P&=
\begin{cases}
\tau_x s_x, \quad \t{(topological)} \\
\tau_y s_x. \quad \t{(trivial)} \\
\end{cases}
\label{eq:parity-choice}
\end{align}
A choice of parity $P=\tau_x s_x$, commuting with PHS ($[C,P]=0$),
leads to a topologically non-trivial insulator, as explained later with Clifford algebras.
We can choose either of parity symmetries by adding appropriate terms to the Hamiltonian.

Block diagonalization with respect to $S_z$ becomes clear,
if we change bases as
$
(\sigma_x s_z,\sigma_y s_z, s_x, s_y,\sigma_z s_z)
\to
(\sigma_x s_x,\sigma_y s_x,\sigma_z s_x, s_y, s_z)
$,
\begin{align}
 \mathcal{H}(k)&=-n_x(k) \sigma_x \tau_z s_z - n_y(k) \sigma_y s_z + n_z(k) \sigma_z, \n
 S_z&=s_z.
\end{align}
The block Hamiltonian with $s_z=-1$ is given by
\begin{align}
 \mathcal{H}(k)&=n_x(k) \sigma_x \tau_z + n_y(k) \sigma_y + n_z(k) \sigma_z,
\label{eq:Sz-block}
\end{align}
characterized by a CP symmetry with $U_{\mathrm{CP}}=\tau_x\sigma_x$.
This corresponds to a non-trivial CP symmetric TI 
given in Eq. (\ref{choice_U_CP})
with $\eta=+1$.
In (\ref{eq:Sz-block}), the mass term $\sigma_z$ 
is the unique mass compatible with the CP symmetry.
Thus Hamiltonians with different signs of the unique mass term are topologically distinct.
If we try to double the system where doubled 2 by 2 degrees of freedom is 
described by Pauli matrices $\rho$,
allowed mass terms are not unique since we have 
$\sigma_x \tau_x \rho_y$
and $\sigma_x \tau_y \rho_y$ in addition to $\sigma_z$.
Then we can adiabatically connect two states in the doubled system
by appropriately rotating in the space of mass term,
which indicates that the classification 
of CP symmetric TI in 2D is $\mathbb{Z}_2$.

Next we show that the above classification for class D accompanied with spin  U(1) and parity $P$ is equivalent to  that in class AII,
by adopting Clifford algebras classification for Dirac models.
The original classification for class D in $d$-dimensions is given by a Clifford algebra \cite{Morimoto2013}
\begin{align}
\{\gamma_0, C,CJ,J\gamma_1,\ldots,J\gamma_d \}
\end{align}
and its extension problem with respect to the mass term $\gamma_0$ is
\begin{align}
Cl_{d,2}\to Cl_{d,3},
\end{align}
where the topological index is given by $\pi_0(R_{2-d})$,
especially $\mathbb{Z}$ for $d=2$.
The spin U(1) symmetry ($S_z$) and the parity symmetry ($P$), satisfying (\ref{eq:Sz-and-P})
and $[C,P]=0$, can be included in the Clifford algebra as
\begin{align}
\{\gamma_0, C,CJ,J\gamma_1,\ldots,J\gamma_d,J\gamma_1 P,\gamma_1 P S_z \},
\end{align}
for which
the extension problem for $\gamma_0$ is written as
\begin{align}
Cl_{d,4}\to Cl_{d,5},
\end{align}
 and the classification is given by $\pi_0(R_{4-d})$.
In our example in $d=2$, we have $\mathbb{Z}_2$ topological number.
Above classification is the same as that for class AII in $d$-dimensions,
which shows that CP symmetric TI and TR symmetric TI are equivalent in the level of Dirac models.
Indeed, the effective TRS is given from CP symmetry and a kinetic gamma matrix as (\ref{eq:effecitve-TRS}).

On the other hand, in the case of the parity symmetry anti-commuting with PHS ($\{C,P\}=0$),
the relevant Clifford algebra is
\begin{align}
\{\gamma_0, C,CJ,J\gamma_1,\ldots,J\gamma_d,\gamma_1 P,J\gamma_1 P S_z \},
\end{align}
and the extension problem for $\gamma_0$ is 
\begin{align}
Cl_{d+2,2}\to Cl_{d+2,3}.
\end{align}
Then the topological invariant is $\pi_0(R_{-d})$,
where we have a trivial insulator for $d=2$ as $\pi_0(R_{-2})=0$,
that is equivalent to class AI in $d=2$.
This is the reason why we have a trivial insulator if we choose 
a parity symmetry
$P=\tau_y s_x$ in Eq.\ (\ref{eq:parity-choice}).

While we have so far discussed the system in class D with $S_z$ and $P$ in two dimensions,
we note that a system in three dimensions possesses non-trivial $\mathbb{Z}_2$ topological invariant.
This is interesting,
 since the original class D system in three dimensions is trivial and a block-diagonalized system with $S_z$ (class A system) is also trivial,
while the CP symmetry gives rise to a non-trivial insulator.

\subsection{Topological classification of other symmetries}
\label{Topological classification of other symmetries}

In a similar manner as CP symmetric TIs,
we can define PT symmetric TIs.
PT symmetry can be defined by a unitary operator $U_{\mathrm{PT}}$ satisfying
\begin{align}
 U_{\mathrm{PT}} \mathcal{H}^*(k_1,-k_2,\ldots,-k_d) U_{\mathrm{PT}}^{-1}= 
 \mathcal{H}(k_1,k_2,\ldots,k_d).
\label{eq:PT-symmetry}
\end{align}
Then PT symmetric TI is a topological insulator that
does not possess P nor T symmetry 
but is characterized with a combined PT symmetry.
In an analogous way for CP symmetric TI,
a classification of PT symmetric TI in $d$-dimensions is obtained 
by considering a system
 with TRS in $d-1$-dimensions 
containing a topological defect with co-dimension $1$.
We assume that classifying space for the TR symmetric TI
with $\tilde T=U_{\mathrm{PT}}K$ in 0-dimensions is $R_q$.
We have $q=0$ for $(U_{\mathrm{PT}}K)^2=+1$, 
and $q=4$ for $(U_{\mathrm{PT}}K)^2=-1$.
Then the classification for PT symmetric TI is given by
\begin{align}
\pi_1(R_{q-(d-1)})\simeq \pi_0(R_{q+2-d}).
\end{align}
Non-trivial PT symmetric TIs are found in 2-dimensional systems in class AI or AII with a reflection symmetry,
where TRS and P are broken by a diagonalization with respect to some unitary symmetry but the combined PT remains,
which is characterized by a non-trivial topological number $\mathbb{Z}$.
A shift of classifying space by $2$ is interpreted as a change of an effective symmetry class into that with PHS,
which is again consistent with the CPT theorem. 


\section{Topological CPT theorem and topological CPT-equivalent symmetry classes}
\label{Topological CPT theorem and topological CPT-equivalent symmetry classes}

Actually, as we discussed in previous sections, such CPT-equivalence holds "topologically" for more general symmetry classes (not just for cases discussed in the last sections). In this section, we discuss the "topological CPT theorem" and topological CPT-equivalent symmetry classes in noninteracting fermionic systems.

Combining symmetries $T$, $C$, and $P$, we define the CPT symmetry by an unitary operator $W$  satisfying
\begin{align}
 W\mathcal{H}(\tilde{k})W^{-1}
 =-\mathcal{H}(k),
\quad W^2=1, \label{CPT_sym}
\end{align}
where $\mathcal{H}(k)$ is a $d$-dimensional ($d\geq 1$) single particle Hamiltonian and $\tilde{k}=(-k_1, k_2, \dotsc, k_d)$.
Note that here, as $W$ is unitary, $W^2$ can alway be fixed to be 1 by the redefinition 
$W'=e^{i\alpha}W$ with any phase factor (such redefinition is also accompanied by changing 
the commutation relations with other existed symmetries at the same time). Now, if the system already 
has some symmetries, adding the CPT symmetry constraint [Eq. (\ref{CPT_sym})] on $\mathcal{H}$ would or would not change the classifying space or the topological classification with respect to existing symmetries. That is, in the latter case, there exists a CPT operator $W=W_0$ such that the system transforms  "topologically-trivially" under $W_0$. Therefore, we have the following statement:

{\it {\bf Topological CPT theorem for noninteracting fermionic systems:} 
Let $\{g_i\}$ be a set of symmetries (can be a null set) composed of $T$, $C$, $P$, 
and/or their combinations.  Then for non-interacting fermionic systems there is a "trivial" CPT operator $W=W_0$,
which anticommutes with $T$ and $P$ and commutes with $C$ 
(from which other commutation relations between $g_i$ and $W_0$ can also be deduced),
such that the system with symmetries $\{g_i\}$ and the system of with symmetries $\{g_i, W_0\}$ possess the same classifying space or topological classification.
}

The proof of the above theorem is straightforward as we consider the Dirac Hamiltonian (\ref{Dirac_Hamiltonian}) (the idea here is similar to the discussion in Sec. \ref{Dirac model and CPT theorem}) :
\begin{align}
 \mathcal{H}(k)=m\gamma_0 + \sum_{i=1}^d k_i \gamma_i, \nonumber
\end{align}
where $\gamma_i$'s are anti-commuting gamma matrices. The symmetries $T$, $C$, and $P$  
(if present) satisfy
\begin{align}
 &[T,\gamma_0]=0, \quad \{T,\gamma_{i\neq 0}\}=0, \n
 &\{C,\gamma_0\}=0, \quad [C,\gamma_{i\neq 0}]=0, \n
 &\{P,\gamma_1\}=0, \quad [P,\gamma_{i\neq 1}]=0, 
\end{align}
while the CPT symmetry $W$ satisfies
\begin{align}
\{W,\gamma_0\}=0, \quad [W,\gamma_{i=1}]=0, \quad \{W,\gamma_{i\neq 0, 1}\}=0.
\end{align}
Define $M=\gamma_1W $,
we then have $[M,\gamma_i]=0\ \forall i$ and thus $M$ is an unitary symmetry commuting with $\mathcal{H}$:
\begin{align}
[M,\mathcal{H}]=0, \quad M^2=1.
\end{align}
For the system with symmetries $\{g_i\}$, composed of $T$, $C$, $P$, and/or their combinations, if the additional symmetry $M=M_0$ commutes with all $g_i$, or equivalently if $W= W_0$ satisfies (if $\{g_i\}$ includes some of the following symmetries)
\begin{align}
&\{W_0,T\}=0, \quad [W_0,C]=0, \quad \{W_0,P\}=0, \n
&\{W_0,CP\}=0, \quad [W_0,TP]=0, \quad \{W_0,TC\}=0,
\label{trivial_CPT_commutation_relations}
\end{align}
we can block diagonalize $\mathcal{H}$ with respect to $M$ such that all symmetries $g_i$ are still preserved in each eigenspace of $M$. Therefore, the symmetry class and hence the classification would not change as the symmetry $M_0$ or $W_0$ is added to the original set of symmetries $\{g_i\}$ of the system. This completes the proof.

We would like to point out that,  though the topological CPT theorem is "proved" (or argued) by considering the Dirac model (as a representative model of Clifford algebras that capture the topology of classifying spaces), which seems obviously to be invariant under a (trivial) CPT symmetry because of its Lorentz invariance, the same conclusion can be reached by more (mathematically) rigorous ways, such as topological K-theory, which is irrelevant to Lorentz invariance. Actually, this is what we did (in Sec. \ref{Classification of CP symmetric TI in arbitrary dimension}) in the discussion for equivalence between T and CP, and C and PT, as part of topological CPT theorem discussed here, using K-theory in a way similar to topological defects discussed in Ref. \onlinecite{teo-kane10}.

\
\begin {table}[ht]
\centering
\begin{tabular}{|cccc|}
  \hline
  Class & $S_{\mathrm{M},\mathrm{TC}}$ & $S_{\mathrm{W},\mathrm{TC}}$ & Classifying space  \\ 
  \hline 
  AIII & $-$ & $+$ & $C_0$  \\ 
  \hline\hline
  Class & $S_{\mathrm{M},\mathrm{T}}$ or $S_{\mathrm{M},\mathrm{C}}$ & $S_{\mathrm{W},\mathrm{T}}$ or $S_{\mathrm{W},\mathrm{C}}$ & Classifying space  \\
  \hline
  AI, AII & $-$ & $+$ & $C_0$  \\
  \hline
  D, C & $-$ & $-$ & $C_0$  \\ 
  \hline\hline
Class & $(S_{\mathrm{M},\mathrm{T}}, S_{\mathrm{M},\mathrm{C}})$ & $(S_{\mathrm{W},\mathrm{T}}, S_{\mathrm{W},\mathrm{C}})$ & Classifying space  \\
  \hline
  \multirow{3}{*}{BDI, CII} & $(-, +)$ & $(+, +)$ & $R_q \rightarrow R_{q+1}$  \\
   & $(+, -)$ & $(-, -)$ & $R_q \rightarrow R_{q-1}$ \\
   & $(-, -)$ & $(+, -)$ & $C_1$ \\
  \hline
  \multirow{3}{*}{DIII, CI} & $(+, -)$ & $(-, -)$ & $R_q \rightarrow R_{q+1}$ \\
   & $(-, +)$ & $(+, +)$ & $R_q \rightarrow R_{q-1}$  \\
   & $(-, -)$ & $(+, -)$ & $C_1$ \\
  \hline
\end{tabular}
\caption{Classification of AZ symmetry classes in the presence of a nontrivial CPT symmetry $W$ or an additional unitary symmetry $M$ (commuting with $\mathcal{H}$) in zero dimension. $S_{\mathrm{M},\mathrm{g}}$ and $S_{\mathrm{W},\mathrm{g}}$ dictate commutation $(+)$ or anticommutation$(-)$ relation of symmetry $g$, which can be $T$, $C$, or $TC$. $"R_q"$ in the last column denotes the original classifying spaces for the corresponding symmetry classes (before adding $M$ or $W$).}
\label{tab_classification_nontrivial_CPT_free_fermion}
\end{table}


\
\begin {table*}[ht]
\centering
\begin{tabular}{|c||c|c|}
  \multicolumn{2}{ l }{(a)} \\
  \hline
  CPT-equiv. sym. classes "generated" from AZ classes by trivial CPT &  $C_q$or$R_q$ & $\pi_0(C_q \mbox{or} R_q)$  \\ 
  \hline\hline
  "None" (A),  $\Gamma_{+}(CPT)$ & $C_0$ & $\mathbb{Z}$ \\
  \hline
   $\Gamma_{+}(TC)$ (AIII), $\Gamma_{+}(P)$, $\Gamma^{-}_{++}(TC, P)$ & $C_1$ & 0 \\
   \hline\hline
   $\Gamma_{+}(T)$ (AI), $\Gamma_{-}(CP)$, $\Gamma^{+}_{+-}(T, CP)$ & $R_0$ & $\mathbb{Z}$ \\
   \hline
   $\Gamma^{+}_{++}(T, C)$ (BDI), $\Gamma^{-}_{++}(C, P)$, $\Gamma^{+}_{++}(T, P)$, 
   $\Gamma^{+-+}_{+++}(T, C, P)$, $\Gamma^{+}_{-+}(CP, PT)$
   & $R_1$ & $\mathbb{Z}_2$ \\
   \hline
   $\Gamma_{+}(C)$ (D), $\Gamma_{+}(PT)$, $\Gamma^{+}_{++}(C, PT)$ & $R_2$ & $\mathbb{Z}_2$ \\
   \hline
   $\Gamma^{+}_{-+}(T, C)$ (DIII), $\Gamma^{+}_{++}(C, P)$, $\Gamma^{-}_{-+}(T, P)$, 
   $\Gamma^{++-}_{-++}(T, C, P)$, $\Gamma^{+}_{++}(CP, PT)$
   & $R_3$ & 0 \\
   \hline
   $\Gamma_{-}(T)$ (AII), $\Gamma_{+}(CP)$, $\Gamma^{+}_{-+}(T, CP)$ & $R_4$ & $\mathbb{Z}$ \\
   \hline
   $\Gamma^{+}_{--}(T, C)$ (CII), $\Gamma^{-}_{-+}(C, P)$, $\Gamma^{+}_{-+}(T, P)$, 
   $\Gamma^{+-+}_{--+}(T, C, P)$, $\Gamma^{+}_{+-}(CP, PT)$
   & $R_5$ & 0 \\
   \hline
   $\Gamma_{-}(C)$ (C), $\Gamma_{-}(PT)$, $\Gamma^{+}_{--}(C, PT)$ & $R_6$ & 0 \\
   \hline
   $\Gamma^{+}_{+-}(T, C)$ (CI), $\Gamma^{+}_{-+}(C, P)$, $\Gamma^{-}_{++}(T, P)$, 
   $\Gamma^{++-}_{+-+}(T, C, P)$, $\Gamma^{+}_{--}(CP, PT)$
   & $R_7$ & 0 \\
  \hline 
\end{tabular}

\begin{tabular}{|c||c|c|}
  \multicolumn{2}{ l }{(b)} \\
  \hline
  Other sym. classes "generated" from AZ classes by nontrivial CPT &  $C_q$or$R_q$  & $\pi_0(C_q \mbox{or} R_q)$ \\ 
  \hline\hline
   $\Gamma^{+}_{++}(TC, P)$, $\Gamma^{+}_{++}(T, CP)$, $\Gamma^{+}_{--}(T, CP)$, 
   $\Gamma^{+}_{+-}(C, PT)$, $\Gamma^{+}_{-+}(C, PT)$ & $C_0$ & $\mathbb{Z}$ \\
  \hline
   $\Gamma^{++-}_{+++}(T, C, P)$, $\Gamma^{+-+}_{-++}(T, C, P)$, $\Gamma^{++-}_{--+}(T, C, P)$,
   $\Gamma^{+-+}_{+-+}(T, C, P)$ & $C_1$ & 0 \\
   \hline\hline
   $\Gamma^{+++}_{+-+}(T, C, P)$, $\Gamma^{+--}_{+++}(T, C, P)$ & $R_0$ & $\mathbb{Z}$ \\
   \hline
   $\Gamma^{+++}_{+++}(T, C, P)$, $\Gamma^{+--}_{-++}(T, C, P)$ & $R_2$ & $\mathbb{Z}_2$ \\
   \hline
   $\Gamma^{+++}_{-++}(T, C, P)$, $\Gamma^{+--}_{--+}(T, C, P)$ & $R_4$ & $\mathbb{Z}$ \\
   \hline
   $\Gamma^{+++}_{--+}(T, C, P)$, $\Gamma^{+--}_{+-+}(T, C, P)$ & $R_6$ & 0 \\
   \hline
\end{tabular}
\caption{Classification of TIs and TSCs for non-interacting fermion systems with symmetry classes composed of $T$, $C$, $P$, and/or their combinations in zero dimension. This can be obtained by adding the CPT symmetry (either trivial or nontrivial ones) to the AZ classes: (a) CPT-equivalent symmetry classes "generated" from AZ classes by trivial CPT; (b) Other symmetry classes "generated" from AZ classes by nontrivial CPT (based on the result in Table \ref{tab_classification_nontrivial_CPT_free_fermion}). In this table we have
fixed $[A_i, A_j]=0$ and $U_i^2=1$ (other choices are equivalent), where $A_{i/j}$ and $U_i$ represent antiunitary and unitary symmetries, respectively. Classification in arbitrary dimensions $d$ is given by $\pi_0(C_{q-d})$ or $\pi_0(R_{q-d})$, as deduced from zero-dimensional classifying spaces $C_q$ or $R_q$ by K-theory.}
\label{tab_classification_C,P,T_free_fermion}
\end{table*}


Based on topological CPT theorem, some symmetry classes, defined as topological CPT-equivalent symmetry classes here, possess the same classification. For example, symmetry classes
\begin{align}
\Gamma_{-}(T),\ \Gamma_{+}(CP),\ \Gamma^{-}_{-+}(T, CP),
\label{CPT_eq_classes_AII}
\end{align}
all have the same classification. Here we use notations
\begin{align} 
&\Gamma_{S_{\mathrm{g}}}(g),\ \Gamma^{S_{\mathrm{g}_1, \mathrm{g}_2}}_{S_{\mathrm{g}_1} S_{\mathrm{g}_2}}(g_1, g_2), \n
&\Gamma^{S_{\mathrm{g}_1, \mathrm{g}_2} S_{\mathrm{g}_2, \mathrm{g}_3} S_{\mathrm{g}_3, \mathrm{g}_1} }_{S_{\mathrm{g}_1} S_{\mathrm{g}_2} S_{\mathrm{g}_3} }(g_1, g_2, g_3),
\end{align}
to denote the symmetry classes composed of $\{g_i\}$, with signs $S_{\mathrm{g}_i, \mathrm{g}_j}$ dictating the commutation ($+$) or anticommutation ($-$) relation between $g_i$ and $g_j$.  (\ref{CPT_eq_classes_AII}) can be deduced from the following CPT-equivalent symmetry classes :
\begin{align}
\Gamma_{-}(T)&\circeq\Gamma^{-}_{-+}(T, W_0)=\Gamma^{-}_{-+}(T, W_0T)=\Gamma^{-}_{++}(W_0, W_0T) \n
&\circeq\Gamma_{+}(W_0T)=\Gamma_{+}(CP),
\end{align}
where "$\circeq$" represents the CPT-equivalence relations for symmetry classes. Similarly, as another example, symmetry classes 
\begin{align}
&\Gamma^{+}_{-+}(T, C),\ \Gamma^{+}_{++}(C, P),\ \Gamma^{-}_{-+}(T, P),\ \Gamma^{++-}_{-++}(T, C, P), \n
&\Gamma^{-}_{++}(CP, PT)=\Gamma^{-}_{-+}(TC, PT)=\Gamma^{-}_{-+}(TC, CP), 
\end{align}
all have the same classification. In Refs.\ \onlinecite{Chiu2013, Morimoto2013} the first four symmetry classes are denoted respectively as classes DIII, D$+R_+$, AII$+R_-$, and DIII$+R_{-+}$, which have the same zero-dimensional classifying space $R_3$ and thus the same classification in any dimension (using K-theory). Moreover, it can be checked that, from the results of Refs.\ \onlinecite{Chiu2013, Morimoto2013}, the topological CPT theorem indeed holds. 

We note that a natural choice of $C$, $P$, and $T$ for spin-1/2 fermions 
leads to a trivial CPT as expected.
This can be explicitly seen in the CPT-equivalence of class DIII [$\Gamma^{+}_{-+}(T, C)$] and class DIII+$R_{-+}$ [$\Gamma^{++-}_{-++}(T, C, P)$].
For spin-1/2 fermions, the symmetry operators are given as
\begin{align}
C&=\sigma_x K,& P&=s_x, & T&=is_y K,
\end{align}
where $\sigma_i, s_i$ are Pauli matrices acting on particle-hole and spin degrees of freedom. 
The parity symmetry $P$ is a reflection along $x$-direction and involves a $\pi$-rotation of spin around $x$-axis,
which is denoted by $R_{-+}$ in classification in Refs.\ \onlinecite{Chiu2013, Morimoto2013}.
Then if we consider CPT symmetry given as $W=-iCPT=\sigma_x s_z$,
$W$ satisfies commutation relations in Eq.~(\ref{trivial_CPT_commutation_relations}) and is a trivial CPT.
Thus an addition of a trivial CPT $W$ changes class DIII [$\Gamma^{+}_{-+}(T, C)$] to class DIII+$R_{-+}$ 
[$\Gamma^{++-}_{-++}(T, C, P)$],
 while it does not change the topological classification.

On the other hand, adding a nontrivial CPT symmetry $W$ [which can be represented as a combination of a trivial CPT symmetry and some (onsite) order-two unitary symmetry commuting with $\mathcal{H}$ such that $W$ changes the commutation relations (\ref{trivial_CPT_commutation_relations})] to a symmetry class would change the original classifying space. Using the result of Ref.\ \onlinecite{Morimoto2013}, we can directly obtain the change of classifying spaces for AZ symmetry classes in the presence of extra unitary symmetry $M=\gamma_1W$ (commuting with  $\mathcal{H}$). The result is summarized  in TABLE \ref{tab_classification_nontrivial_CPT_free_fermion}.

The complete classification of TIs and TSCs  (and TCIs and TCSCs if spatial symmetries such as $P$ or $CP$ are present) for non-interacting fermionic systems with T, C, P, and/or their combinations, instead of studying these symmetries separately, can also be obtained by the one with symmetry classes "AZ+CPT" [in Refs.\ \onlinecite{Chiu2013, Morimoto2013} classification for symmetry classes "AZ+P (or reflection R)" have been discussed, but some combined symmetries like CP are not included there]. Two cases are involved: (a) CPT-equivalent symmetry classes "generated" from AZ classes by a trivial CPT symmetry; (b) Other symmetry classes "generated" from AZ classes by nontrivial CPT symmetries (based on the result in Table \ref{tab_classification_nontrivial_CPT_free_fermion}). The result is summarized in Table \ref{tab_classification_C,P,T_free_fermion}.

Generally, we can also reverse an odd number of spatial coordinates as the parity $P$ and the corresponding CPT symmetry $W$, with $\tilde{k}=(-k_1,\dotsc, -k_{2n+1}, k_{2n+2}, \dotsc, k_d)$ in (\ref{CPT_sym}). In this situation, the "effective" unitary symmetry $M$ can be defined as $M=i^n\gamma_1\dotsm\gamma_{2n+1}W$,
and the commutation relations between the trivial CPT symmetry $W_0$ and other symmetries (\ref{trivial_CPT_commutation_relations})  will also change if $n$ is odd (only commutation relations with antiunitary symmetries such as $T$ and $C$ will change). Nevertheless, previous discussions on the case for $n=0$ (the same for even $n$) can be straightforwardly applied to the case for odd $n$.

As related to the results in this section, a similar but more general discussion can also be found in Ref. \onlinecite{Shiozaki2014}. The CPT symmetry defined in (\ref{CPT_sym}) here is one kind of order-two spatial symmetries defined there (on a system without defects). Therefore, classification of AZ classes in the presence of either trivial CPT (related by topological CPT theorem) or nontrivial CPT (result shifts of the classifying spaces shown in TABLE \ref{tab_classification_nontrivial_CPT_free_fermion}) discussed here can also be deduced from the general properties of the K-groups for the additional order-two spatial symmetries, as derived in Ref. \onlinecite{Shiozaki2014}.

\section{Classification of 2D interacting SPT phases: K-matrix formulation}
\label{Classification of 2D interacting SPT phases: K-matrix formulation}

In the previous sections we have discussed topological phases protected by T, C, P and/or corresponding combined symmetries and classification related by topological CPT theorem in non-interacting fermionic systems. Actually, such CPT-equivalence is expected to hold even for interacting systems of either fermions or bosons, as the original CPT theorem applies to Lorentz invariant quantum field theories with interactions. As a simple but instructive demonstration, in this section we discuss interacting SPT phases (without topological order) in two dimensions by using (Abelian) K-matrix Chern-Simons theory.


\subsection{Bulk and edge K-matrix theories incorporated with symmetries}
\label{Bulk and edge Chern-Simons theories incorporated with symmetries}

We begin with the bulk K-matrix action
$S_{\mathrm{bulk}}= \int dt d^2{\bf x}\, \mathcal{L}_{\mathrm{bulk}}$,
$\mathcal{L}_{\mathrm{bulk}}= \mathcal{L}^0_{\mathrm{bulk}} + 
\mathcal{L}^{\mathrm{ex}}_{\mathrm{bulk}}$:
\begin{align}
 \mathcal{L}_{\mathrm{bulk}}^0&=\frac{1}{4\pi} \epsilon^{\mu\nu\lambda} K_{IJ}a_{I\mu}\partial_{\nu}a_{J\lambda},  
 \nonumber \\
 \mathcal{L}_{\mathrm{bulk}}^{\mathrm{ex}}&=-
 \frac{ e Q_I }{2\pi}
 \epsilon^{\mu\nu\lambda} 
 A_{\mu}\partial_{\nu}a_{I\lambda} 
-\frac{s S_I}{2\pi}\epsilon^{\mu\nu\lambda} B_{\mu}\partial_{\nu}a_{I\lambda} 
\label{K-matrix_bulk}, 
\end{align}
where $a_{\mu}$ represents the $N$-flavors of dynamical Chern-Simons (CS) gauge fields, 
$A_{\mu}$ and $B_{\mu}$ are the external gauge potentials coupling to the electric charges and spin degrees of freedom (along some quantization axis),
$K$ is an integer-valued $N\times N$ matrix (symmetric and invertible), and $Q$ and $S$ are integer-valued $N$-components vectors representing electric charges (in unit of the electric charge $e$) and spin charges  (in unit of the spin charge $s$), respectively.
The currents in the bulk are
\begin{align}
J^{\mu}_c = \frac{e}{2\pi}\epsilon^{\mu\nu\lambda} Q_I \partial_{\nu}a_{I\lambda}, 
\quad 
J^{\mu}_s =\frac{s}{2\pi}\epsilon^{\mu\nu\lambda} S_I \partial_{\nu}a_{I\lambda},
\end{align}
where $J_c$ and $J_s$ are the total charge and spin currents, respectively.

In the bulk, we have the transformation laws under symmetries such as TRS ($\mathcal{T}$), 
PHS ($\mathcal{C}$), and PS ($\mathcal{P}$) in the $x$-direction [$g_{\mu\nu}=\mathrm{diag}\,(+,-,-)$]:
\begin{align}
\mathcal{T}&: J^{\mu}_c \rightarrow g_{\mu\nu}J^{\nu}_c, 
\quad 
J^{\mu}_s \rightarrow -g_{\mu\nu}J^{\nu}_s, 
\quad (t,{\bf x})\to (-t, {\bf x}), 
\nonumber \\
\mathcal{C}&:  J^{\mu}_c \rightarrow -J^{\mu}_c, \quad 
J^{\mu}_s \rightarrow -J^{\mu}_s,  
\nonumber \\
\mathcal{P}&: 
J^{\mu}_c \rightarrow \tilde{J}^{\mu}_c, \quad
J^{\mu}_s \rightarrow -\tilde{J}^{\mu}_s,\quad
x^{\mu}\to \tilde{x}^{\mu}, 
\label{T,C,P_trans_of_J}
\end{align}
where we have defined $\tilde{X}^{\mu} \equiv (X^0, -X^1, X^2)^T$ for any vector $X^{\mu}$. 
We assume that the gauge fields $a^{\mu}$ 
(flavor index is suppressed) obey the following transformation laws: 
\begin{align}
\mathcal{T}a^{\mu}(t,{\bf x})\mathcal{T}^{-1}&=g_{\mu\nu}U_{\mathrm{T}}a^{\nu}(-t,{\bf x}),
\nonumber \\
\mathcal{C}a^{\mu}(t,{\bf x})\mathcal{C}^{-1}&=U_{\mathrm{C}}a^{\mu}(t,{\bf x}),
\nonumber \\
\mathcal{P}a^{\mu}(x)\mathcal{P}^{-1}&=U_{\mathrm{P}}\tilde{a}^{\mu}(\tilde{x}), \label{T,C,P_trans_of_a}
\end{align}
where $U_{\mathrm{T}}, U_{\mathrm{C}},$ and $U_{\mathrm{P}}$ are 
integer-valued $N\times N$ matrices, 
then we can find these matrices of transformations by the symmetries of the theory. 
However,
the above symmetry transformation law
does not fully specify the symmetry properties of charged excitations. \cite{LevinStern2012}

A convenient way to complete the description of the symmetries is to consider the action at the edge
$S_{\mathrm{edge}}= \int dt dx\, \mathcal{L}_{\mathrm{edge}}$,
$\mathcal{L}_{\mathrm{edge}}= \mathcal{L}^0_{\mathrm{edge}} + 
\mathcal{L}^{\mathrm{ex}}_{\mathrm{edge}}$:
\begin{align}
\mathcal{L}_{\mathrm{edge}}^0&=
\frac{1}{4\pi}\left(
K_{IJ}\partial_t\phi_I\partial_x\phi_{J}-V_{IJ}\partial_x\phi_I\partial_x\phi_{J} 
\right),  
\nonumber \\
\mathcal{L}_{\mathrm{edge}}^{\mathrm{ex}}&=
\frac{e}{2\pi}\epsilon^{\mu\nu} Q_I \partial_{\mu}\phi_I A_{\nu}+\frac{s}{2\pi}\epsilon^{\mu\nu} S_I \partial_{\mu}\phi_I B_{\nu},
\label{K-matrix_edge} 
\end{align}
which is derived from the usual bulk-edge correspondence of the bulk Chern-Simons theory (\ref{K-matrix_bulk}). Now the currents in the edge theory are
\begin{align}
j^{\mu}_c= \frac{e}{2\pi}\epsilon^{\mu\nu} Q_I \partial_{\nu}\phi_I,
\quad
j^{\mu}_s =\frac{s}{2\pi}\epsilon^{\mu\nu} S_I \partial_{\nu}\phi_I.
\end{align}
Under $\mathcal{T}$, $\mathcal{C}$, and $\mathcal{P}$, 
the edge currents transform similarly as the bulk currents. 
The transformation law for the bosonic fields $\phi_I$ 
is translated from the gauge fields $a^{\mu}$ 
(\ref{T,C,P_trans_of_a}), 
with additional (constant) phases: 
\begin{align}
\mathcal{T}\phi(t, x)\mathcal{T}^{-1}&=-U_{\mathrm{T}}\phi(-t,x)+\delta \phi_{\mathrm{T}},
\nonumber \\
\mathcal{C}\phi(t, x)\mathcal{C}^{-1}&=U_{\mathrm{C}}\phi(t,x)+\delta \phi_{\mathrm{C}},
\nonumber \\
\mathcal{P}\phi(t, x)\mathcal{P}^{-1}&=U_{\mathrm{P}}\phi(t,-x)+\delta \phi_{\mathrm{P}}. 
\label{T,C,P_trans_of_phi}
\end{align}
The minus sign in front of $U_{\mathrm{T}}$ is just a convention for a antiuntary operator.
For the edge theory (\ref{K-matrix_edge}) with a general symmetry group $G$ 
that has elements as combinations of $\mathcal{T}$, $\mathcal{C}$, and $\mathcal{P}$, and/or U(1) symmetries, we have 
\begin{align}
 \mathcal{G}{S}_{\mathrm{edge}}\mathcal{G}^{-1}={S}_{\mathrm{edge}}, \ \forall\mathcal{G}\in G, \label{S_edge_with_symG}
\end{align}
with the chiral boson fields transformed as 
\begin{align}
&\mathcal{G}\phi\mathcal{G}^{-1}=\alpha_{\mathrm{G}} U_{\mathrm{G}}\phi+\delta \phi_{\mathrm{G}},  \ \forall\mathcal{G}\in G, \label{G_trans}
\end{align}
where $\alpha_{\mathrm{G}}=1\ (-1)$ represents an unitary (antiunitary) operator $\mathcal{G}$.
Specifically, for TRS, PHS, and PS, 
(\ref{S_edge_with_symG}) gives the constraints for the matrices $U_{\mathrm{T}}$, $U_{\mathrm{C}}$, $U_{\mathrm{P}}$, and charge and spin vectors $Q$, $S$
\begin{align}
 \text{TRS}\, &: \  U_{\mathrm{T}}^TKU_{\mathrm{T}}=-K, \n
& \left(I_N+ U_{\mathrm{T}}^T\right)Q=0,  
\quad \left(I_N-U_{\mathrm{T}}^T \right)S=0, \n
\text{PHS}\, &: \  U_{\mathrm{C}}^TKU_{\mathrm{C}}=K, \n
&  \left(I_N+ U_{\mathrm{C}}^T\right)Q=0,  
 \quad 
 \left(I_N+U_{\mathrm{C}}^T \right)S=0, \n
 \text{PS}\, &: \  U_{\mathrm{P}}^TKU_{\mathrm{P}}=-K, \n
&  \left(I_N+ U_{\mathrm{P}}^T\right)Q=0,
\quad
  \left(I_N-U_{\mathrm{P}}^T \right)S=0,  \label{K-matrix_sym_constraints}
\end{align}
where $I_N$ is the $N\times N$ identity matrix. Cases of the combined symmetries like $\mathcal{CP}$ are straightforward. 

%
For the charge and spin U(1) symmetries of the system,
\begin{align}
\mathcal{U}_c\phi(t, x)\mathcal{U}_c^{-1}&=\phi(t,x)+\delta \phi_c,
\nonumber \\
\mathcal{U}_s\phi(t, x)\mathcal{U}_s^{-1}&=\phi(t,x)+\delta \phi_s,
\label{Uc/s_trans}
\end{align}
where
$\mathcal{U}_c\equiv e^{i\theta_c \int dx\, j^0_c/e}$ 
and $\mathcal{U}_s\equiv e^{i\theta_s \int dx\, j^0_s/s}$ 
are the charge and spin U(1) transformations, respectively,
and 
the corresponding phase shifts are given by
\begin{align}
\delta \phi_c &=\theta_cK^{-1}Q, 
\quad
\delta \phi_s =\theta_sK^{-1}S. \label{delta_phi_U}
\end{align}

On the other hand, the phases $\delta \phi$ in 
Eq.\ (\ref{G_trans}) are determined by how the local quasiparticle excitations, which are described by normal-ordered vertex operators $\ddagger e^{il^T\phi} \ddagger =\ddagger e^{i\Lambda^TK\phi}\ddagger \equiv \ddagger e^{i\Theta(\Lambda)}\ddagger $, with $l=K\Lambda$ and $\Lambda$ being integer $N$-components vectors, under the symmetry transformations. That is, the transformation law for $\Theta(\Lambda)$ 
is determined by the algebraic relations of the underlying symmetry operators. To classify these discrete $Z_2$ symmetries for interacting systems (beyond the single-particle picture),
we constrain the symmetry operators by the following algebraic relations: 
\begin{align}
&\mathcal{G}_i^2=S_{\mathrm{G}_i}^{N_f}, \quad \forall \mathcal{G}_i\in G; \nonumber\\
&\mathcal{G}_i\mathcal{G}_j\mathcal{G}_i^{-1}\mathcal{G}_j^{-1}=S_{\mathrm{G}_i,\mathrm{G}_j}^{N_f}, 
\quad \forall \mathcal{G}_i, \mathcal{G}_j\in G, 
\label{algebraic_relation}
\end{align}
where $S$ has values $\pm1$. In a bosonic system, the operator $S^{N_f}$ (subscript omitted) is just the identity 1. In a fermionic system, $S^{N_f}$ can be either the identity $1$ or the fermion number parity operator $\mathcal{P}_f\equiv (-1)^{N_f}$ 
(i.e., symmetries are realized projectively)
, where $N_f$ is the total fermion number operator. Since all $\mathcal{T}$, $\mathcal{C}$, and $\mathcal{P}$ (and of course the combined symmetries) commute with $\mathcal{P}_f$, we have $S_{\mathrm{G}_1,\mathrm{G}_2}=S_{\mathrm{G}_2,\mathrm{G}_1}$ for any two symmetry operators $\mathcal{G}_1$ and $\mathcal{G}_2$. 

In the presence of U(1) symmetries, the algebraic relations (\ref{algebraic_relation}) for fermionic systems might be "gauge equivalent" through the redefinition of the discrete symmetry $\mathcal{G}$ to $\mathcal{U}_{\alpha}\mathcal{G}$, where $\mathcal{U}_{\alpha}$ can be charge or spin U(1) with some phase $\alpha$. Denoting $\mathcal{G}$ and $\widetilde{\mathcal{G}}$ the discrete symmetries with the relations to $\mathcal{U}_{\alpha}$ as $\mathcal{G}\mathcal{U}_{\alpha}\mathcal{G}^{-1}=\mathcal{U}_{\alpha}$ and $\widetilde{\mathcal{G}}\mathcal{U}_{\alpha}\widetilde{\mathcal{G}}^{-1}=\mathcal{U}^{-1}_{\alpha}$ (specific symmetries are discussed in Appendix \ref{Algebraic relations of symmetry operators}), respectively, we have
\begin{align}
&\mathcal{G}^2 \rightarrow \mathcal{U}^2_{\alpha}\mathcal{G}^2, \quad 
\widetilde{\mathcal{G}}^2 \rightarrow \widetilde{\mathcal{G}}^2, \n
&\mathcal{G}_i\mathcal{G}_j\mathcal{G}_i^{-1}\mathcal{G}_j^{-1} \rightarrow \mathcal{G}_i\mathcal{G}_j\mathcal{G}_i^{-1}\mathcal{G}_j^{-1}, \n
&\widetilde{\mathcal{G}}_i\widetilde{\mathcal{G}}_j\widetilde{\mathcal{G}}_i^{-1}\widetilde{\mathcal{G}}_j^{-1} \rightarrow \mathcal{U}^2_{\alpha_i}\mathcal{U}^{-2}_{\alpha_j}\widetilde{\mathcal{G}}_i\widetilde{\mathcal{G}}_j\widetilde{\mathcal{G}}_i^{-1}\widetilde{\mathcal{G}}_j^{-1}, \n
&\mathcal{G}_i\widetilde{\mathcal{G}}_j\mathcal{G}_i^{-1}\widetilde{\mathcal{G}}_j^{-1} \rightarrow \mathcal{U}^2_{\alpha_i}\mathcal{G}_i\widetilde{\mathcal{G}}_j\mathcal{G}_i^{-1}\widetilde{\mathcal{G}}_j^{-1}, 
\label{fermion_U(1)_gauge_redundancy}
\end{align}
as we redefine $\mathcal{G}_i$, $\widetilde{\mathcal{G}}_j$ to $\mathcal{U}_{\alpha_i}\mathcal{G}_i$, $\mathcal{U}_{\alpha_j}\widetilde{\mathcal{G}}_j$. Therefore, the signs $S_{\mathrm{G}_i}$ and $S_{\mathrm{G}_i,\mathrm{G}_j}$ that characterize the symmetry group $G$ can be fixed to be either $1$ or $-1$ (for fermionic systems) if appropriate phases $\alpha$'s are chosen, as the U(1) symmetry is present. In such cases, some symmetry groups with different algebraic relations might correspond to the same physical SPT phase. For bosonic systems, on the other hand, such U(1) gauge redundancy of symmetry operators arises in a more subtle way, since the algebraic relations between symmetries on bosons are "trivial" (as the signs $S=1$).

From the symmetry constraints (\ref{S_edge_with_symG}) and (\ref{algebraic_relation}), we can determine how the chiral boson fields transform under the symmetry group $G$, i.e., the data $\{ U_{\mathrm{G}},  \delta \phi_{\mathrm{G}}\}$. To be more explicit, see Appendix \ref{Calculations of symmetry transformations and SPT phases in K-matrix theories}. Note that we also have the (gauge) equivalence for the forms of these symmetry transformations\cite{Lu2012, LevinStern2012} for all physically equivalent K-matrix theories:
\begin{align} 
\{U_{\mathrm{G}}, \delta\phi_{\mathrm{G}}\} &\rightarrow   \{X^{-1}U_{\mathrm{G}}X, \ X^{-1}\left(\delta\phi_{\mathrm{G}}- \alpha_{\mathrm{G}}\Delta\phi + U_{\mathrm{G}}\Delta\phi \right)\}, \nonumber \\
&\text{if} \ X \in GL(N,\mathbb{Z}),\quad \det(X)=\pm 1,
\end{align}
where $\alpha_{\mathrm{G}}=1 \ (-1)$ if $\mathcal{G}$ is an unitary (antiunitary) operator. This means we can choose some $X$ and $\Delta\phi$ to fix $\{U_{\mathrm{G}_i}, \delta\phi_{\mathrm{G}_i}\}$ to the inequivalent forms of transformations.

\paragraph{Statistical phase factors of vertex operators 
under symmetry transformation}

The edge theory (\ref{K-matrix_edge}) is quantized according to the equal-time commutators
\begin{align}
 \label{CCR_K-matrix_1}
[\phi^I(t, x), \phi^J(t, x')]=-i\pi\left[
 (K^{-1})^{IJ}\text{sgn}(x-x')+\Theta^{IJ} \right],
\end{align}
where the Klein factor
\begin{align}
\Theta^{IJ}:= (K^{-1})^{IK} \left[\text{sgn}(K-L)(K_{KL}+Q_KQ_L) \right](K^{-1})^{LJ}
\end{align}
is included to ensure that local excitations satisfy the proper commutation relations when $x\neq x'$ and $I\neq J$:
\begin{align}
&
[(K \phi)_I(t, x), 
(K\phi)_J(t, x')]
\nonumber \\
&\quad
=-i\pi\text{sgn}(I-J)Q_IQ_J+2\pi i N_{IJ}, \label{CCR_K-matrix_2}
\end{align}
where $N_{IJ}$ is the component of some integer matrix.

For any local quasiparticle excitation 
$\ddagger \exp{i\Lambda^TK\phi}\ddagger$, 
with
$\Lambda^TK\phi=
\sum_I \Lambda_I(K\phi)_I\equiv \sum_I \theta_I$, the symmetry transformation $\mathcal{G}$ acts as
\begin{align}
&\quad 
\mathcal{G}
 \ddagger 
 e^{i\Lambda^TK\phi}
 \ddagger 
 \mathcal{G}^{-1} 
=\mathcal{G}
\ddagger e^{\sum_I i\theta_I}
\ddagger \mathcal{G}^{-1} 
\nonumber \\
&=
\mathcal{G}\ddagger 
{\prod_I}'e^{i\theta_I}\cdot 
e^{-\frac{1}{2}\sum_{I<J} [i\theta_I, i\theta_J]} 
\ddagger \mathcal{G}^{-1} 
\nonumber \\
&\equiv  
\ddagger 
{\prod_I}'e^{\mathcal{G}i\theta_I\mathcal{G}^{-1}}
\ddagger 
\cdot e^{-\frac{1}{2}\sum_{I<J} [\mathcal{G}i\theta_I\mathcal{G}^{-1},\mathcal{G} i\theta_J\mathcal{G}^{-1}]} 
\cdot 
e^{ i \Delta \phi^{\Lambda}_{\mathrm{G}} }
\nonumber \\
&=
\ddagger 
e^{\sum_I\mathcal{G} i\theta_I\mathcal{G}^{-1}
+i\Delta\phi^{\Lambda}_\mathrm{G} }
\ddagger ,
\end{align}
where we have used the Baker-Campbell-Hausdorff formula 
(with the commutator $[i\theta_I, i\theta_J]$ being a $c$-number), 
the ordered-product "${\prod}'_I$" is defined 
as an ordered product 
in the ascending order of indices, 
and
\begin{align}
i\Delta\phi^{\Lambda}_\mathrm{G}
\equiv \frac{1}{2}\sum_{I<J} \left([\mathcal{G}i\theta_I\mathcal{G}^{-1},\mathcal{G} i\theta_J\mathcal{G}^{-1}]- \mathcal{G}[i\theta_I, i\theta_J]\mathcal{G}^{-1}\right), \label{statistical_phase_1}
\end{align}
which can be deduced from the commutator (\ref{CCR_K-matrix_2}).Note that we keep the form $\mathcal{G}[i\theta_I, i\theta_J]\mathcal{G}^{-1}$ even if $[i\theta_I, i\theta_J]$ is a $c$-number, since in general $\mathcal{G}$ can be an antiunitary operator 
(e.g. TRS).
On the other hand, 
\begin{align}
\mathcal{G}e^{i\Lambda^TK\phi} \mathcal{G}^{-1}&=e^{\mathcal{G}i\Lambda^TK\phi\mathcal{G}^{-1}}=e^{\mathcal{G}\left(\sum_I i\theta_I\right)\mathcal{G}^{-1}},
\end{align}
so we have
\begin{align}
\mathcal{G}\left(i\Lambda^TK\phi\right)\mathcal{G}^{-1}
&=
\sum_I\mathcal{G} i\theta_I\mathcal{G}^{-1}
+i\Delta\phi^{\Lambda}_\mathrm{G}  \mod 2\pi i. 
\label{statistical_phase_2}
\end{align}
This means the way the operator $\mathcal{G}$ acts on the chiral boson field $\phi$ is not always linear, because some nontrivial phase  
$\Delta\phi^{\Lambda}_\mathrm{G}$ ($\neq 2n\pi$) might arise. 
In bosonic systems, the phase is always the multiple of $2\pi $, corresponding to the Bose statistics, and thus we can ignore it (in this case $\mathcal{G}$ is linear in $\phi$). 
In fermionic systems,
however, we must be careful with the phase, 
which might be nontrivial, 
because of the Fermi statistics.

  For the unitary operator 
  $\mathcal{G}=\mathcal{G}_e$ ($e$ means "identity element") 
  that has the form of the identity or the fermion number parity operator $\mathcal{P}_f$ 
  [such as $\mathcal{G}_i^2$ and $\mathcal{G}_i\mathcal{G}_j\mathcal{G}_i^{-1}\mathcal{G}_j^{-1}$ in Eq.\ (\ref{algebraic_relation})], 
  we have $\mathcal{G}_ei\theta_j\mathcal{G}_e^{-1}=i\theta_j+$ const. 
  In this case the phase $i\Delta\phi^{\Lambda}_{\mathrm{G}_e}$ vanishes (in the sense of mod $2\pi i$) and thus such $\mathcal{G}_e$ is linear on $\phi$. This fact tells us that instead of specifying the transformation properties $\mathcal{G}_e(i\Lambda^TK\phi)\mathcal{G}_e^{-1}$ for all local quasiparticle excitations, we can solely consider the transformations $\mathcal{G}_ei\phi\mathcal{G}_e^{-1}$ to determine the phase $\delta \phi$, as defined in (\ref{G_trans}), under symmetry transformations.

\subsection{Edge stability criteria for SPT Phases}
\label{Edge Stability and Criteria for SPT Phases}

In this section, we briefly discuss the edge stability and criteria for SPT phases. 
\cite{Neupert2011, LevinStern2012, Lu2012}

The general terms of interactions (perturbations from the tunneling and scattering  process of local excitations) for the 1+1D edge theory are the bosonic condensations:
\begin{align} 
{S}^{\mathrm{int} }_{\mathrm{edge}}=&\sum^{\mathrm{bosonic}}_{\Lambda}
\int dtdx\n
&\,
U_{\Lambda}(t, x)\cos
\left[ 
 \Lambda^TK\phi(t, x) + \alpha_{\Lambda}(t, x) \right]. 
 \label{interaction_edge}
\end{align}
Note that an integer vector $\Lambda=(\Lambda_1, \cdots, \Lambda_N)^T$ are bosonic (i.e. excitation $\ddagger e^{i\Lambda^TK\phi} \ddagger$ is a boson) if $\Lambda$ satisfy $\pi\Lambda^TK\Lambda=0 \mod 2\pi$. In the discussion of this paper we assume the coupling $U_{\Lambda}$ is a constant (independent of $t$ and $x$). In the absence of any symmetry, a collection of bosonic $\{\Lambda_a\}$, which satisfies Haldane's null vector condition\cite{Haldane1995}
\begin{align} 
\Lambda^T_aK\Lambda_b=0, \ \forall a, b=1,\dotsc, N/2
\label{Haldane's_null_vector}
\end{align}
(here $N$ is even since we focus on 
the K-matrix with equal numbers of positive and negative eigenvalues), can condense (be localized) with various (classical) expectation values by adding the corresponding ${S}^{\mathrm{int} }_{\mathrm{edge}}$ to $S_{\mathrm{edge}}$. That is, the edge can be gapped by such perturbations (thus the gapless edge modes are unstable), and the phase is (topologically) trivial.

Such gapping mechanism might be forbidden by symmetries, resulting nontrivial SPT phases, which can not be transformed adiabatically from the trivial phase within the symmetry constraints, even in the presence of interactions. That is, if any possible interaction ${S}^{\mathrm{int} }_{\mathrm{edge}}$ can not be added to the edge theory (\ref{K-matrix_edge}) without breaking some symmetry, both explicitly and spontaneously, then the system manifests a nontrivial SPT phase (protected by these symmetries). To be more specific, one wants to check that whether the following conditions are all satisfied:

(i) There exists symmetry preserving ${S}^{\mathrm{int} }_{\mathrm{edge}}$ with a set of Haldane's null vector $\{\Lambda_a\}$.

(ii) All edge states can be gapped without breaking any symmetry spontaneously. This can be checked whether all the elementary bosonic variables $\{v^T_a\phi\}$, with 
\begin{align} 
v_a=\frac{l_a}{gcd(l_{a,1}, l_{a,2}, \dotsc, l_{a,N})}, \ \forall a,
\label{elementary_bosonic_variables}
\end{align}
which are generated from any collections of linear combinations of $\{l^T_a\phi=(K\Lambda)^T_a\phi\}$, condense without breaking any symmetry.

If both conditions are satisfied, the phase is (topologically) trivial. Otherwise, the phase is a SPT phase.

%

\subsection{Classification of SPT phases by K-matrix thoeries: CPT-equivalent SPT phases and dual SPT phases}

By studying the stability/gappability of the 1D edge of K-matrix theories, we can classify 2D interacting SPT phases with T, C, P,  the combined symmetries, and/or U(1) symmetries, for either bosonic or fermionic systems. Here we consider non-chiral SPT phases described by K-matrices with even dimensions ($N$ is even) in the absence of topological order ($|\det K|=1$): the canonical forms of generic K-matrix are
\begin{align}
K=\sigma_x\oplus\sigma_x\oplus\cdots\oplus\sigma_x=I_{N/2}\otimes\sigma_x
\label{integer_boson_K}
\end{align}
in bosonic systems, and
\begin{align}
K=\sigma_z\oplus\sigma_z\oplus\cdots\oplus\sigma_z=I_{N/2}\otimes\sigma_z
\label{integer_fermion_K}
\end{align}
in fermionic systems.\cite{LevinStern2012, Lu2012, Wang2012, Lu2013}

Similar to the case of non-interacting fermionic systems discussed in the previous sections, we also have CPT-equivalence among interacting bosonic and fermionic SPT phases with these discrete symmetries. Combining symmetries $\mathcal{T}$, $\mathcal{C}$, $\mathcal{P}$, we define the CPT symmetry by an antiunitary operator $\mathcal{W}$, 
\begin{align}
\mathcal{W}\phi(t, x)\mathcal{W}^{-1}=-U_{\mathrm{W}}\phi(-t, -x)+\delta\phi_{\mathrm{W}},
\label{CPT_trans}
\end{align}
which satisfies
\begin{align}
\mathcal{W}{S}_{\mathrm{edge}}\mathcal{W}^{-1}&={S}_{\mathrm{edge}} \n
\Rightarrow\quad U_{\mathrm{W}}^TKU_{\mathrm{W}}&=K, \n
\quad\left(I_N+ U_{\mathrm{W}}^T\right)Q=0, \quad &\left(I_N+U_{\mathrm{W}}^T \right)S=0.
\label{K-matrix_CPT_constraints}
\end{align}
Imposing $\mathcal{W}$ to the system with some existed symmetries would or would not change the classification of the original (interacting) topological phase. As the latter case, there exists a "trivial" CPT operator $\mathcal{W}=\mathcal{W}_0$ such that the 1D edge theory with any gapping interactions ${S}^{\mathrm{int} }_{\mathrm{edge}}$ [with a set of Haldane's null vectors (\ref{Haldane's_null_vector})] is invariant under $\mathcal{W}_0$, and thus the corresponding 2D topological phase would not be "additionally" protected by the presence of such trivial CPT symmetry. Therefore, we have the following statement:

{\it {\bf Topological CPT theorem for interacting fermionic and bosonic non-chiral SPT phases in two dimensions:} 
Let $\{\mathcal{G}_i\}$ be a set of symmetries (can be a null set) composed of $\mathcal{T}$, $\mathcal{C}$, $\mathcal{P}$, 
the combined symmetries, and/or any (order-two) onsite unitary symmetries [including U(1) symmetries].  
Then for 2D interacting fermionic or bosonic systems (in the absence of topological order) 
described by K-matrix theories, 
there exists a "trivial" CPT operator $\mathcal{W}=\mathcal{W}_0$ , 
with $\mathcal{W}^2_0$ being identity operator and relations to other symmetries (if present) as 
$\mathcal{W}_0\mathcal{G}_i\mathcal{W}_0^{-1}\mathcal{G}_i^{-1}=1$ if $U_{\mathrm{G}_i}^TKU_{\mathrm{G}_i}=K$ 
and
$\mathcal{W}_0\mathcal{G}_i\mathcal{W}_0^{-1}\mathcal{G}_i^{-1}=(-1)^{N_f}$ if $U_{\mathrm{G}_i}^TKU_{\mathrm{G}_i}=- K$
for fermionic systems (for bosonic systems the algebraic relations is "trivial"),
such that the topological phases protected by $\{\mathcal{G}_i\}$ 
and the topological phases protected by $\{\mathcal{G}_i, \mathcal{W}_0\}$ possess the same classification.
}

  The proof of the above theorem is left to Appendix \ref{Proof of topological CPT theorem for interacting fermionic and bosonic non-chiral SPT phases in two dimensions}. Now, from this theorem we can define CPT-equivalent symmetry groups/classes or SPT phases for interacting bosonic and fermionic systems, as we did similarly for non-interacting fermionic systems in Sec. \ref{Topological CPT theorem and topological CPT-equivalent symmetry classes}. As an example again, the (bosonic or fermionic) topological phases protected by both TRS and charge U(1) symmetry \cite{LevinStern2009, Neupert2011, LevinStern2012} and the topological phases protected by both CP and charge U(1) symmetry \cite{Hsieh2014}  possess the same classification, even in the presence of interactions (to be more precise, we have the CPT-equivalent symmetry classes $\{\mathcal{T}, \mathcal{U}_c|\ \mathcal{T}^2=(\pm 1)^{N_f}\} \circeq \{\mathcal{CP}, \mathcal{U}_c|\ \mathcal{CP}^2=(\mp 1)^{N_f}\}$ for interacting fermionic systems, where "$\circeq$" represents the CPT-equivalence relations). 

Through the trivial CPT symmetry $\mathcal{W}_0$,  any nontrivial CPT symmetry $\mathcal{W}$ can be expressed as the combination of $\mathcal{W}_0$ and some onsite unitary $Z_2$ symmetry $\mathcal{M}$. So imposing a nontrivial CPT symmetry to a system is identical to imposing such unitary symmetry to this system, which might change the classification of the original SPT phases with existed symmetries.

Besides CPT-equivalence for SPT phases, there are other "dualities" between SPT phases: classification of the $\{\mathcal{G}_i\}$-protected topological phases and the $\{\widetilde{\mathcal{G}}_i\}$-protected phases are the same, where $\{\mathcal{G}_i\}$ and  $\{\widetilde{\mathcal{G}}_i\}$ are dual symmetries (see the discussion later). An example is the T-duality between the topological phases protected by CP and $\mathrm{U}(1)_c$ and the topological phases protected by P and $\mathrm{U}(1)_s$, as we discussed for the non-interacting fermionic systems in previous sections. In K-matrix formalism, this can be observed by the transformation law for the two U(1) currents under the discrete $Z_2$ symmetries (\ref{T,C,P_trans_of_J}). We can see that the way $J_{c/s}^{\mu}$ transforms under $\mathcal{CP}$ is the same as the way $J_{s/c}^{\mu}$ transforms under $\mathcal{P}$. In general, we have the following duality between these discrete symmetries as we exchange charge and spin U(1) symmetries:
\begin{align}
&\mathcal{T} \leftrightarrow \mathcal{TC}, \quad\mathcal{P} \leftrightarrow \mathcal{CP}, \n
\quad\mathcal{C} \leftrightarrow \mathcal{C}, &\quad\mathcal{PT} \leftrightarrow \mathcal{PT}, \quad\mathcal{CPT} \leftrightarrow \mathcal{CPT}.
\label{TCP_duality_K-matrix}
\end{align}

Actually, for the bulk K-matrix theory (\ref{K-matrix_bulk}) [the same prospect for the edge theory (\ref{K-matrix_edge}) by the bulk-edge correspondence] we can rewrite it as

\begin{align}
&\mathcal{L}_{\mathrm{bulk}}=\frac{1}{4\pi} \epsilon^{\mu\nu\lambda} \widetilde{K}_{IJ}\widetilde{a}_{I\mu}\partial_{\nu}\widetilde{a}_{J\lambda}
-\widetilde{J}^{\mu}_cA_{\mu}-\widetilde{J}^{\mu}_sB_{\mu}, \n
&\widetilde{K}=X^TKX,\quad \widetilde{a}=X^{-1}a, \quad \widetilde{Q}=X^TQ, \quad\widetilde{S}=X^TS, \n
&X \in GL(N,\mathbb{Z}), \quad \det(X)=\pm 1.
\label{dual_K-matrix_bulk}
\end{align}
There are two interpretations for Eq. (\ref{dual_K-matrix_bulk}), corresponding to physical equivalent theories described in different ways [passive or active transformation by $X\in GL(N,\mathbb{Z})$]:

(i) It is nothing but field redefinitions (change of basis); the relabeled gauge fileds $\widetilde{a}_I$ describe the same degrees of freedom as $a_I$. 

(ii) The gauge fileds $a_I$ are transformed to the dual gauge fields $\widetilde{a}_I$, which characterize different degrees of freedom (as $X$ is not the identity matrix) as the original ones (e.g. charge-vortex duality). The dual theory describes the same physical system.

As symmetries are present, in description (i) $\{\mathcal{G}_i\}$ (defined to be on $a_I$) and $\{\widetilde{\mathcal{G}}_i\}$ (defined to be on $\widetilde{a}_I$) are identical, while in description (ii) $\{\mathcal{G}_i\}$ and $\{\widetilde{\mathcal{G}}_i\}$ "look" different (e.g. $\mathcal{P}$ and $\mathcal{CP}$), as they act on different degrees of freedom. However, they both describe the same "symmetry of the system".

On the other hand, if we "rotate" every term in (\ref{K-matrix_bulk}) by $X$ except the gauge fields $a_I$ [i.e. remove the tilde of $a$ in (\ref{dual_K-matrix_bulk})], we will obtain a dual theory that describes a different physical system or SPT phase. For example, if we take $X=K$ for the K-matrix described by (\ref{integer_boson_K}) or (\ref{integer_fermion_K}), the dual theory, obtained from a theory with symmetries $\{\mathcal{G}_i\}$,  will describe a system with dual symmetries $\{\widetilde{\mathcal{G}}_i\}$ by the correspondence (\ref{TCP_duality_K-matrix} together with charge-spin exchange.
Since the criteria for arguing a SPT phase is independent of how we choose the gauge $X$ and how we label the field operators, the dual SPT phase has exactly the same classification as the original SPT phase.

In the following subsections, we give a complete classification for K-matrix theories with T, C, P, the combined symmetries, and/or U(1) symmetries, for both bosonic and fermionic systems. We can see that CPT-equivalence and T-duality hold exactly through the classification tables for 2D interacting SPT phases.

\subsubsection{K-matrix classification of bosonic non-chiral SPT phases}

For bosonic K-matrix theories with T, C, P, the combined symmetries, and/or U(1) symmetries, 
it is sufficient to implement the non-chiral short-range entangled states by just considering 
the $2\times 2 $ K-matrix with determinant $\det(K)=(-1)^{\mbox{dim}(K)/2}=-1$. From the canonical form of bosonic K-matirx (\ref{integer_boson_K}) we have $K=\sigma_x$. The detail for calculating symmetry transformations and their corresponding SPT phases is left to Appendix  \ref{Calculations of symmetry transformations and SPT phases in K-matrix theories}. Here we summarize the results in TABLE \ref{tab_boson_nonchiral}.

In TABLE \ref{tab_boson_nonchiral} we show classification of 2D bosonic non-chiral SPT phases for $\{K, Q, S\}=\{\sigma_x, (0, 1)^T, (1, 0)^T\}$, as we focus on cases with vanishing $Q^TKQ$ and $S^TKS$ (so bosonic quantum Hall systems are not included in our discussion here). There are some remarks for TABLE \ref{tab_boson_nonchiral}:

\
\begin {table}[th]
\centering
\begin{tabular}{|c||c|c|c|}
  \hline
\multirow{3}{*}{Sym. group} & \multicolumn{3}{ |c| }{Classification of 2D bosonic non-chiral}  \\ 
  & \multicolumn{3}{ |c| }{ SPT phases } \\
\cline{2-4} &  No U(1)'s   & $\mathrm{U}(1)_{c}$ is present & $\mathrm{U}(1)_{s}$ is present \\
  \hline \hline 
  $Z_2^{\mathrm{T}}$ & 0 &  $\mathbb{Z}_2$ & 0  \\
  \cline{1-4} 
  $Z_2^{\mathrm{C}}$ & 0 & 0  & 0  \\
  \cline{1-4}
  $Z_2^{\mathrm{P}}$ & 0 & 0 & $\mathbb{Z}_2$ \\
  \cline{1-4} 
$Z_2^{\mathrm{CP}}$ & 0 &  $\mathbb{Z}_2$ & 0  \\
  \cline{1-4} 
  $Z_2^{\mathrm{PT}}$ & 0 & 0  & 0  \\
  \cline{1-4}
  $Z_2^{\mathrm{TC}}$ & 0 & 0 & $\mathbb{Z}_2$ \\
  \cline{1-4} 
$Z_2^{\mathrm{CPT}}$ & $\mathbb{Z}_2$ &  $\mathbb{Z}_2$ & $\mathbb{Z}_2$\\ 
  \cline{1-4} 
$Z_2^{\mathrm{T}}\times Z_2^{\mathrm{C}}$  & $\mathbb{Z}_2$ &  $\mathbb{Z}_2$ & $\mathbb{Z}_2$ \\ 
  \cline{1-4}
$Z_2^{\mathrm{C}}\times Z_2^{\mathrm{P}}$  & $\mathbb{Z}_2$ &  $\mathbb{Z}_2$ & $\mathbb{Z}_2$ \\ 
  \cline{1-4} 
$Z_2^{\mathrm{P}}\times Z_2^{\mathrm{T}}$  & $\mathbb{Z}_2$ & $\mathbb{Z}_2$ & $\mathbb{Z}_2$ \\ 
  \cline{1-4}
$Z_2^{\mathrm{CP}}\times Z_2^{\mathrm{PT}}$ & $\mathbb{Z}_2$ &  $\mathbb{Z}_2$ & $\mathbb{Z}_2$ \\ 
  \cline{1-4}
$Z_2^{\mathrm{T}}\times Z_2^{\mathrm{CP}}$  & $\mathbb{Z}_2^2$ & $\mathbb{Z}_2^2$ & $\mathbb{Z}_2$ \\ 
  \cline{1-4}
$Z_2^{\mathrm{C}}\times Z_2^{\mathrm{PT}}$  & $\mathbb{Z}_2$ &  $\mathbb{Z}_2$ & $\mathbb{Z}_2$ \\ 
  \cline{1-4}
$Z_2^{\mathrm{P}}\times Z_2^{\mathrm{TC}}$  & $\mathbb{Z}_2^2$ & $\mathbb{Z}_2$ & $\mathbb{Z}_2^2$ \\ 
  \cline{1-4}
$Z_2^{\mathrm{T}}\times Z_2^{\mathrm{C}}\times Z_2^{\mathrm{P}}$  & $\mathbb{Z}_2^4$ &  $\mathbb{Z}_2^2$ & $\mathbb{Z}_2^2$ \\ 
  \cline{1-4}
\end{tabular}
\caption{Classification of 2D interacting bosonic non-chiral SPT phases with symmetry groups generated by $\mathcal{T}$, $\mathcal{C}$, $\mathcal{P}$, and/or U(1) symmetries.  Each nontrivial SPT phase in this table is implemented by a $2\times2$ K-matrix: $\{K, Q, S\}=\{\sigma_x, (0, 1)^T, (1, 0)^T\}$. Classification shown in this table has removed U(1) gauge redundancy.}
\label{tab_boson_nonchiral}
\end{table}

(i) For each symmetry group listed in TABLE \ref{tab_boson_nonchiral}, except groups $Z_2^{\mathrm{C}}$ and $Z_2^{\mathrm{PT}}$, there are multiple choices for physically inequivalent realizations of the symmetries, which are not characterized by the bosonic algebraic relations. For example, in symmetry group $Z_2^{\mathrm{CP}}$ the CP symmetries can be represented by $\{U_{\mathrm{CP}}, \delta \phi_{\mathrm{CP}}\}= \{-\sigma_z, (0, 0)^T\}$ or $\{-\sigma_z, (0, \pi)^T\}$, as they are physically inequivalent in the absence U(1) symmetries. All inequivalent choices should be considered in each symmetry group, and their corresponding nontrivial SPT phases will form an Abelian group. 

(ii) When U(1) symmetry [either $\mathrm{U}(1)_{c}$ or $\mathrm{U}(1)_{s}$] is present, there might be gauge redundancy among symmetry transformations (as discussed in Sec. \ref{Bulk and edge Chern-Simons theories incorporated with symmetries}). For the example in (i), the two representations of $\mathcal{CP}$ are gauge equivalent when $\mathrm{U}(1)_{s}$ is present, as we can redefine $(\mathcal{CP})'=\mathcal{U}_s(\alpha)\cdot\mathcal{CP}$ with a phase $\alpha=\pi$ to change one representation to another. The classification shown in TABLE \ref{tab_boson_nonchiral} has removed such U(1) gauge redundancy.

(iii) CPT-equivalence: At first glance CPT-equivalence seem violated in TABLE \ref{tab_boson_nonchiral}. For example, SPT phases with $Z_2^{\mathrm{CPT}}$ symmetry are characterized by a $\mathbb{Z}_2$ instead of a trivial classification, which is resulted in (non-chiral) SPT phases without any symmetries. Actually, both trivial  and nontrivial CPT symmetries can be realized from $Z_2^{\mathrm{CPT}}= Z_2^{\mathrm{W}_0\mathrm{M}}$, where $\mathcal{W}_0$ is the trivial CPT and  $\mathcal{M}$ is some onsite unitary $Z_2$ symmetry. 
The nontrivial $\mathbb{Z}_2$ SPT phase here is protected by the nontrivial CPT symmetry $\mathcal{W}_0\cdot\mathcal{M}$ with $\mathcal{M}$ represented by $\{U_{\mathrm{M}}, \delta \phi_{\mathrm{M}}\}= \{I_2, (\pi, \pi)^T\}$. Therefore, $Z_2^{\mathrm{CPT}}$ is CPT-equivalent to $Z_2^{\mathrm{M}}$, which identically gives the $\mathbb{Z}_2$ classification.\cite{Lu2012} As implied from the topological CPT theorem, adding $\mathcal{W}_0$ to some symmetry group $G$ will not change the classification of SPT phases by $G$. Similar argument applies for other symmetry groups that result nontrivial CPT symmetries (by combining the symmetries), such as $Z_2^{\mathrm{T}}\times Z_2^{\mathrm{CP}}$, which is CPT-equivalent to $Z_2^{\mathrm{T}}\times Z_2^{\mathrm{M}}$ and $Z_2^{\mathrm{CP}}\times Z_2^{\mathrm{M}}$ (both have $\mathbb{Z}_2^2$ classification; the former case is discussed in Ref. \onlinecite{Lu2012}). On the other hand, if the symmetry groups related by CPT relations do not possess nontrivial CPT symmetries, such as $\{Z_2^{\mathrm{T}}, Z_2^{\mathrm{CP}}\}$ and $\{Z_2^{\mathrm{T}}\times Z_2^{\mathrm{C}}, Z_2^{\mathrm{C}}\times Z_2^{\mathrm{P}}, Z_2^{\mathrm{P}}\times Z_2^{\mathrm{T}},  Z_2^{\mathrm{CP}}\times Z_2^{\mathrm{PT}}\}$,  they must have the same classification.

(iv) Finally, we can also see T-duality holds exactly between related symmetry groups [with the correspondence (\ref{TCP_duality_K-matrix})]
in TABLE \ref{tab_boson_nonchiral}.

\
\begin {table*}[tb]
\centering
\begin{tabular}{|c||c|c|c|c|c|}
  \hline
  \multirow{2}{*}{Sym.} & \multicolumn{3}{ |c| }{Symmetry groups for 2D nontrivial fermionic non-chiral SPT phases } & Top.  & \multirow{2}{*}{Non-int.} \\ \cline{2-4} &  No U(1)'s  & $\mathrm{U}(1)_{c}$ is present & $\mathrm{U}(1)_{s}$ is present & class. &\\
  \hline \hline 
  $\mathcal{T}$ & - &  $G_-(\mathcal{U}_c, \mathcal{T})$ & - & $\mathbb{Z}_2$ & $\Gamma_{-}(T)$\\
  \cline{1-6} 
  $\mathcal{C}$ & - & - & - & - & -\\
  \cline{1-6}
  $\mathcal{P}$ & - & - & $G_+(\mathcal{U}_s, \mathcal{P})$ &$\mathbb{Z}_2$ & $\ddagger\widetilde{\Gamma}_{+}(CP)$\\
  \cline{1-6} 
  $\mathcal{CP}$ & - &  $G_+(\mathcal{U}_c, \mathcal{CP})$ & - & $\mathbb{Z}_2$ & $\Gamma_{+}(CP)$\\
  \cline{1-6} 
  $\mathcal{PT}$ & - & - & - & - & -\\
  \cline{1-6}
  $\mathcal{TC}$ & - & - & $G_-(\mathcal{U}_s, \mathcal{TC})$ &$\mathbb{Z}_2$ & $\ddagger\widetilde{\Gamma}_{-}(T)$\\
  \cline{1-6} 
$\mathcal{CPT}$ & $G_+(\mathcal{CPT})$ &  $G_+(\mathcal{U}_c, \mathcal{CPT})$ & $G_+(\mathcal{U}_s, \mathcal{CPT})$ & $\mathbb{Z}_4$ &   $\Gamma_{+}(CPT)$ \\ 
  \cline{1-6} 
$\mathcal{T, C}$  & $G^+_{-+}(\mathcal{T}, \mathcal{C})$ &  $G^+_{-+}(\mathcal{U}_c, \mathcal{T}, \mathcal{C})$ & $G^+_{-+}(\mathcal{U}_s, \mathcal{T}, \mathcal{C})$ &$\mathbb{Z}_2$ & $\Gamma^{+}_{-+}(T, C)$ \\
  \cline{1-6}
$\mathcal{C, P}$  & $G^+_{++}(\mathcal{C}, \mathcal{P})$ &  $G^+_{++}(\mathcal{U}_c, \mathcal{C}, \mathcal{P})$ & $G^+_{++}(\mathcal{U}_s, \mathcal{C}, \mathcal{P})$ &$\mathbb{Z}_2$ & $\Gamma^{+}_{++}(C, P)$ \\
  \cline{1-6}
$\mathcal{P, T}$  & $G^-_{+-}(\mathcal{P}, \mathcal{T})$ &  $G^-_{+-}(\mathcal{U}_c, \mathcal{P}, \mathcal{T})$ & $G^-_{+-}(\mathcal{U}_c, \mathcal{P}, \mathcal{T})$ &$\mathbb{Z}_2$ & $\Gamma^{-}_{-+}(T, P)$ \\
  \cline{1-6}
$\mathcal{CP, PT}$  & $G^+_{++}(\mathcal{CP}, \mathcal{PT})$ &  $G^+_{++}(\mathcal{U}_c, \mathcal{CP}, \mathcal{PT})$ & $G^+_{++}(\mathcal{U}_s, \mathcal{CP}, \mathcal{PT})$ &$\mathbb{Z}_2$ & $\Gamma^{+}_{++}(CP, PT)$ \\
  \cline{1-6}
\multirow{4}{*}{$\mathcal{T, CP}$} & $G^+_{-+}(\mathcal{T}, \mathcal{CP})$  &  $G^+_{-+}(\mathcal{U}_c, \mathcal{T}, \mathcal{CP})$ & $\downarrow$  & $\mathbb{Z}_2$ & $\Gamma^{+}_{-+}(T, CP)$ \\
  \cline{2-6} 
&\multirow{2}{*}{$G^+_{++}(\mathcal{T}, \mathcal{CP})$, $G^+_{--}(\mathcal{T}, \mathcal{CP})$} & $G^+_{++}(\mathcal{U}_c, \mathcal{T}, \mathcal{CP})$, & \multirow{2}{*}{$G^+_{++}(\mathcal{U}_s, \mathcal{T}, \mathcal{CP})$} & \multirow{2}{*}{$\mathbb{Z}_4$} & $\Gamma^{+}_{++}(T, CP)$, \\
& & $G^+_{--}(\mathcal{U}_c,\mathcal{T}, \mathcal{CP})$ &  &  & $\Gamma^{+}_{--}(T, CP)$ \\
\cline{1-6} 
 $\mathcal{C,PT}$ & $G^+_{++}(\mathcal{C,PT})$ &  $G^+_{++}(\mathcal{U}_c, \mathcal{C,PT})$ & $G^+_{++}(\mathcal{U}_s, \mathcal{C,PT})$ & $\mathbb{Z}_4$ & $\Gamma^{+}_{++}(C, TP)$\\ 
  \cline{1-6} 
\multirow{4}{*}{$\mathcal{P, TC}$} & $G^+_{+-}(\mathcal{P}, \mathcal{TC})$  & $\downarrow$ & $G^+_{+-}(\mathcal{U}_s, \mathcal{P}, \mathcal{TC})$ & $\mathbb{Z}_2$  & $\widetilde{\Gamma}^{+}_{-+}(T, CP)$\\
  \cline{2-6} 
&\multirow{2}{*}{$G^+_{++}(\mathcal{P}, \mathcal{TC})$, $G^+_{--}(\mathcal{P}, \mathcal{TC})$} & \multirow{2}{*}{$G^+_{++}(\mathcal{U}_c, \mathcal{P}, \mathcal{TC})$} & $G^+_{++}(\mathcal{U}_s, \mathcal{P}, \mathcal{TC}), $ & \multirow{3}{*}{$\mathbb{Z}_4$} & \multirow{2}{*}{$\Gamma^{+}_{++}(TC, P)$} \\
& & & $G^+_{--}(\mathcal{U}_s,\mathcal{P}, \mathcal{TC})$ & & 
\\
\cline{1-6}   
\multirow{5}{*}{$\mathcal{T, C, P}$} & $G^{-++}_{+++}(\mathcal{T}, \mathcal{C}, \mathcal{P})$, $G^{-+-}_{-++}(\mathcal{T}, \mathcal{C}, \mathcal{P})$, &  &  &\multirow{3}{*}{$\mathbb{Z}_2$}& \multirow{3}{*}{$\Gamma^{++-}_{-++}(T, C, P)$} \\
&$G^{++-}_{-++}(\mathcal{T}, \mathcal{C}, \mathcal{P})$, $G^{+--}_{-++}(\mathcal{T}, \mathcal{C}, \mathcal{P})$,
&  $G^{++-}_{-++}(\mathcal{U}_c, \mathcal{T}, \mathcal{C}, \mathcal{P})$ 
& $G^{++-}_{-++}(\mathcal{U}_s, \mathcal{T}, \mathcal{C}, \mathcal{P})$ & &\\
& $G^{+-+}_{-+-}(\mathcal{T}, \mathcal{C}, \mathcal{P})$& & & &\\
  \cline{2-6} 
&$G^{+++}_{+++}(\mathcal{T}, \mathcal{C}, \mathcal{P})$, $G^{---}_{++-}(\mathcal{T}, \mathcal{C}, \mathcal{P})$,  & $G^{+++}_{+++}(\mathcal{U}_c, \mathcal{T}, \mathcal{C}, \mathcal{P})$, & $G^{+++}_{+++}(\mathcal{U}_s, \mathcal{T}, \mathcal{C}, \mathcal{P})$, & \multirow{2}{*}{$\mathbb{Z}_4$}& $\Gamma^{+++}_{+++}(T, C, P)$, \\
&$G^{---}_{-++}(\mathcal{T}, \mathcal{C}, \mathcal{P})$, $G^{+++}_{-+-}(\mathcal{T}, \mathcal{C}, \mathcal{P})$ 
& $G^{+--}_{-++}(\mathcal{U}_c, \mathcal{T}, \mathcal{C}, \mathcal{P})$ 
& $G^{---}_{++-}(\mathcal{U}_s, \mathcal{T}, \mathcal{C}, \mathcal{P})$ 
& & $\Gamma^{+--}_{-++}(T, C, P)$ \\
\cline{1-6} 
\end{tabular}
\caption{Classification of 2D interacting nontrivial fermionic non-chiral SPT phases with symmetry groups generated by $\mathcal{T}$, $\mathcal{C}$, $\mathcal{P}$, the combined symmetries, and/or U(1) symmetries. In this table we do not consider the case for $S_{\mathrm{C}}=-1$ [$\mathcal{C}^2=(-1)^{N_f}$] and $S_{\mathrm{PT}}=-1$ [$(\mathcal{PT})^2=(-1)^{N_f}$], so each nontrivial SPT phase in this table can be implemented by a $2\times 2$ $K$-matirx: $\{K, Q, S\}=\{\sigma_z, (1, -1)^T, (1, 1)^T\}$. Symmetry groups shown in this table have removed U(1) gauge redundancy. Classification is obtained for deconfined fermionic SPT phases with perturbative interactions. The last column shows relevant symmetry classes represented in single particle Hamiltonians (non-interacting fermionic systems) from TABLE \ref{tab_classification_C,P,T_free_fermion}; symmetry groups with additional $\mathrm{U}(1)_{s}$ can be realized in BdG systems with $S_z$ conservation, and here we list $\widetilde{\Gamma}_{+}(CP)$ and $\widetilde{\Gamma}_{-}(T)$ (noted by $\ddagger$) as examples ($\widetilde{\Gamma}$ indicates "T-dual" to $\Gamma$).} 
\label{tab_fermion_nonchiral}
\end{table*}

\subsubsection{K-matrix classification of fermionic SPT phases}

For fermionic K-matrix theories with T, C, P,
the combined symmetries, and/or U(1) symmetries, we can also implement the non-chiral short-range entangled  states by considering the $2\times 2 $ K-matrix, 
except for cases of $\mathcal{C}^2=(-1)^{N_f}$ and $\mathcal{(PT)}^2=(-1)^{N_f}$, which must be realized at least by a $4\times 4 $ K-matrix. From the canonical form of fermionic K-matirx (\ref{integer_fermion_K}) we have $K=\sigma_z$. To discuss the classification of the non-chiral SPT phases with symmetries specified by fermionic algebraic relations, for convenience we use the following notation
\begin{align} 
&G_{S_{\mathrm{G}}}(\mathcal{G}), \ G_{S_{\mathrm{G}}}(\mathcal{U}_{c/s}, \mathcal{G}), \
G^{S_{\mathrm{G}_1, \mathrm{G}_2}}_{S_{\mathrm{G}_1} S_{\mathrm{G}_2}}(\mathcal{G}_1, \mathcal{G}_2), 
\nonumber\\
&G^{S_{\mathrm{G}_1, \mathrm{G}_2}}_{S_{\mathrm{G}_1} S_{\mathrm{G}_2}}(\mathcal{U}_{c/s},\mathcal{G}_1, \mathcal{G}_2), \ 
G^{S_{\mathrm{G}_1, \mathrm{G}_2} S_{\mathrm{G}_2, \mathrm{G}_3} S_{\mathrm{G}_3, \mathrm{G}_1} }_{S_{\mathrm{G}_1} S_{\mathrm{G}_2} S_{\mathrm{G}_3} }(\mathcal{G}_1, \mathcal{G}_2, \mathcal{G}_3), \nonumber\\
&G^{S_{\mathrm{G}_1, \mathrm{G}_2} S_{\mathrm{G}_2, \mathrm{G}_3} S_{\mathrm{G}_3, \mathrm{G}_1} }_{S_{\mathrm{G}_1} S_{\mathrm{G}_2} S_{\mathrm{G}_3} } (\mathcal{U}_{c/s},\mathcal{G}_1, \mathcal{G}_2, \mathcal{G}_3)
\label{fermionic_sym_group_notation}
\end{align}
to denote fermionic symmetry groups, where $\mathcal{G}_i$ are the symmetry operators,
signs $S_{\mathrm{G}_i}$ and $S_{\mathrm{G}_i, 
\mathrm{G}_j}$ are defined in Eq.\ (\ref{algebraic_relation}), 
and $\mathcal{U}_{c/s}$ is the charge/spin U(1) symmetry. 
The detail for calculating symmetry transformations and their corresponding SPT phases is left to Appendix  \ref{Calculations of symmetry transformations and SPT phases in K-matrix theories}. Here we summarize the results in Table \ref{tab_fermion_nonchiral}. 

In TABLE \ref{tab_fermion_nonchiral} we show classification of 2D fermionic non-chiral SPT phases for $\{K, Q, S\}=\{\sigma_z, (1, -1)^T, (1, 1)^T\}$. Here we focus on deconfined fermionic SPT phases obtained from perturbing non-interacting fermions. Classification of confined fermionic SPT phases with bosonic degrees of freedom (such as bosonic Cooper pairs formed by fermions) can be described by the bosonic SPT phases discussed in the last subsections. There are some remarks for TABLE \ref{tab_fermion_nonchiral}:

(i) As discussed in Sec. \ref{Bulk and edge Chern-Simons theories incorporated with symmetries}, the nontrivial statistical phase factors might be present when symmetries act on the bosonized fields of fermions (due to Fermi statistics). Using Eqs. (\ref{statistical_phase_1}) and the commutations relations of chiral bosons (\ref{CCR_K-matrix_2}) we can determine the statistical phase factors for different symmetries on local quasiparticle excitations $\ddagger \exp{i\Lambda^TK\phi}\ddagger$. For example, for bosonic excitations $\Lambda_{\pm}\equiv (1, \pm 1)^T$ we have  
\begin{align}
&\Delta\phi^{\Lambda_{\pm}}_\mathrm{T}=\Delta\phi^{\Lambda_{\pm}}_\mathrm{C}=\Delta\phi^{\Lambda_{\pm}}_\mathrm{TC}=0 \mod 2\pi; \\ &\Delta\phi^{\Lambda_{\pm}}_\mathrm{P}=\Delta\phi^{\Lambda_{\pm}}_\mathrm{CP}=\Delta\phi^{\Lambda_{\pm}}_\mathrm{PT}=
\Delta\phi^{\Lambda_{\pm}}_\mathrm{CPT}=\pi \mod 2\pi. \nonumber 
\end{align}
We must be careful about these extra phase factors (which might cause sign changes) 
when we analyze the invariance of condensed (local) bosonic field variables under symmetry transformations. 
It would affect the way we determine the SPT phases with correct fermionic algebraic relations
[signs $S$'s in Eq.\ (\ref{fermionic_sym_group_notation})].

(ii) Contrast to bosonic systems, for each symmetry group listed in TABLE \ref{tab_fermion_nonchiral}, except groups $G_{\pm}(\mathcal{CPT})$ and $G^{\pm}_{++}(\mathcal{C,PT})$, there is only one physically inequivalent realization of the set of symmetries, which has been characterized (or fixed) by fermionic algebraic relations. For example, in symmetry group $G^{S_{\mathrm{T}, \mathrm{CP}}}_{S_{\mathrm{T}} S_{\mathrm{CP}}}(\mathcal{T}, \mathcal{CP})$ the symmetries are represented by  
\begin{align}
&U_{\mathrm{T}}=\sigma_x, \ \delta\phi_{\mathrm{T}}=\pi(0, \eta_{\mathrm{T}})^T; \nonumber \\
&U_{\mathrm{CP}}=-\sigma_x, \ \delta \phi_{\mathrm{CP}}=
\pi\left(n+\frac{\eta_{\mathrm{T, CP}}}{2}, n+\eta_{\mathrm{CP}}+\frac{\eta_{\mathrm{T, CP}}}{2}\right)^T;\nonumber\\ 
&\eta_{\mathrm{T}},\ \eta_{\mathrm{CP}},\ \eta_{\mathrm{T,CP}},\ n=0, 1.  \nonumber
\end{align}
As the parameters $\eta$'s specify the fermionic symmetry group, the "internal" parameter $n$ can always be fixed (to 0, say) by redefining $(\mathcal{CP})'=(-1)^{N_f}\cdot\mathcal{CP}$, since the fermion parity is conserved.  On the other hand, there are two physically inequivalent realizations of symmetries (specified by some "internal" parameters) in $G_{\pm}(\mathcal{CPT})$ and $G^{\pm}_{++}(\mathcal{C,PT})$.

(iii) There is gauge redundancy for specifying fermionic symmetry groups in the presence of U(1) symmetry [either $\mathrm{U}(1)_{c}$ or $\mathrm{U}(1)_{s}$]. We can remove such gauge redundancy based on the algebraic relations between discrete symmetries and U(1) symmetries (\ref{fermion_U(1)_gauge_redundancy}).  For example, groups $G^+_{++}(\mathcal{U}_s, \mathcal{T}, \mathcal{CP})$, $G^+_{+-}(\mathcal{U}_s, \mathcal{T}, \mathcal{CP})$, $G^+_{-+}(\mathcal{U}_s, \mathcal{T}, \mathcal{CP})$, $G^+_{--}(\mathcal{U}_s, \mathcal{T}, \mathcal{CP})$ are all $\mathrm{U}(1)_{s}$ gauge equivalent.

(iv) CPT-equivalence: From the topological CPT theorem, classification of fermionic SPT phases with symmetry groups generated by $\{\mathcal{G}_i\}$ and by $\{\mathcal{G}_i, \mathcal{W}_0\}$ are equivalent. Here the trivial CPT symmetry $\mathcal{W}_0$ (with $\mathcal{W}^2_0=1$) satisfies the fermionic algebraic relations $S_{\mathrm{W}_0,\mathrm{G}_i}=\epsilon$ for $U_{\mathrm{G}_i}KU^{-1}_{\mathrm{G}_i}=\epsilon K$, when it is included to a symmetry group. Some examples can be found in TABLE \ref{tab_fermion_nonchiral}:
\begin{align}
&G_-(\mathcal{U}_c, \mathcal{T})\circeq G_+(\mathcal{U}_c, \mathcal{CP})\circeq G^-_{-+}(\mathcal{U}_c, \mathcal{T}, \mathcal{CP}), \n
&G^+_{-+}(\mathcal{T}, \mathcal{C})\circeq G^+_{++}(\mathcal{C}, \mathcal{P}) \circeq G^-_{+-}(\mathcal{P}, \mathcal{T}) \n
&\circeq G^+_{++}(\mathcal{CP}, \mathcal{PT})\circeq G^{++-}_{-++}(\mathcal{T}, \mathcal{C}, \mathcal{P}),
\end{align}
where $\circeq$ represents the CPT-equivalence relations for symmetry groups and the last symmetry group in the first line is $\mathrm{U}(1)_{c}$ gauge equivalent to $G^+_{-+}(\mathcal{U}_c, \mathcal{T}, \mathcal{CP})$ in TABLE \ref{tab_fermion_nonchiral}. On the other hand, since a nontrivial CPT can be represented by by $\mathcal{W}=\mathcal{W}_0\cdot\mathcal{M}$ for some onsite unitary $Z_2$ symmetry $\mathcal{M}$, we also have the CPT-equivalence among symmetry groups generated by $\{\mathcal{G}_i, \mathcal{M}\}$ and by $\{\mathcal{G}_i, \mathcal{W}\}$. This happens when symmetries in a group can combine to nontrivial CPT symmetries. For example, $\mathcal{M}$ can be the spin parity (chiral $Z_2$ parity) $(-1)^{N_L}$ or $(-1)^{N_R}$ so that we have
\begin{align}
&G_+(\mathcal{CPT})\circeq G_+(\mathcal{M}),\n
&G^+_{++}(\mathcal{T}, \mathcal{CP})\circeq G^-_{++}(\mathcal{T}, \mathcal{M})\circeq G^-_{++}(\mathcal{CP}, \mathcal{M})\n
&\circeq G^+_{--}(\mathcal{T}, \mathcal{CP})\circeq G^-_{-+}(\mathcal{T}, \mathcal{M})\circeq G^-_{-+}(\mathcal{CP}, \mathcal{M}).
\end{align}
Topological phases protected by these symmetry groups are all characterized by $\mathbb{Z}_4$ classification (while they are all characterized by $\mathbb{Z}$ classification for non-interacting fermions). 
Therefore, due to CPT-equivalence, imposing the nontrivial CPT symmetry effectively enforces the spin parity, resulting the same classification of either non-interacting or interacting SPT phases.
This provides connections between TSCs protected by (nontrivial) CPT and TSCs protected by spin parity discussed in Ref. \onlinecite{Ryu2012}, 
and also between TSCs protected by T and P (the same as by T and CP since C is trivial for Majorana fermions)
and TSCs protected by T and spin parity, as discussed in Refs. \onlinecite{Yao2013} and \onlinecite{Qi2013}, respectively. (Actually, all the above TCSs possess $\mathbb{Z}_8$ instead $\mathbb{Z}_4$ classification. The difference comes from the fact that Majorana edge modes of TCSs have half-integer center charge, while the edges of K-matrix Chern-Simons theory we study here have integer center charge. Nevertheless, the above argument is in a consistent and reasonable way, as indicated in Ref. \onlinecite{Lu2012}.)

(v) As a comparison, for each symmetry group with nontrivial SPT phases in TABLE \ref{tab_fermion_nonchiral}, we also list the relevant symmetry classes represented in the single-particle Hamiltonians from TABLE \ref{tab_classification_C,P,T_free_fermion}. As each $\mathbb{Z}_2$ classification is unchanged, each (non-chiral) $\mathbb{Z}$ classification changes to $\mathbb{Z}_4$ classification from non-interacting to interacting topological phases.

(vi) Like bosonic theories, T-duality also holds exactly between related symmetry groups [with the correspondence (\ref{TCP_duality_K-matrix})] in TABLE \ref{tab_fermion_nonchiral}.

\section{Discussion}

We have gone through topological classification problems in the presence of parity symmetry
with emphasis on duality (equivalence) relations among various topological phases.  

One issue which we did not discuss is possible physical realizations of these topological systems considered in this paper.
While we leave detailed discussion on this issue for the future,  
a few comments are in order; 
CP symmetric systems are rather exotic in condensed matter context, but we have shown that,
through T-duality, they have representation in terms of parity symmetric BdG systems with $S_z$ conservation,
which may be more realizable. 
On the other hand, with fine tuning, CP symmetric systems may be realized 
in electron-hole coupled systems (like excitons); 
As seen in this example, CPT-equivalence and T-duality allow us to explore topological phases not listed in
conventional Altland-Zirnbauer classes. 
Other interesting examples to explore are 
insulators with $TP$ or $\{T, P\}$ symmetry, 
which are dual to known topological superconducting phases.

Another issue which we have not discussed is 
a relation of these topological classification to quantum anomalies. 
The boundary (edge) theories that we discussed in analyzing topological classification  
are not possible to 
gap out in the presence of symmetry conditions (i.e., ``protected'' by the symmetries).
These theories should not exist as an isolated system but should be realized only as a 
boundary of a bulk topological system. 
In other words, these theories should be, in the presence of an appropriate set of symmetry conditions,
anomalous or inconsistent.  
We plan to visit  
possible anomalies that pertain to topological phases discussed in this paper   
in a forth coming publication. 
For the cases of topological phases protected by CP symmetry and charge U(1) symmetry,
partial discussion on 
a quantum anomaly that underlies the topological classification is given in 
Ref.\ \onlinecite{Hsieh2014}.

\acknowledgements 

We thank 
Gil Young Cho 
for discussion and
collaborations in closely related works.

\appendix 

\section{Entanglement spectrum and effective symmetries}
\label{Entanglement spectrum and effective symmetries}

Let us consider a tight-binding Hamiltonian,
\begin{align}
H 
&=
\sum_{r,r'} 
\sum_{\alpha,\alpha'}
\psi^{\dag}_{\alpha}(r)\, 
\mathcal{H}^{\alpha \alpha'}(r,r')\, 
\psi^{\ }_{\alpha'}(r')
\nonumber \\
&=
\sum_{I,I'} 
\psi^{\dag}_I\, 
\mathcal{H}_{II'}
\psi_{I'}, 
\label{tb Hamiltonian}
\end{align}
where 
$\psi_{\alpha}(r)
$ 
($\alpha=1,\ldots, N_f$)
is an $N_f$-component 
fermion annihilation operator,
and index $r=(r_1,r_2,\ldots, r_d)$ labels a site
on a $d$-dimensional lattice.
In the second line
in Eq.\ (\ref{tb Hamiltonian}), 
we have used a more compact notation
with the collective index
$I=(r,\alpha)$, etc. 
Each block in the single particle Hamiltonian 
$\mathcal{H}(r,r')$ is 
an $N_f\times N_f$ matrix,
satisfying the hermiticity condition
$\mathcal{H}^{\dag}(r',r) 
=\mathcal{H}(r,r')$, 
and we assume the total size of the 
single particle Hamiltonian
$\mathcal{H}_{II'}$
is
$N_{\mathrm{tot}}\times N_{\mathrm{tot}}=N_f V\times N_f V$,
where $V$ is the total number of lattice sites. 
The components in $\psi(r)$
can describe, 
{\it e.g.,}
orbitals or spin degrees of freedom,
as well as
different sites within a crystal unit cell
centered at $r$.
With a canonical transformation, 
the $N_{\mathrm{tot}}\times N_{\mathrm{tot}}$ Hamiltonian 
can be diagonalized as
\begin{align}
&
U^\dagger\mathcal{H}U=\textrm{diag}(E_A),
\quad
A=1,\cdots, N_{\mathrm{tot}}, 
\nonumber \\
 &
 \mbox{with}
 \quad 
U=(\vec{u}_1,\ldots,\vec{u}_{N_{\mathrm{tot}}}),
\end{align}
where $\vec{u}_A$ is the $A$-th eigenvector 
with the eigenenergy $E_A$. 
Under this canonical transformation, 
the fermionic operator 
$\psi_I=\psi_{i\alpha}(r)$ 
can be expressed into 
fermionic operator $\chi_{A}$ as
\begin{equation}
\psi_I = \sum_{A=1}^{N_{\mathrm{tot}}} U_{IA} \chi_A.
\end{equation}
Through out the paper, 
we consider
situations where
there is a spectral gap in the single particle Hamiltonian
and 
the fermi level is located within the spectral gap. 
Then the ground state 
$|\Psi_G\rangle$
at zero temperature can be expressed 
as 
\begin{align}
|\Psi_G\rangle=
\prod_{A =1}^{N_{\mathrm{occ}}}
\chi^{\dagger}_A |0\rangle, 
\end{align}
where we assume the eigenvalues
$E_{A}$ for $A=1,\ldots,N_{\mathrm{occ}}<N_{\mathrm{tot}}$
are below the Fermi level.

\subsection{Entanglement spectrum} 
\label{subsec Entanglement spectrum}

We bipartition the total Hilbert space 
into two subspaces, which we call ``$L$'' and ``$R$''.
The discussion below is valid for an arbitrary bipartitioning;
we will later focus on the case where 
the two subspaces are associated to two spatial regions
of the total system, 
which are adjacent to each other.
We are interested in the entanglement entropy 
and spectrum for the ground state $|\Psi_G\rangle$
with the bipartitioning specified by 
the subsystems $L$ and $R$.

In a free fermion system, 
the entanglement spectrum can be directly obtained 
from its correlation matrix 
(equal-time correlation function)
\cite{Peschel2003} 
\begin{align}
C_{IJ} &=
\langle \Psi_G|\psi^\dagger_I \psi^{\ }_J |\Psi_G\rangle. 
\end{align}
In terms of the eigen wavefunctions, 
the correlation matrix $C_{IJ}$ can be written as
\begin{align}
C_{IJ} &=
\sum_{A=1}^{N_{\mathrm{occ}}}U^*_{I A}U_{JA}
=
\sum_{A=1}^{N_{\mathrm{occ}}} (\vec{u}_{A})_I^*(\vec{u}_{A})_J.
\end{align}
One then verifies 
the correlation matrix is a projection operator,
as it satisfies
\begin{align}
C^2 = C.
\end{align}
Thus, all eigenvalues of the correlation matrix
$C_{IJ}$ are either 0 or 1. 
For later purposes, we define
\begin{align}
Q_{IJ}
: =
1-2C_{IJ},
\end{align}
which has $\pm 1$ as its eigenvalues.

For the total system divided
into two subsystems $L$ and $R$, 
we introduce the following block structure, 
\begin{eqnarray}
C
=
\left(
\begin{array}{cc}
C^{\ }_L & C^{\ }_{LR} \\
C_{RL} & C^{\ }_R
\end{array}
\right),
\quad
C^{\ }_{RL}=C^{\dag}_{LR}. 
\label{block str, correl mat} 
\end{eqnarray}
Then the set of eigenvalues $\{\xi_\nu\}$ 
of $C_L$ 
is the entanglement spectrum. 
Similarly, 
the set of eigenvalues of 
\begin{align}
Q_L : =1-2C_L 
\end{align}
is $\{1-2\xi_\nu\}$,
with $-1\le 1-2\xi_\nu \le +1$. 
We refer $C_L$ and $Q_L$
as the entanglement Hamiltonian.
(These terminologies may not be entirely precise,
since $\ln (1/\xi_{\mu}-1)$ may better be
entitled to be called the entanglement energy.)

We now derive the algebraic relations obeyed
these blocks, by making use of $C^2 = C$. 
\cite{Turner2012, Hughes2011}
Then, 
\begin{align}\label{cc} 
C^2_L -C_L &=
- C^{\ }_{LR}C_{RL}, 
\nonumber \\Q_L C_{LR} &= -C_{LR} Q_R, 
\nonumber \\
C_{RL} Q_L  &= - Q_R C_{RL},
\nonumber \\
C^2_R
-
C_R 
&=
-C_{RL} C_{LR}, 
\end{align}
where
$Q_R : =1-2C_R$.
This algebraic structure, inherent to the correlation matrix
(the entanglement Hamiltonian), is 
quite analogous to supersymmetric quantum mechanics (SUSY QM). 
To see this, 
define, 
\begin{align}
&
S_L 
= 1- Q^2_L,
\quad
S_R = 
1- Q^2_R,
\nonumber \\
&
A^+ = A^{\dag} = 2C_{LR},
\quad
A^- = A = 2C_{RL},
\end{align}
where note that $S_{L/R}$
are positive semidefinite, and bounded as 
$0\le S_{L/R}\le 1$.  
One then verifies 
\begin{align}
&
S_L A^+
=
A^+ S_R, 
\quad
S_R A^- 
=
A^-  S_L, 
\nonumber \\
&
S_L = A^+ A^-,
\quad
S_R = A^- A^+. 
\end{align}
This is the standard setting of SUSY QM.
Furthering defining 
\begin{align}
 \mathcal{S}
&=
\left(
\begin{array}{cc}
S_L  & 0 \\
0 & S_R
\end{array}
\right),
\quad
\mathcal{Q} =
\left(
\begin{array}{cc}
0 & 0 \\
A & 0 
\end{array}
\right),
\quad
\mathcal{Q}^{\dag} =
\left(
\begin{array}{cc}
0 & A^{\dag} \\
0 & 0 
\end{array}
\right),
\end{align}
They satisfy
the SUSY algebra:
\begin{align}
&
[\mathcal{S}, \mathcal{Q}] 
= [\mathcal{S}, \mathcal{Q}^{\dag}]= 0,
\nonumber \\
&
\{\mathcal{Q}, \mathcal{Q}^{\dag}\} = \mathcal{S},
\quad
\{\mathcal{Q}, \mathcal{Q}\}
=
\{\mathcal{Q}^{\dag}, \mathcal{Q}^{\dag}\}
=0. 
\end{align}
Observe that the above SUSY algebra is true 
for any quadratic fermionic Hamiltonian and
for any choice of partitioning.

One can also prove
a ``chiral symmetry'':
define
\begin{align}
 &
\Lambda  :=
2i \left(
\begin{array}{cc}
 & C_{LR}\\
 -C_{RL} & 
\end{array}
\right)
=
i (\mathcal{Q}^{\dag}-\mathcal{Q}),
\quad
\Lambda^{\dag}= \Lambda, 
\nonumber \\
&
\Lambda \Lambda^{\dag}
=
\Lambda^2 
=
-4
\left(
\begin{array}{cc}
 C^2_L -C_L & \\
 & C^2_R -C_R
\end{array}
\right)
=
\mathcal{S}
\end{align}
Then, 
the $Q$-matrix satisfies
a chiral symmetry
\begin{align}
& Q \Lambda =- \Lambda Q. 
\end{align}
This effective chiral symmetry can be combined 
with other physical symmetries. 
E.g., CP symmetry. 

\subsection{Properties
  of entanglement spectrum with symmetries}
\label{sec:prop_symm}

We now focus on the case where
the dimensions of the two Hilbert spaces $L$ and $R$
are the same.
(In this case, in general, there is no zero mode of $S$ expected from SUSY.)
In addition,
we will consider the cases where there is a (discrete) 
symmetry which relates (or: ``intertwines'') 
the two Hilbert spaces. 
As in the case of the symmetry protected topological phases,
such cases arise when 
there is a (discrete) symmetry in the total system before bipartitioning,
and when the bipartitioning is consistent with the symmetry.

Let us consider a symmetry operation $\mathcal{O}$
that acts on the fermion operator as follows:
\begin{align}
\mathcal{O}\psi^{\ }_I \mathcal{O}^{\dag}= O^{\ }_{IJ} \psi^{\dag}_J,
\end{align}
where $O$ is an $N_fV \times N_fV$ unitary matrix. 
The system is invariant under the symmetry operation when
$\mathcal{O} H\mathcal{O}^{\dag}= H$,
{\it i.e.,} 
\begin{align}
O^{\dag} \mathcal{H}^T O = -\mathcal{H}. 
\end{align}
This symmetry property of the Hamiltonian
is inherited by the correlation matrix, 
\begin{align}
 O^{\dag} Q^* O &= -Q, 
\label{eq:symm}
\end{align}
Defining a block structure
as in
Eq.\ (\ref{block str, correl mat}),
we have
\begin{align}
O =
\left(
\begin{array}{cc}
O_L & O_{LR} \\
O_{RL} & O_R
\end{array}
\right).
\end{align}

Our focus below is the case where the symmetry operation 
intertwines the $L$ and $R$ Hilbert spaces. 
In other words, we can naturally categorize discrete symmetries
into two groups;
Firstly, there are symmetry operations which act on $L$ and $R$ Hilbert spaces
independently.
If the bipartitioning is done in the manner that
respects the locality
of the system,
these include {\it local} symmetry operations such as
time-reversal symmetry, and spin-rotation symmetry, etc.
On the other hand, certain spatial symmetries such as
reflection, inversion, and (discrete) spatial rotations
can exchange (intertwines) the two sub Hilbert spaces.
Focusing on the latter situations,
we thus assume
the following off-diagonal form:
\begin{align}
O =
\left(
\begin{array}{cc}
0 & O_{LR} \\
O_{RL} & 0 
\end{array}
\right),
\quad 
O^{\ }_{LR} O^{\dag}_{LR}
=
O^{\ }_{RL} O^{\dag}_{RL}
=
1. 
\end{align}

Combining $O$ and the chiral symmetry $\Lambda$, 
\begin{align}
 Q \Lambda O^{\dag} =  \Lambda O^{\dag} Q^{*}. 
\end{align}
In particular, 
\begin{align}
 Q_L C_{LR} O^{\dag}_{LR} &= C_{LR} O^{\dag}_{LR} Q^{*}_L,
 \nonumber \\
 Q_R C_{RL} O^{\dag}_{RL} &= C_{RL} O^{\dag}_{RL} Q^{*}_R. 
\end{align}
These show that the $Q$-matrix and its diagonal blocks 
$Q_{L,R}$ obey an
effective time-reversal symmetry.

\section{Calculations of symmetry transformations and SPT phases in K-matrix theories}
\label{Calculations of symmetry transformations and SPT phases in K-matrix theories}

\subsection{Algebraic relations of symmetry operators}
\label{Algebraic relations of symmetry operators}

As mentioned in the text, we consider a symmetry group generated by $Z_2$ symmetry operators ($\mathcal{T}$, $\mathcal{C}$, $\mathcal{P}$, and the combined symmetries in our discussion) with the following algebraic relations:
\begin{align}
&\mathcal{G}_i^2=S_{\mathrm{G}_i}^{N_f}, \ \forall \ \text{discrete} \ \mathcal{G}_i\in G; \nonumber\\
&\mathcal{G}_i\mathcal{G}_j\mathcal{G}_i^{-1}\mathcal{G}_j^{-1}=S_{\mathrm{G}_i,\mathrm{G}_j}^{N_f}, \ \forall \ \text{discrete} \ \mathcal{G}_i, \mathcal{G}_j\in G, 
\label{algebraic_relation_Appendix}
\end{align}
where $S$ has values $\pm1$. In a bosonic system, the operator $S^{N_f}$ (subscript omitted) is just the identity 1. In a fermionic system, $S^{N_f}$ can be the identity $1$ or the fermion number parity operator $\mathcal{P}_f\equiv (-1)^{N_f}$, where $N_f$ is the total fermion number operator. Note that  $S_{\mathrm{G}_1,\mathrm{G}_2}=S_{\mathrm{G}_2,\mathrm{G}_1}$ for any two symmetry operators $\mathcal{G}_1$ and $\mathcal{G}_2$.


On the other hand, algebraic relations between $Z_2$ symmetries and U(1) symmetries are described as follows. The total charge operator
\begin{align}
N_c\equiv\int dx j^0_c=\frac{e}{2\pi}Q_I \int dx \partial_{x}\phi_I (t,x)=eN_f.
\end{align}
and the corresponding charge U(1) transformation $\mathcal{U}_c\equiv e^{i\theta_c N_c/e}$ 
satisfy the following relations:  
\begin{align}
\mathcal{G}N_c\mathcal{G}^{-1}&=N_c, \ \text{for } \mathcal{G}=\mathcal{T}, \ \mathcal{P}, \ \mathcal{TP};  \n
\mathcal{G}N_c\mathcal{G}^{-1}&=-N_c, \ \text{for } \mathcal{G}=\mathcal{C}, \ \mathcal{TC}, \ \mathcal{CP}, \ \mathcal{CPT};   \n
\mathcal{G}\mathcal{U}_c\mathcal{G}^{-1}&=\mathcal{U}_c, \ \text{for } \mathcal{G}=\mathcal{P}, \ \mathcal{TC}, \ \mathcal{CPT}; \n
\mathcal{G}\mathcal{U}_c\mathcal{G}^{-1}&=\mathcal{U}_c^{-1}, \ \text{for } \mathcal{G}=\mathcal{T}, \ \mathcal{C}, \ \mathcal{CP}, \ \mathcal{TP}.
\label{algebraic_relation_chargeU(1)_Appendix}
\end{align}
Similarly, the total spin operator
\begin{align}
N_s\equiv\int dx j^0_s=\frac{s}{2\pi}S_I \int dx \partial_{x}\phi_I (t,x)
\end{align}
and the corresponding spin U(1) transformation $\mathcal{U}_s\equiv e^{i\theta_s N_s/s}$ satisfy
\begin{align}
\mathcal{G}N_s\mathcal{G}^{-1}&=N_s, \ \text{for } \mathcal{G}=\mathcal{TC}, \ \mathcal{CP}, \ \mathcal{TP};  \n
\mathcal{G}N_s\mathcal{G}^{-1}&=-N_s, \ \text{for } \mathcal{G}=\mathcal{T}, \ \mathcal{C}, \ \mathcal{P},  \ \mathcal{CPT};   \n
\mathcal{G}\mathcal{U}_s\mathcal{G}^{-1}&=\mathcal{U}_s, \ \text{for } \mathcal{G}=\mathcal{T}, \ \mathcal{CP}, \ \mathcal{CPT}; \n
\mathcal{G}\mathcal{U}_s\mathcal{G}^{-1}&=\mathcal{U}_s^{-1}, \ \text{for } \mathcal{G}= \mathcal{C}, \ \mathcal{P}, \ \mathcal{TC}, \ \mathcal{TP}.
\label{algebraic_relation_spinU(1)_Appendix}
\end{align}

\
\begin {table}[ht]
\centering
\begin{tabular}{|c||c|}
  \hline
  Sym. & Transformations (boson: $K=\sigma_x$)\\
  \hline \hline 
  $\mathcal{T}$ & $U_{\mathrm{T}}=\sigma_z$, $\delta \phi_{\mathrm{T}}=n_{\mathrm{T}} \pi {v} $\\
  \hline
  $\mathcal{C}$ & $U_{\mathrm{C}}=-I_2$, $\delta \phi_{\mathrm{C}}=0 $\\
  \hline
  $\mathcal{P}$ & $U_{\mathrm{P}}=\sigma_z$, $\delta \phi_{\mathrm{P}}=n_{\mathrm{P}} \pi {u} $\\
  \hline
  $\mathcal{CP}$ & $U_{\mathrm{CP}}=-\sigma_z$, $\delta \phi_{\mathrm{CP}}=n_{\mathrm{CP}}\pi {v} $ \\
  \hline
  $\mathcal{PT}$ & $U_{\mathrm{PT}}=I_2$, $\delta \phi_{\mathrm{PT}}=0 $\\
  \hline
  $\mathcal{TC}$ & $U_{\mathrm{TC}}=-\sigma_z$, $\delta \phi_{\mathrm{TC}}=n_{\mathrm{TC}} \pi {u} $\\
  \hline
  $\mathcal{CPT}$ & $U_{\mathrm{CPT}}=-I_2$, $\delta \phi_{\mathrm{CPT}}=n_{\mathrm{CPT}}\pi{u}+m_{\mathrm{CPT}}\pi{v} $\\
  \cline{1-2}
\multirow{2}{*}{$\mathcal{T, C}$}  & $U_{\mathrm{T}}=\sigma_z$, $\delta \phi_{\mathrm{T}}=n_{\mathrm{T}}\pi{u}+m_{\mathrm{T}}\pi{v} $\\ 
& $U_{\mathrm{C}}=-I_2$, $\delta \phi_{\mathrm{C}}=0 $\\
  \cline{1-2}
\multirow{2}{*}{$\mathcal{C, P}$}  & $U_{\mathrm{C}}=-I_2$, $\delta \phi_{\mathrm{C}}=0 $\\ 
& $U_{\mathrm{P}}=\sigma_z$, $\delta \phi_{\mathrm{P}}=n_{\mathrm{P}}\pi{u}+m_{\mathrm{P}}\pi{v} $\\
  \cline{1-2}
  \multirow{2}{*}{$\mathcal{P, T}$}  & $U_{\mathrm{P}}=\sigma_z$, $\delta \phi_{\mathrm{P}}=n_{\mathrm{P}}\pi {u} $\\ 
& $U_{\mathrm{T}}=\sigma_z$, $\delta \phi_{\mathrm{T}}=n_{\mathrm{T}} \pi {v} $\\
  \cline{1-2}
\multirow{2}{*}{$\mathcal{CP, PT}$}  & $U_{\mathrm{CP}}=-\sigma_z$, $\delta \phi_{\mathrm{CP}}=n_{\mathrm{CP}}\pi{u}+m_{\mathrm{CP}}\pi{v} $\\ 
& $U_{\mathrm{PT}}=I_2$, $\delta \phi_{\mathrm{PT}}=0 $\\
  \cline{1-2}
\multirow{2}{*}{$\mathcal{T, CP}$}  & $U_{\mathrm{T}}=\sigma_z$, $\delta \phi_{\mathrm{T}}=n_{\mathrm{T}} \pi {v} $\\ 
& $U_{\mathrm{CP}}=-\sigma_z$, $\delta \phi_{\mathrm{CP}}=n_{\mathrm{CP}}\pi{u}+m_{\mathrm{CP}}\pi{v} $\\
  \cline{1-2}
\multirow{2}{*}{$\mathcal{C, PT}$}  & $U_{\mathrm{C}}=-I_2$, $\delta \phi_{\mathrm{C}}=0 $\\ 
& $U_{\mathrm{PT}}=I_2$, $\delta \phi_{\mathrm{PT}}=n_{\mathrm{PT}}\pi{u}+m_{\mathrm{PT}}\pi{v} $\\
  \cline{1-2}
\multirow{2}{*}{$\mathcal{P, TC}$}  & $U_{\mathrm{P}}=\sigma_z$, $\delta \phi_{\mathrm{P}}=n_{\mathrm{P}} \pi {u} $\\ 
& $U_{\mathrm{TC}}=-\sigma_z$, $\delta \phi_{\mathrm{TC}}=n_{\mathrm{TC}}\pi{u}+m_{\mathrm{TC}}\pi{v} $\\
  \cline{1-2}
\multirow{3}{*}{$\mathcal{T, C, P}$}  & $U_{\mathrm{T}}=\sigma_z$, $\delta \phi_{\mathrm{T}}=n_{\mathrm{T}}\pi{u}+m_{\mathrm{T}}\pi{v} $\\ 
& $U_{\mathrm{C}}=-I_2$, $\delta \phi_{\mathrm{C}}=0 $\\
& $U_{\mathrm{P}}=\sigma_z$, $\delta \phi_{\mathrm{P}}=n_{\mathrm{P}}\pi{u}+m_{\mathrm{P}}\pi{v} $\\
  \cline{1-2}
\end{tabular}
\caption{(Gauge inequivalent) symmetry transformations in different classes for bosonic K-matrix theories with $K=\sigma_x$. In this table: ${u}=(1, 0)^T$, ${v}=(0, 1)^T$, and all $n$'s and $m$'s have the value 0 or 1.} 
\label{tab_boson_sym_trans}
\end{table}

\
\begin {table*}[ht]
\centering
\begin{tabular}{|c||c|}
  \hline
  Sym. & Transformations (fermion: $K=\sigma_z$)\\
  \hline \hline 
  $\mathcal{T}$ & $U_{\mathrm{T}}=\sigma_x$, $\delta \phi_{\mathrm{T}}=\eta_{\mathrm{T}} \pi {v} $\\
  \hline
  $\mathcal{C}$ & $U_{\mathrm{C}}=-I_2$, $\delta \phi_{\mathrm{C}}=0 $\\
  \hline
  $\mathcal{P}$ & $U_{\mathrm{P}}=\sigma_x$, $\delta \phi_{\mathrm{P}}=\eta_{\mathrm{P}} \pi {v} $\\
  \hline
  $\mathcal{CP}$ & $U_{\mathrm{CP}}=-\sigma_x$, $\delta \phi_{\mathrm{CP}}=\eta_{\mathrm{CP}} \pi {v} $ \\
  \hline
  $\mathcal{PT}$ & $U_{\mathrm{PT}}=I_2$, $\delta \phi_{\mathrm{PT}}=0 $\\
  \hline
  $\mathcal{TC}$ & $U_{\mathrm{TC}}=-\sigma_x$, $\delta \phi_{\mathrm{TC}}=\eta_{\mathrm{TC}} \pi {v} $\\
  \hline
  $\mathcal{CPT}$ & $U_{\mathrm{CPT}}=-I_2$, $\delta \phi_{\mathrm{CPT}}=\left(n+\frac{\eta_{\mathrm{CPT}}}{2}\right)\pi{u}+\left(m+\frac{\eta_{\mathrm{CPT}}}{2}\right)\pi{v} $\\
  \cline{1-2}
\multirow{2}{*}{$\mathcal{T, C}$}  & $U_{\mathrm{T}}=\sigma_x$, $\delta \phi_{\mathrm{T}}=\left(n+\frac{\eta_{\mathrm{T}}-\eta_{\mathrm{T, C}}}{2}\right)\pi{u}+\left(n-\frac{\eta_{\mathrm{T}}+\eta_{\mathrm{T, C}}}{2}\right)\pi{v} $\\ 
& $U_{\mathrm{C}}=-I_2$, $\delta \phi_{\mathrm{C}}=0 $\\
  \cline{1-2}
\multirow{2}{*}{$\mathcal{C, P}$}  & $U_{\mathrm{C}}=-I_2$, $\delta \phi_{\mathrm{C}}=0 $\\ 
& $U_{\mathrm{P}}=\sigma_x$, $\delta \phi_{\mathrm{P}}=\left(n+\frac{\eta_{\mathrm{P}}-\eta_{\mathrm{C, P}}}{2}\right)\pi{u}+\left(n+\frac{\eta_{\mathrm{P}}+\eta_{\mathrm{C, P}}}{2}\right)\pi{v} $\\
  \cline{1-2}
  \multirow{2}{*}{$\mathcal{P, T}$}  & $U_{\mathrm{P}}=\sigma_x$, $\delta \phi_{\mathrm{P}}=\eta_{\mathrm{P}} \pi {v} $\\ 
& $U_{\mathrm{T}}=\sigma_x$, $\delta \phi_{\mathrm{T}}=\eta_{\mathrm{T}} \pi {v} $\\
  \cline{1-2}
\multirow{2}{*}{$\mathcal{CP, PT}$}  & $U_{\mathrm{CP}}=-\sigma_x$, $\delta \phi_{\mathrm{CP}}=\left(n+\frac{\eta_{\mathrm{CP}}+\eta_{\mathrm{CP, PT}}}{2}\right)\pi{u}+\left(n-\frac{\eta_{\mathrm{CP}}-\eta_{\mathrm{CP, PT}}}{2}\right)\pi{v} $\\ 
& $U_{\mathrm{PT}}=I_2$, $\delta \phi_{\mathrm{PT}}=0 $\\
  \cline{1-2}
\multirow{2}{*}{$\mathcal{T, CP}$}  & $U_{\mathrm{T}}=\sigma_x$, $\delta \phi_{\mathrm{T}}=\eta_{\mathrm{T}} \pi {v} $\\ 
& $U_{\mathrm{CP}}=-\sigma_x$, $\delta \phi_{\mathrm{CP}}=\left(n+\frac{\eta_{\mathrm{T, CP}}}{2}\right)\pi{u}+\left(n+\eta_{\mathrm{CP}}+\frac{\eta_{\mathrm{T, CP}}}{2}\right)\pi{v} $\\
  \cline{1-2}
\multirow{2}{*}{$\mathcal{C, PT}$}  & $U_{\mathrm{C}}=-I_2$, $\delta \phi_{\mathrm{C}}=0 $\\ 
& $U_{\mathrm{PT}}=I_2$, $\delta \phi_{\mathrm{PT}}=\left(n+\frac{\eta_{\mathrm{C, PT}}}{2}\right)\pi{u}+\left(m+\frac{\eta_{\mathrm{C, PT}}}{2}\right)\pi{v} $\\
  \cline{1-2}
\multirow{2}{*}{$\mathcal{P, TC}$}  & $U_{\mathrm{P}}=\sigma_x$, $\delta \phi_{\mathrm{P}}=\eta_{\mathrm{P}} \pi {v} $\\ 
& $U_{\mathrm{TC}}=-\sigma_x$, $\delta \phi_{\mathrm{TC}}=\left(n+\frac{\eta_{\mathrm{P, TC}}}{2}\right)\pi{u}+\left(n+\eta_{\mathrm{TC}}-\frac{\eta_{\mathrm{P, TC}}}{2}\right)\pi{v} $\\
  \cline{1-2}
\multirow{3}{*}{$\mathcal{T, C, P}$}  & $U_{\mathrm{T}}=\sigma_x$, $\delta \phi_{\mathrm{T}}=\left(n+\frac{\eta_{\mathrm{T}}-\eta_{\mathrm{T, C}}}{2}\right)\pi{u}+\left(n-\frac{\eta_{\mathrm{T}}+\eta_{\mathrm{T, C}}}{2}\right)\pi{v} $\\ 
& $U_{\mathrm{C}}=-I_2$, $\delta \phi_{\mathrm{C}}=0 $\\
& $U_{\mathrm{P}}=\sigma_x$, $\delta \phi_{\mathrm{P}}=\left(n+\frac{\eta_{\mathrm{P}}-\eta_{\mathrm{C, P}}}{2}\right)\pi{u}+\left(n+\frac{\eta_{\mathrm{P}}+\eta_{\mathrm{C, P}}}{2}\right)\pi{v} $\\
  \cline{1-2}
\end{tabular}
\caption{(Gauge inequivalent) symmetry transformations in different classes for fermionic K-matrix theories with $K=\sigma_z$. In this table: (i) ${u}=(1, 0)^T$, ${v}=(0, 1)^T$, and all $\eta$'s [defined in (\ref{def_eta})], $n$'s, and $m$'s have the value 0 or 1 ; (ii) here we do not consider cases for $\eta_{\mathrm{C}}=1$ and $\eta_{\mathrm{PT}}=1$, which must be realized in the theory with a $4\times 4$ $K$-matrix at least; (iii) $\eta_{\mathrm{P, T}}=0$ in symmetry classes $\{\mathcal{P, T}\}$ and $\{\mathcal{C, P, T}\}$, as deduced from the equations of identity element (\ref{eqs_identity_Z_2_Appendix}).} 
\label{tab_fermion_sym_trans}
\end{table*}

\subsection{Equations of identity elements}
\label{Equations of identity elements}
In the basis of chiral boson fields, the constraints (\ref{algebraic_relation_Appendix}) give the following equations of identity elements (for bosonic systems we just take all $S^{N_f}=1$ in the following expressions):
\begin{align}
\label{eqs_identity_Z_2_Appendix} 
&\mathcal{G}_i^2=S_{\mathrm{G}_i}^{N_f}:
U_{\mathrm{G}_i}^2=I_N, \  (I_N+\alpha_{\mathrm{G}_i}U_{\mathrm{G}_i})\delta\phi_{\mathrm{G}_i}=\eta_{\mathrm{G}_i}\pi t_N;  \n
&\mathcal{G}_i\mathcal{G}_j\mathcal{G}_i^{-1}\mathcal{G}_j^{-1}=S_{\mathrm{G}_i,\mathrm{G}_j}^{N_f} \text{ or }
\left(\mathcal{G}_i\mathcal{G}_j\right)^2=\left(S_{\mathrm{G}_i}S_{\mathrm{G}_j}S_{\mathrm{G}_i,\mathrm{G}_j}\right)^{N_f}: \n
&(U_{\mathrm{G}_i}U_{\mathrm{G}_j})^2=I_N, \\
& (I_N+\alpha_{\mathrm{G}_i}\alpha_{\mathrm{G}_j}U_{\mathrm{G}_j}U_{\mathrm{G}_i})
\left(\delta\phi_{\mathrm{G}_j}+U_{\mathrm{G}_j}\delta\phi_{\mathrm{G}_i}\right)=\eta_{\mathrm{G}_i,\mathrm{G}_j}\pi t_N, \nonumber
\end{align}
where $\alpha_{\mathrm{G}}=1\ (-1)$ represents an unitary (antiunitary) operator $\mathcal{G}$, $t_N\equiv (1, \cdots, 1)^T$ is an $N$-component vector, and the numbers $\eta_{\mathrm{G}}$ and $\eta_{\mathrm{G}_1, \mathrm{G}_2}$ via the relations to $S_{\mathrm{G}}$ and $S_{\mathrm{G}_1, \mathrm{G}_2}$:
\begin{align}
e^{i\pi\eta_{\mathrm{G}}}=S_{\mathrm{G}},\ e^{i\pi\eta_{\mathrm{G}_1, \mathrm{G}_2}}=S_{\mathrm{G}_1}S_{\mathrm{G}_2}S_{\mathrm{G}_1, \mathrm{G}_2}. \label{def_eta}
\end{align}
Again for the bosonic systems we just take all $\eta=0$. Note that in the above (and the following) equations all phases are mod $2\pi$.

The constraints including the U(1) symmetry (\ref{algebraic_relation_chargeU(1)_Appendix}) and (\ref{algebraic_relation_spinU(1)_Appendix}) also become
\begin{align}
&\mathcal{G}\mathcal{U}_A\mathcal{G}^{-1}=\mathcal{U}_A^{\pm1} 
\text{ or } 
\mathcal{G}\mathcal{U}_A^{\mp1}\mathcal{G}\mathcal{U}_A=\mathcal{G}^2=S_{\mathrm{G}}^{N_f}: \n
&(I_N+\alpha_{\mathrm{G}}U_{\mathrm{G}})\delta\phi_{\mathrm{G}}+
(I_N\mp\alpha_{\mathrm{G}}U_{\mathrm{G}})\delta\phi_A
=\eta_{\mathrm{G}}\pi t_N \n
&\Rightarrow (I_N\mp\alpha_{\mathrm{G}}U_{\mathrm{G}})\delta\phi_A=0,
\label{eqs_identity_U(1)_Appendix}
\end{align}
where the label $A$ represents charge or spin U(1) symmetries. As examples, for T, C, and P symmetries we have the corresponding U(1) symmetries (if present) satisfying
\begin{align}
 \text{TRS}\, &: \left(I_N- U_{\mathrm{T}}\right)\delta\phi_c=0,  
\quad \left(I_N+U_{\mathrm{T}} \right)\delta\phi_s=0, \n
\text{PHS}\, &: \left(I_N+ U_{\mathrm{C}}\right)\delta\phi_c=0,  
 \quad 
 \left(I_N+U_{\mathrm{C}} \right)\delta\phi_s=0, \n
 \text{PS}\, &: \left(I_N- U_{\mathrm{P}}\right)\delta\phi_c=0,
\quad
  \left(I_N+U_{\mathrm{P}} \right)\delta\phi_s=0,  \label{T, C, P_U(1)_constraints_Appendix}
\end{align}
respectively. Identifying $\delta\phi_c=\theta_cK^{-1}Q$ and $\delta\phi_s=\theta_sK^{-1}S$,
(\ref{T, C, P_U(1)_constraints_Appendix}) exactly corresponds to (\ref{K-matrix_sym_constraints}), which are equations of symmetry constraints for gauged K-matrix Chern-Simons theories (coupled to external gauge fields).


\begin {table*}[ht]
\centering
\begin{tabular}{|c||c|c|c|c|c|c|c|}
  \hline
  \multirow{2}{*}{Sym. group} & \multirow{2}{*}{Parameters} & \multicolumn{6}{ |c| }{Nontrivial bosonic non-chiral SPT phases and their classification}   \\ 
\cline{3-8} 
&  (from TABLE \ref{tab_boson_sym_trans}) & \multicolumn{2}{c|}{No U(1)'s}  & \multicolumn{2}{c|}{$\mathrm{U}(1)_c$ is present} & \multicolumn{2}{c|}{$\mathrm{U}(1)_s$ is present} \\
  \hline \hline 
  $Z_2^{\mathrm{T}}$ & $[n_{\mathrm{T}}]$ & - & 0  &  $[1]$ & $\mathbb{Z}_2$ & - & 0    \\
  \hline 
  $Z_2^{\mathrm{C}}$ & - & - & 0 & - & 0 & - & 0  \\
  \hline
  $Z_2^{\mathrm{P}}$ & $[n_{\mathrm{P}}]$ & - & 0 & - & 0 & $[1]$ & $\mathbb{Z}_2$\\
  \hline 
$Z_2^{\mathrm{CP}}$ & $[n_{\mathrm{CP}}]$ & - & 0 &  $[1]$ & $\mathbb{Z}_2$ & - & 0  \\
  \hline 
  $Z_2^{\mathrm{PT}}$ & - & - & 0 & - & 0 & - & 0 \\
  \hline
  $Z_2^{\mathrm{TC}}$ & $[n_{\mathrm{TC}}]$ & - & 0 & - & 0 & $[1]$ & $\mathbb{Z}_2$\\
  \hline 
$Z_2^{\mathrm{CPT}}$ & $[n_{\mathrm{CPT}}, m_{\mathrm{CPT}}]$ & $[1, 1]$ & $\mathbb{Z}_2$ & $[1, 1]$ & $\mathbb{Z}_2$ & $[1, 1]$ & $\mathbb{Z}_2$ \\ 
  \hline 
$Z_2^{\mathrm{T}}\times Z_2^{\mathrm{C}}$ & $[n_{\mathrm{T}}, m_{\mathrm{T}}]$ & $[1,1]$ & $\mathbb{Z}_2$ &  $[1, 1]$ & $\mathbb{Z}_2$ & $[1, 1]$ & $\mathbb{Z}_2$ \\ 
\hline 
 $Z_2^{\mathrm{C}}\times Z_2^{\mathrm{P}}$ &  $[n_{\mathrm{P}}, m_{\mathrm{P}}]$ & $[1,1]$ & $\mathbb{Z}_2$ &  $[1, 1]$ & $\mathbb{Z}_2$ & $[1, 1]$ & $\mathbb{Z}_2$ \\ 
\hline 
 $Z_2^{\mathrm{P}}\times Z_2^{\mathrm{T}}$ & $[n_{\mathrm{P}}, n_{\mathrm{T}}]$ & $[1,1]$ & $\mathbb{Z}_2$ &  $[1, 1]$ & $\mathbb{Z}_2$ & $[1, 1]$ & $\mathbb{Z}_2$ \\ 
\hline 
$Z_2^{\mathrm{CP}}\times Z_2^{\mathrm{PT}}$ & $[n_{\mathrm{CP}}, m_{\mathrm{PT}}]$ & $[1,1]$ & $\mathbb{Z}_2$ &  $[1, 1]$ & $\mathbb{Z}_2$ & $[1, 1]$ & $\mathbb{Z}_2$ \\ 
\hline 
\multirow{2}{*}{$Z_2^{\mathrm{T}}\times Z_2^{\mathrm{CP}}$} & \multirow{2}{*}{$[n_{\mathrm{T}}, n_{\mathrm{CP}}, m_{\mathrm{CP}}]$} & $[0, 1, 1], [1, 1, 0]$, & \multirow{2}{*}{$\mathbb{Z}_2^2$} & $[0, 1, 1], [1, 1, 0]$, & \multirow{2}{*}{$\mathbb{Z}_2^2$} & \multirow{2}{*}{$[0, 1, 1]$} &\multirow{2}{*}{$\mathbb{Z}_2$}\\
& & $[1, 1, 1]$ & & $[1, 1, 1]$ &  & & \\
\hline 
 $Z_2^{\mathrm{C}}\times Z_2^{\mathrm{PT}}$ & $[n_{\mathrm{PT}}, m_{\mathrm{PT}}]$ & $[1,1]$  & $\mathbb{Z}_2$ &  $[1,1]$ & $\mathbb{Z}_2$ & $[1,1]$ & $\mathbb{Z}_2$ \\ 
  \hline 
\multirow{2}{*}{$Z_2^{\mathrm{P}}\times Z_2^{\mathrm{TC}}$} & \multirow{2}{*}{$[n_{\mathrm{P}}, n_{\mathrm{TC}}, m_{\mathrm{TC}}]$} & $[0, 1, 1], [1, 0, 1]$,  & \multirow{2}{*}{$\mathbb{Z}_2^2$} & \multirow{2}{*}{$[0, 1, 1]$} & \multirow{2}{*}{$\mathbb{Z}_2$} & $[0, 1, 1], [1, 0, 1]$, &\multirow{2}{*}{$\mathbb{Z}_2^2$}\\
& &  $[1, 1, 1]$ & & & & $[1, 1, 1]$  &\\
\hline 
\multirow{4}{*}{$Z_2^{\mathrm{T}}\times Z_2^{\mathrm{C}}\times Z_2^{\mathrm{P}}$} & \multirow{4}{*}{$[n_{\mathrm{T}}, m_{\mathrm{T}}, n_{\mathrm{P}}, m_{\mathrm{P}}]$} & 15 nontrivial phases & \multirow{4}{*}{$\mathbb{Z}_2^4$} & \multirow{2}{*}{$[0, 0, 1, 1], [0, 1, 1, 0]$,}  & \multirow{5}{*}{$\mathbb{Z}_2^2$} & \multirow{2}{*}{$[0, 0, 1, 1], [1, 0, 0, 1]$,} &\multirow{4}{*}{$\mathbb{Z}_2^2$}\\
& & generated by & &  & & & \\
& & $[0, 0, 1, 1], [0, 1, 1, 0]$, & & \multirow{2}{*}{$[0, 1, 1, 1]$} & & \multirow{2}{*}{$[1, 0, 1, 1]$} & \\
& & $[1, 0, 0, 1], [1, 1, 0, 0]$ & &  & &  & \\
\hline 
\end{tabular}
\caption{Nontrivial bosonic non-chiral SPT phases and their classification with symmetry groups generated by $\mathcal{T}$, $\mathcal{C}$, $\mathcal{P}$, the combined symmetries, and/or U(1) symmetries, implemented by a $2\times2$ K-matrix: $\{K, Q, S\}=\{\sigma_x, (0, 1)^T, (1, 0)^T\}$. Results are based on symmetry transformations described in TABLE \ref{tab_boson_sym_trans}.  Here we use $[n, m, \cdots]$ to label SPT phases  (parameters $n$'s and $m$'s shown here are not specified explicitly in symmetry groups in bosonic systems). Classification shown in this table has removed U(1) gauge redundancy. TABLE \ref{tab_boson_nonchiral} is constructed directly from this table.}
\label{tab_boson_SPT}
\end{table*}

\begin {table*}[ht]
\centering
\begin{tabular}{|c||c|c|c|c|c|}
  \hline
  \multirow{2}{*}{Sym.} & \multirow{2}{*}{Parameters} & \multicolumn{3}{ |c| }{Sym. groups for nontrivial fermionic non-chiral SPT phases } & Top.  \\
 \cline{3-5} 
& (from TABLE \ref{tab_fermion_sym_trans}) & No U(1)'s  & $\mathrm{U}(1)_c$ is present & $\mathrm{U}(1)_s$ is present & class.\\
  \hline \hline 
  $\mathcal{T}$ & $\eta_{\mathrm{T}}$ & - &  $1$ & - & $\mathbb{Z}_2$ \\
  \hline 
  $\mathcal{C}$ & - & - & - & - & -\\
  \cline{1-6}
  $\mathcal{P}$ & $\eta_{\mathrm{P}}$ & - & - & $0$ &$\mathbb{Z}_2$\\
  \hline 
$\mathcal{CP}$ & $\eta_{\mathrm{CP}}$ & - &  $0$ & - & $\mathbb{Z}_2$ \\
  \hline 
  $\mathcal{PT}$ & - & - & - & - & - \\
  \hline
  $\mathcal{TC}$ & $\eta_{\mathrm{TC}}$ & - & - & $1$ &$\mathbb{Z}_2$\\
  \hline 
$\mathcal{CPT}$ & $\eta_{\mathrm{CPT}}$ & $0$ & 0 & 0 & $\mathbb{Z}_4$ \\ 
  \hline 
$\mathcal{T,C}$ & $(\eta_{\mathrm{T}}, \eta_{\mathrm{T,C}})$ & $(1,1)$ &  $(1,1)$ & $(1,1)$ & $\mathbb{Z}_2$ \\ 
\hline 
 $\mathcal{C,P}$ &  $(\eta_{\mathrm{P}}, \eta_{\mathrm{C,P}})$ & $(0,0)$ &  $(0,0)$ & $(0,0)$ & $\mathbb{Z}_2$ \\ 
\hline 
$\mathcal{P,T}$ & $(\eta_{\mathrm{P}}, \eta_{\mathrm{T}})$ & $(0,1)$ &  $(0,1)$ & $(0,1)$ & $\mathbb{Z}_2$ \\ 
\hline 
$\mathcal{CP,PT}$ & $(\eta_{\mathrm{CP}}, \eta_{\mathrm{CP,PT}})$ & $(0,0)$ &  $(0,0)$ & $(0,0)$ & $\mathbb{Z}_2$ \\ 
\hline
\multirow{2}{*}{$\mathcal{T, CP}$} & \multirow{2}{*}{$(\eta_{\mathrm{T}}, \eta_{\mathrm{CP}}, \eta_{\mathrm{T,CP}})$} & $(1, 0, 1)$  & $(1, 0, 1)$ &  $\downarrow$ & $\mathbb{Z}_2$\\
  \cline{3-6} 
& &$(0, 0, 0), (1, 1, 0)$ & $(0, 0, 0), (1, 1, 0)$ & $(0, 0, 0)$ & $\mathbb{Z}_4$\\
\hline 
 $\mathcal{C,PT}$ & $\eta_{\mathrm{C,PT}}$ & $0$  &  0 & 0 & $\mathbb{Z}_4$ \\ 
  \hline 
\multirow{2}{*}{$\mathcal{P, TC}$} & \multirow{2}{*}{$(\eta_{\mathrm{P}}, \eta_{\mathrm{TC}}, \eta_{\mathrm{P,TC}})$} & $(0, 1, 1)$  & $\downarrow$ & $(0, 1, 1)$ & $\mathbb{Z}_2$\\
  \cline{3-6} 
& &$(0, 0, 0), (1, 1, 0)$ & $(0, 0, 0)$ & $(0, 0, 0), (1, 1, 0)$ & $\mathbb{Z}_4$\\
\hline    
\multirow{5}{*}{$\mathcal{T, C, P}$} & \multirow{5}{*}{$(\eta_{\mathrm{T}}, \eta_{\mathrm{P}}, \eta_{\mathrm{T,C}}, \eta_{\mathrm{C,P}})$} & $(0, 0, 1, 0), (1, 0, 0, 0)$, &  \multirow{3}{*}{$(1, 0, 1, 0)$} & \multirow{3}{*}{$(1, 0, 1, 0)$} &\multirow{3}{*}{$\mathbb{Z}_2$}\\
& & $(1, 0, 1, 0), (1, 0, 1, 1)$, &   &  &\\
& & $(1, 1, 1, 0)$ & & & \\
  \cline{3-6} 
& & $(0, 0, 0, 0), (0, 1, 1, 0)$,  & $(0, 0, 0, 0)$, & $(0, 0, 0, 0)$, & \multirow{2}{*}{$\mathbb{Z}_4$}\\
& &$(1, 0, 0, 1), (1, 1, 1, 1)$ & $(1, 0, 1, 1)$ & $(0, 1, 1, 0)$ &\\
\hline 
\end{tabular}
\caption{Symmetry groups for nontrivial fermionic non-chiral SPT phases and topological classification, by considering $\mathcal{T}$, $\mathcal{C}$, $\mathcal{P}$, the combined symmetries, and/or U(1) symmetries for a $2\times2$ K-matrix: $\{K, Q, S\}=\{\sigma_z, (1, -1)^T, (1, 1)^T\}$. Results are based on symmetry transformations described in TABLE \ref{tab_fermion_sym_trans}. Parameters $\eta$'s shown here can be transferred to the signs $S$'s that characterize symmetry groups in fermionic systems by Eq. (\ref{def_eta}) (note that $\eta_{\mathrm{C}}$, $\eta_{\mathrm{PT}}$, and $\eta_{\mathrm{P,T}}$, which are all zero, are not specified as parameters). Symmetry groups and classification shown in this table have removed U(1) gauge redundancy. TABLE \ref{tab_fermion_nonchiral} is constructed directly from this table. }
\label{tab_fermion_SPT}
\end{table*}

In this appendix, we use these constraint equations to find the way that the chiral boson fields transform under $\mathcal{T}$, $\mathcal{C}$, $\mathcal{P}$, the combined symmetries, and/or the U(1) symmetries. Note that we also have the gauge equivalence for the forms of these symmetry transformations:
\begin{align} 
\{U_{\mathrm{G}}, \delta\phi_{\mathrm{G}}\} \ \rightarrow& \  \{X^{-1}U_{\mathrm{G}}X, \ X^{-1}\left(\delta\phi_{\mathrm{G}}- \alpha_{\mathrm{G}}\Delta\phi + U_{\mathrm{G}}\Delta\phi \right)\}, \nonumber \\
&\text{if} \ X \in GL(N,\mathbb{Z}), \ X^TKX=K,
\end{align}
where $\alpha_{\mathrm{G}}=1 \ (-1)$ if $\mathcal{G}$ is an unitary (antiunitary) operator. This means we can choose some $X$ and $\Delta\phi$ to fix $\{U_{\mathrm{G}_i}, \delta\phi_{\mathrm{G}_i}\}$ to the inequivalent forms of transformations.

Here we just consider the case of $2\times 2$ K matrix ($K=\sigma_x$ for a bosonic system and a $K=\sigma_z$ for fermionic system). All gauge inequivalent solutions for discrete symmetry transformations in different classes are summarized in TABLE \ref{tab_boson_sym_trans} for bosonic systems, and in TABLE \ref{tab_fermion_sym_trans} for fermionic systems, respectively.

\subsection{Discussion of SPT phases}
\label{Discussion of SPT phases}

Using the criteria for SPT phases developed 
in Sec. \ref{Edge Stability and Criteria for SPT Phases}, 
we can find nontrivial SPT phases and thus topological classification 
for both bosonic and fermionic systems, 
based on the symmetry transformations described 
in TABLEs \ref{tab_boson_sym_trans} and \ref{tab_fermion_sym_trans}. 
The results are summarized in TABLEs \ref{tab_boson_SPT} and \ref{tab_fermion_SPT}, 
which directly derive TABLEs 
\ref{tab_boson_nonchiral} and \ref{tab_fermion_nonchiral} in the text, respectively. 

To accomplish TABLEs \ref{tab_boson_SPT} and \ref{tab_fermion_SPT}, we give explicit calculations for SPT phases with some examples. In the following discussions, bosonic systems are realized by $\{K, Q, S\}=\{\sigma_x, (0, 1)^T, (1, 0)^T\}$, while fermionic systems are realized by $\{K, Q, S\}=\{\sigma_z, (1, -1)^T, (1, 1)^T\}$. The discussion for finding group structures of SPT phases follows Ref. \onlinecite{Lu2012}.

\subsubsection{Bosonic SPT phases with symmetry group $Z_2^{\mathrm{CP}}$}

From Table \ref{tab_boson_sym_trans}, the symmetry transformations for $\mathcal{CP}$ are given by
\begin{align}
&U_{\mathrm{CP}}=-\sigma_z, \  \delta\phi_{\mathrm{CP}}= \pi(0, n_{\mathrm{CP}})^T, \
n_{\mathrm{CP}}=0, 1.
\label{boson_CP_Appendix}
\end{align}

Without U(1) symmetries, the symmetry invariant perturbations can be either (for any values of $n_{\mathrm{CP}}$)
\begin{align} 
\mathcal{S}^{c.p.}_{\mathrm{edge}}= \sum_{l\in\mathbb{Z}}A_l\int dtdx\cos\left(2l\phi_2+ \alpha_l \right)
 \label{boson_c.p._Appendix}
\end{align}
or
\begin{align} 
\mathcal{S}^{s.p.}_{\mathrm{edge}}= \sum_{l\in\mathbb{Z}}B_l\int dtdx\cos\left(l\phi_1+k_l\pi \right),
\label{boson_s.p._Appendix}
\end{align}
where $A_l,\ B_l,\ \alpha_l\in\mathbb{R}$, $k_l\in\mathbb{Z}$, and $c.p.$ ($s.p.$) stands for "charge preserved" ("spin preserved") perturbations, i.e., invariant under $\mathrm{U}(1)_c$ [$\mathrm{U}(1)_s$].  In this case, we can always condense $\phi_1$ without breaking $\mathcal{CP}$:
\begin{align} 
\langle \phi_1\rangle \overset{\mathcal{CP}}{\longrightarrow} 
- \langle \phi_1\rangle \mod 2\pi,
\label{phi_1_underCP_Appendix}
\end{align}
say, $\langle\phi_1\rangle$ has expectation value 0 or $\pi$ (depending on how we choose $\mathcal{S}^{s.p.}_{\mathrm{edge}}$). Thus there are just trivial phases in the absence of U(1) symmetries.

In addition to $\mathcal{CP}$, if we now include the charge U(1) symmetry
\begin{align}
\mathcal{U}_c: \delta\phi_c=\theta_cK^{-1}Q=q_c\theta_c(1, 0)^T, \
\theta_c \in \mathbb{R},
\label{boson_chargeU(1)_trans_Appendix}
\end{align}
then only the charge preserved perturbations (bosonic variable $\phi_2$) are allowed to add to the system (to condense). Since
\begin{align} 
\langle \phi_2\rangle \overset{\mathcal{CP}}{\longrightarrow}  
\langle \phi_2\rangle+n_{\mathrm{CP}}\pi \mod 2\pi,
\label{phi_2_underCP_Appendix}
\end{align}
we find that $n_{\mathrm{CP}}=1$ ($n_{\mathrm{CP}}=0$) corresponds to a nontrivial (trivial) SPT phase, as the edge can not (can) be gapped by $\mathcal{S}^{c.p.}_{\mathrm{edge}}$ without breaking $\mathcal{CP}$.  Denoting $[n_{\mathrm{CP}}]$ the phase with corresponding CP symmetry, we have $[1]\oplus[1]=[0]$, i.e., putting (adding) two copies of $n_{\mathrm{CP}}=1$ phases $\{\phi_1^k, \phi_2^k, k=1,2\}$ together we can gap out the edge by condensing $\{\phi_1^1-\phi_1^2, \phi_2^1+\phi_2^2\}$ without spontaneously breaking CP and charge U(1) symmetries. Therefore, bosonic SPT phases with $\mathrm{U}(1)_c\rtimes Z_2^{\mathrm{CP}}$ are classified by a $\mathbb{Z}_2$ group: $[n_{\mathrm{CP}}=0]$ and $[n_{\mathrm{CP}}=1]$ correspond to (two) elements of this $\mathbb{Z}_2$ group. 

On the other hand, if we include the spin U(1) symmetry instead [no charge U(1)]
\begin{align}
\mathcal{U}_s: \delta\phi_s=\theta_sK^{-1}S=q_s\theta_s(0, 1)^T, \
\theta_s \in \mathbb{R},
\label{boson_spinU(1)_trans_Appendix}
\end{align}
then only the the spin preserved perturbations (bosonic variable $\phi_1$) are allowed to add to the system (to condense). In this case, there are just trivial phases as we can alway condense $\phi_1$ (\ref{phi_1_underCP_Appendix}).

\subsubsection{Bosonic SPT phases with symmetry group $Z_2^{\mathrm{T}}\times Z_2^{\mathrm{C}}\times Z_2^{\mathrm{P}}$}

From Table \ref{tab_boson_sym_trans}, the symmetry transformations for $\{\mathcal{T, C, P}\}$ are given by
\begin{align}
&U_{\mathrm{T}}=\sigma_z, \  \delta\phi_{\mathrm{T}}= \pi(n_{\mathrm{T}}, m_{\mathrm{T}})^T, \
n_{\mathrm{T}},\ m_{\mathrm{T}}=0, 1; \n
&U_{\mathrm{C}}=-I_2, \  \delta\phi_{\mathrm{C}}= 0; \n
&U_{\mathrm{P}}=\sigma_z, \  \delta\phi_{\mathrm{P}}= \pi(n_{\mathrm{P}}, m_{\mathrm{P}})^T, \
n_{\mathrm{P}},\ m_{\mathrm{P}}=0, 1.
\label{boson_T_C_P_Appendix}
\end{align}

Without U(1) symmetries, the symmetry invariant perturbations can be either (for any values of $n$'s) $\mathcal{S}^{c.p.}_{\mathrm{edge}}$ (\ref{boson_c.p._Appendix}) or $\mathcal{S}^{s.p.}_{\mathrm{edge}}$ (\ref{boson_s.p._Appendix}). However, the edge can not be gapped, as either $\mathcal{S}^{c.p.}_{\mathrm{edge}}$ (\ref{boson_c.p._Appendix}) or $\mathcal{S}^{s.p.}_{\mathrm{edge}}$ is added to the system, without breaking any symmetry for some choices of $n$'s. The following lists the 
nontrivial SPT phases that are denoted as $[n_{\mathrm{T}}, m_{\mathrm{T}}, n_{\mathrm{P}}, m_{\mathrm{P}}]$:
\begin{align}
&[0, 0, 1, 1],\ [0, 1, 1, 0],\ [0, 1, 1, 1], \n
&[1, 0, 0, 1],\ [1, 1, 0, 0],\ [1, 1, 0, 1], \n
&[1, 0, 1, 1],\ [1, 1, 1, 0],\ [1, 1, 1, 1].
\label{boson_T_C_P_nontrivialSPT_1_Appendix}
\end{align}
Putting two identical copies of each phase above together will result a trivial phase, i.e., $[n_{\mathrm{T}}, m_{\mathrm{T}}, n_{\mathrm{P}}, m_{\mathrm{P}}]^2=[\text{trivial}]$. 
On the other hand, all phases above are inequivalent [we can not obtain a trivial phase by putting any two different elements in the set (\ref{boson_T_C_P_nontrivialSPT_1_Appendix}) together].
However, phases shown in (\ref{boson_T_C_P_nontrivialSPT_1_Appendix}) are not all possible nontrivial SPT phases with symmetries $Z_2^{\mathrm{T}}\times Z_2^{\mathrm{C}}\times Z_2^{\mathrm{P}}$; putting different phases above together might result other nontrivial SPT phases that are not listed above. To determine the group structure of these phases, we observe that putting three phases in any vertical or  horizontal line of the array (\ref{boson_T_C_P_nontrivialSPT_1_Appendix}) together will result a trivial phase, or equivalently, we have
\begin{align}
&\oplusto{[0, 0, 1, 1]}{} \oplus \oplusto{[0, 1, 1, 0]}{} = \oplusto{[0, 1, 1, 1]}{} \n
&\equalto{[1, 0, 0, 1]}{} \oplus \equalto{[1, 1, 0, 0]}{} = \equalto{[1, 1, 0, 1]}{} \n
&[1, 0, 1, 1] \oplus [1, 1, 1, 0] = [1, 1, 1, 1].
\label{boson_T_C_P_nontrivialSPT_2_Appendix}
\end{align}
From this fact, we know that all nontrivial phases can be generated by a specific set of four phases, say, $\{[0, 0, 1, 1],\ [0, 1, 1, 0],\ [1, 0, 0, 1],\ [1, 1, 0, 0]\}$ (there are $3\times 3=9$ equivalent choices for these four group generators); there are total fifteen nontrivial SPT phases, which form a $\mathbb{Z}_2^4$ group.

In addition to $\mathcal{T}$, $\mathcal{C}$, and $\mathcal{P}$,  if we now include the charge U(1) symmetry (\ref{boson_chargeU(1)_trans_Appendix}), then only the charge preserved perturbations (bosonic variables) are allowed to add to the system (to condense). Besides those nontrivial phases discussed in the case without U(1) symmetries, there are other nontrivial phases protected by $\mathrm{U}(1)_c$ additionally. Some of the nontrivial phases are  $\mathrm{U}(1)_c$ gauge equivalent to each other, and we can just consider the inequivalent set $[0, 0, 1, 1]$, $[0, 1, 1, 0]$, and $[0, 1, 1, 1]$, which form a $\mathbb{Z}_2^2$ group.

On the other hand, if we include the spin U(1) symmetry (\ref{boson_spinU(1)_trans_Appendix}) instead [no charge U(1)],
then only the spin preserved perturbations (bosonic variable) are allowed to add to the system (to condense). Similar to the discussion for charge U(1), the $\mathrm{U}(1)_s$ gauge inequivalent nontrivial phases can be $[0, 0, 1, 1]$, $[1, 0, 0, 1]$, and $[1, 0, 1, 1]$, which form a $\mathbb{Z}_2^2$ group.

\subsubsection{Fermionic SPT phases with symmetry $\mathcal{CP}$}

From Table \ref{tab_fermion_sym_trans}, the symmetry transformations for $\mathcal{CP}$ are given by
\begin{align}
&U_{\mathrm{CP}}=-\sigma_x, \  \delta\phi_{\mathrm{CP}}= \pi(0, \eta_{\mathrm{CP}})^T, \
\eta_{\mathrm{CP}}=0, 1.
\label{fermion_CP_Appendix}
\end{align}

Without U(1) symmetries, the symmetry invariant perturbations can be either (for any values of $\eta_{\mathrm{CP}}$)
\begin{align} 
\mathcal{S}^{c.p.}_{\mathrm{edge}}= \sum_{l\in\mathbb{Z}}A_l\int dtdx\cos\left[2l(\phi_L-\phi_R)+ \alpha_l \right]
 \label{fermion_CP_c.p._App}
\end{align}
or
\begin{align} 
\mathcal{S}^{s.p.}_{\mathrm{edge}}= \sum_{l\in\mathbb{Z}}B_l\int dtdx\cos\left[2l(\phi_L+\phi_R)+k_l\pi \right], \label{fermion_CP_s.p._App}
\end{align}
where $A_l,\ B_l,\ \alpha_l\in\mathbb{R}$ and $k_l\in\mathbb{Z}$. Note that under $\mathcal{CP}$ the boson fields $l_{\pm}^T\phi\equiv\phi_L\pm\phi_R$ transform as
\begin{align} 
(\mathcal{CP})(l_{\pm}^T\phi)(\mathcal{CP})^{-1}&
=l_{\pm}^T\left(U_{\mathrm{CP}}\phi+\delta\phi_{\mathrm{CP}}\right)+\Delta\phi^{l_{\pm}}_\mathrm{CP} ,
\end{align}
with the statistical phase $\Delta\phi^{l_{\pm}}_\mathrm{CP}=\pi \mod 2\pi$ (while statistical phases in a bosonic K-matrix theory is always trivial; see discussions in Sec. \ref{Bulk and edge Chern-Simons theories incorporated with symmetries}). In this case, we can always condense $l_+\phi$ without breaking $\mathcal{CP}$:
\begin{align} 
\langle l_+\phi\rangle \overset{\mathcal{CP}}{\longrightarrow} - \langle l_+\phi\rangle+(\eta_{\mathrm{CP}}+1)\pi \mod 2\pi,
\label{l_+phi_underCP_Appendix}
\end{align}
say, $\langle l_+\phi\rangle$ has expectation value $(\eta_{\mathrm{CP}}\pm1)\pi/2$ (depending on how we choose $\mathcal{S}^{s.p.}_{\mathrm{edge}}$). Thus there are just trivial phases in the absence of U(1) symmetries.

In addition to $\mathcal{CP}$, if we now include the charge U(1) symmetry
\begin{align}
\mathcal{U}_c: \delta\phi_c=\theta_cK^{-1}Q=q_c\theta_c(1, 1)^T, \
\theta_c \in \mathbb{R},
\label{fermion_chargeU(1)_trans_Appendix}
\end{align}
then only the charge preserved perturbations (bosonic variable $l_-\phi$) are allowed to add to the system (to condense). Since
\begin{align} 
\langle l_-\phi\rangle \overset{\mathcal{CP}}{\longrightarrow}  \langle l_-\phi\rangle+(\eta_{\mathrm{CP}}+1)\pi \mod 2\pi,
\label{l_-phi_underCP_Appendix}
\end{align}
we find that $\eta_{\mathrm{CP}}=0$ ($\eta_{\mathrm{CP}}=1$) corresponds to a nontrivial (trivial) SPT phase, as the edge can not (can) be gapped by $\mathcal{S}^{c.p.}_{\mathrm{edge}}$ without breaking $\mathcal{CP}$. Moreover, SPT phases with $\eta_{\mathrm{CP}}=0$ [group $G_+(\mathcal{U}_c, \mathcal{CP})$] form a $\mathbb{Z}_2$ group. This can be seen if we put two copies of $\eta_{\mathrm{CP}}=0$ phases $\{\phi_L^k, \phi_R^k, k=1,2\}$ together and then gap out the edge by condensing independent bosonic variables, say, $\{\phi_L^1-\phi_R^2, \phi_R^1-\phi_L^2\}$, without spontaneously breaking CP and charge U(1) symmetries.

On the other hand, if we include the spin U(1) symmetry instead [no charge U(1)]
\begin{align}
\mathcal{U}_s: \delta\phi_s=\theta_sK^{-1}S=q_s\theta_s(1, -1)^T, \
\theta_s \in \mathbb{R},
\label{fermion_spinU(1)_trans_Appendix}
\end{align}
then only the the spin preserved perturbations (bosonic variable $l_+\phi$) are allowed to add to the system (to condense). In this case, there are just trivial phases as we can alway condense $l_+\phi$ (\ref{l_+phi_underCP_Appendix}).

\subsubsection{Fermionic SPT phases with symmetries $\{\mathcal{T}, \mathcal{CP} \}$}

From Table \ref{tab_fermion_sym_trans}, the symmetry transformations for $\{\mathcal{T}, \mathcal{CP} \}$ are given by
\begin{align}
&U_{\mathrm{T}}=\sigma_x, \ \delta\phi_{\mathrm{T}}=\pi(0, \eta_{\mathrm{T}})^T; \nonumber \\
&U_{\mathrm{CP}}=-\sigma_x, \ \delta \phi_{\mathrm{CP}}=
\pi\left(n+\frac{\eta_{\mathrm{T, CP}}}{2}, n+\eta_{\mathrm{CP}}+\frac{\eta_{\mathrm{T, CP}}}{2}\right)^T;\nonumber\\ 
&\eta_{\mathrm{T}},\ \eta_{\mathrm{CP}},\ \eta_{\mathrm{T,CP}},\ n=0, 1.  
\label{fermion_T_CP_3_App}
\end{align}

Without U(1) symmetries, the symmetry invariant perturbations can be either (for any values of $\eta$'s)  $\mathcal{S}^{c.p.}_{\mathrm{edge}}$ (\ref{fermion_CP_c.p._App}) or $\mathcal{S}^{s.p.}_{\mathrm{edge}}$ (\ref{fermion_CP_s.p._App}). Note that under $\mathcal{T}/\mathcal{CP}$ the boson fields $l_{\pm}^T\phi\equiv\phi_L\pm\phi_R$ transform as
\begin{align} 
\mathcal{T}(l_{\pm}^T\phi)\mathcal{T}^{-1}&=l_{\pm}^T\left(-U_{\mathrm{T}}\phi+\delta\phi_{\mathrm{T}}\right)+\Delta\phi^{l_{\pm}}_\mathrm{T}, \nonumber\\
(\mathcal{CP})(l_{\pm}^T\phi)(\mathcal{CP})^{-1}&
=l_{\pm}^T\left(U_{\mathrm{CP}}\phi+\delta\phi_{\mathrm{CP}}\right)+\Delta\phi^{l_{\pm}}_\mathrm{CP} ,
\end{align}
with the statistical phases $\Delta\phi^{l_{\pm}}_\mathrm{T}=0 \mod 2\pi$ and $\Delta\phi^{l_{\pm}}_\mathrm{CP}=\pi \mod 2\pi$, respectively.

Label the phases with symmetry transformations (\ref{fermion_T_CP_3_App}) 
as $[\eta_{\mathrm{T}}, \eta_{\mathrm{CP}}, \eta_{\mathrm{T},\mathcal{CP}}, n]$.
Then we want to study the group structure of these phases protected 
by symmetry group $G_{s_{\mathrm{T}}s_{\mathrm{CP}}}^{s_{\mathrm{T},\mathcal{CP}}}$. 
As we are interested, the nontrivial SPT phases, where the edge cannot be gapped out without spontaneously breaking T or CP symmetries, are found to be:

(1) $\mathbb{Z}_2$ classes:
When $(\eta_{\mathrm{T}}, \eta_{\mathrm{CP}}, \eta_{\mathrm{T,CP}}) = (1,0,1)$, which is the symmetry group $G^+_{-+}(\mathcal{T}, \mathcal{CP})$, neither $l_{+}^T\phi=\phi_L+\phi_R$ or $l_{-}^T\phi=\phi_L-\phi_R$ can condense to be invariant under both $\mathcal{T}$ and $\mathcal{CP}$. Thus $[1, 0, 1, n]\neq 0$ ($n$ can be 0 or 1) are nontrivial SPT phase. However, for two copies of this theory with variables $\{\phi_L^k, \phi_R^k, k=1,2\}$, the edge can be gapped by condensing $\{\phi_L^1-\phi_R^2, \phi_R^1-\phi_L^2\}$ without spontaneously breaking any symmetries. So we have $[1, 0, 1, n]^2=[1, 0, 1, n]\oplus[1, 0, 1, n]=0$. On the other hand, it is easy to show $[1, 0, 1, 0]\oplus[1, 0, 1, 1]=0$, which means $[1, 0, 1, 0]^{-1}=[1, 0, 1, 1]$, and thus we have $[1, 0, 1, 0]=[1, 0, 1, 1]$, i.e., these two phases correspond to the same nontrivial SPT phase. Therefore, the topological classification for $G^+_{-+}(\mathcal{T}, \mathcal{CP})$ forms a $\mathbb{Z}_2$ group, generated by the element $[1, 0, 1, 0]=[1, 0, 1, 1]$.

(2) $\mathbb{Z}_4$ classes:
When $(\eta_{\mathrm{T}}, \eta_{\mathrm{CP}}, \eta_{\mathrm{T,CP}}) = (0,0,0)\ [G^+_{++}(\mathcal{T}, \mathcal{CP})]$ or $(1,1,0)\ [G^+_{--}(\mathcal{T}, \mathcal{CP})]$, the family of condensed bosonic fields [or the set of independent elementary bosonic variables (\ref{elementary_bosonic_variables})] are not all invariant under both $\mathcal{T}$ and $\mathcal{CP}$ for one, two, and three copies of the edge theory. Only when we consider four copies of the theory the edge can be gapped out without breaking any symmetry. Thus both these symmetry groups have $\mathbb{Z}_4$ classification. To be more specific, let us consider the case for  symmetry group $G^+_{++}(\mathcal{T}, \mathcal{CP})$. For $G^+_{++}(\mathcal{T}, \mathcal{CP})$ it is easy to show $[0, 0, 0, n]$ for $n=0, 1$ are both nontrivial SPT phases. Now consider two copies of the edge theory with variables $\{\phi_L^k, \phi_R^k, k=1,2\}$, and extended K-matrix $K_2=K\oplus K=\sigma_z\oplus \sigma_z$. Then we find that (either $n=$ 0 or 1), for any independent Haldane null vectors $l_1$ and $l_2$ (which satisfy $l_1^TK_2^{-1}l_1=l_2^TK_2^{-1}l_2=l_1^TK_2^{-1}l_2=0$), there exists an elementary bosonic variable $v_a^T\phi=l_a^T\phi/gcd(l_{a,1}, l_{a,2}, l_{a,3}, l_{a,4})$, where $\phi=(\phi^1, \phi^2)^T$ and $l_a\equiv a_1l_1+a_2l_2$ is some linear combination of $l_1$ and $l_2$, such that $v_a^T\phi$ cannot condense to be invariant under both $\mathcal{T}$ and $\mathcal{CP}$. This  means $[0, 0, 0, n]^2$ for $n=0, 1$ are also nontrivial SPT phases. Then, if we put four $[0, 0, 0, n]$ states with edge variables $\{\phi_L^k, \phi_R^k, k=1,2,3,4 \}$ together, the edge can be gapped out without breaking any symmetry, by localizing the following independent bosonic variables $\{\phi_L^1+\phi_L^2+\phi_R^3+\phi_R^4, \phi_R^1+\phi_R^2+\phi_L^3+\phi_L^4, \phi_L^1+\phi_R^1+\phi_L^3+\phi_R^4, \phi_L^1+\phi_R^1+\phi_R^3+\phi_L^4\}$, i.e., we have $[0, 0, 0, n]^4=0$ for $n=0, 1$. On the other hand, we also have $[0, 0, 0, 0]\oplus[0, 0, 0, 1]=0$, which means $[0, 0, 0, 0]^{-1}=[0, 0, 0, 1]$, and thus we have $[0, 0, 0, 0]^3=[0, 0, 0, 1]$, from the above result. Therefore, all different phases of $G^+_{++}(\mathcal{T}, \mathcal{CP})$ form a $\mathbb{Z}_4$ group, generated by the element $[0, 0, 0, 0]=[0, 0, 0, 1]^{-1}$. Similar analysis can be applied to $G^+_{--}(\mathcal{T}, \mathcal{CP})$.

In addition to $\mathcal{T}$ and $\mathcal{CP}$, if we now include the charge U(1) symmetry (\ref{fermion_chargeU(1)_trans_Appendix}),
then only the the charge preserved perturbations (bosonic variables) are allowed to add to the system (to condense). In this case, we find topological phases protected by $(1, 0, 1)\ [G^+_{-+}(\mathcal{U}_c,\mathcal{T}, \mathcal{CP})]$ and by $(1, 0, 0)\ [G^-_{-+}(\mathcal{U}_c,\mathcal{T}, \mathcal{CP})]$ are classified by $\mathbb{Z}_2$, while those protected by $(0, 0, 0)\ [G^+_{++}(\mathcal{U}_c,\mathcal{T}, \mathcal{CP})]$, by $(1, 1, 0)\ [G^+_{--}(\mathcal{U}_c,\mathcal{T}, \mathcal{CP})]$, by $(0, 0, 1)\ [G^-_{++}(\mathcal{U}_c,\mathcal{T}, by \mathcal{CP})]$, and by $(1, 1, 1)\ [G^-_{--}(\mathcal{U}_c,\mathcal{T}, \mathcal{CP})]$ are classified by $\mathbb{Z}_4$, as we apply similar argument from previous cases. 
We can check the gauge equivalence among these symmetry groups in the presence of $\mathrm{U}(1)_c$ symmetry (See TABLE \ref{tab_fermion_SPT}).

On the other hand, if we include the spin U(1) symmetry (\ref{fermion_spinU(1)_trans_Appendix}) instead [no charge U(1)],
then only the spin preserved perturbations (bosonic variables) are allowed to add to the system (to condense). In this case, there is only $\mathbb{Z}_4$ classification for nontrivial SPT phases, corresponding to symmetry groups $(0, 0, 0)\ [G^+_{++}(\mathcal{U}_s,\mathcal{T}, \mathcal{CP})]$, $(0, 1, 1)\ [G^+_{+-}(\mathcal{U}_s,\mathcal{T}, \mathcal{CP})]$, $(1, 0, 1)\ [G^+_{-+}(\mathcal{U}_s,\mathcal{T}, \mathcal{CP})]$, and $(1, 1, 0)\ [G^+_{--}(\mathcal{U}_s,\mathcal{T}, \mathcal{CP})]$, respectively. Again, there is gauge equivalence among these symmetry groups in the presence of $\mathrm{U}(1)_s$ symmetry (See TABLE \ref{tab_fermion_SPT}).

\section{Proof of topological CPT theorem for interacting non-chiral SPT phases in two dimensions}
\label{Proof of topological CPT theorem for interacting fermionic and bosonic non-chiral SPT phases in two dimensions}

The CPT symmetry $\mathcal{W}$ satisfies [Eqs. (\ref{CPT_trans}) and (\ref{K-matrix_CPT_constraints})]

\begin{align}
&\qquad \mathcal{W}\phi(t, x)\mathcal{W}^{-1}=-U_{\mathrm{W}}\phi(-t, -x)+\delta\phi_{\mathrm{W}}, \n
&U_{\mathrm{W}}^TKU_{\mathrm{W}}=K, \quad\left(I_N+ U_{\mathrm{W}}^T\right)Q=0, \quad \left(I_N+U_{\mathrm{W}}^T \right)S=0,
\end{align}
and
\begin{align}
\mathcal{W}^2=S_{\mathrm{W}}^{N_f}:
 U_{\mathrm{W}}^2=I_N, \  (I_N-U_{\mathrm{W}})\delta\phi_{\mathrm{W}}=\eta_{\mathrm{W}}\pi t_N,
\end{align}
where we have defined an $N$-component vector $t_N\equiv(1,1,\cdots,1)^T$.
We first show that, in the absence of any other symmetries, there exists a CPT operator $\mathcal{W}_0$ such that the 1D edge theory with any gapping interactions ${S}^{\mathrm{int} }_{\mathrm{edge}}$ is invariant under $\mathcal{W}_0$. Remember this must be achieved in two steps: 

(i) $\mathcal{W}_0$ preserves 
\begin{align} 
{S}^{\mathrm{int} }_{\mathrm{edge}}=
\sum^{\mathrm{bosonic}}_{\Lambda\in \mathbb{Z}^N}U_{\Lambda }
\int dtdx\,
\cos
\left( 
 \Lambda^TK\phi + \alpha_{\Lambda}\right)
\end{align} 
with any collections of bosonic vectors $\{\Lambda_a\}$ (i.e., $\pi\Lambda^TK\Lambda=0 \mod 2\pi$) satisfying Haldane's null vector condition (\ref{Haldane's_null_vector}), and

(ii) edge states are gapped without breaking $\mathcal{W}_0$ spontaneously: all elementary bosonic variables $\{v^T_a\phi\}$ [defined in (\ref{elementary_bosonic_variables})] are invariant under $\mathcal{W}_0$.

For a bosonic system with generic K-matrix $K=I_{N/2}\otimes\sigma_x$, a trivial CPT operator $\mathcal{W}_0$ can be chosen as $\{\eta_{\mathrm{W}_0}, U_{\mathrm{W}_0}, \delta\phi_{\mathrm{W}_0}\}=\{0, -I_{N}, 0\}$. Since for a bosonic system the statistical phase factor is trivial ($\Delta\phi^{\Lambda}_{\mathrm{W}_0}=0 \mod 2\pi$), we have $\mathcal{W}_0\Lambda^TK\phi(t, x)\mathcal{W}^{-1}_0=\Lambda^TK\phi(-t, -x)$ for any $\Lambda\in\mathbb{Z}^N$\ and thus any interactions ${S}^{\mathrm{int} }_{\mathrm{edge}}$ and the associated $\{v^T_a\phi\}$ are invariant under $\mathcal{W}_0$.

For a fermionic system with generic K-matrix $K=I_{N/2}\otimes\sigma_z$, a trivial CPT operator $\mathcal{W}_0$ can be chosen as $\{\eta_{\mathrm{W}_0}, U_{\mathrm{W}_0}, \delta\phi_{\mathrm{W}_0}\}=\{0, -I_{N}, t_{N/2}\otimes\chi_L\}$ or $\{0, -I_{N}, t_{N/2}\otimes\chi_R\}$, where $\chi_L\equiv(\pi, 0)^T$ and $\chi_R\equiv(0, \pi)^T$. Due to the nontrivial statistical phase factor that may arise in a fermionic system, under $\mathcal{W}_0$ we have $\mathcal{W}_0\Lambda^TK\phi(t, x)\mathcal{W}^{-1}_0=\Lambda^TK\phi(-t, -x)+\Lambda^TK\delta\phi_{\mathrm{W}_0}+\Delta\phi^{\Lambda}_{\mathrm{W}_0}$. Now we show that, for any bosonic vectors $\Lambda$ satisfying Haldane's null vector criterion, $\Lambda^TK\delta\phi_{\mathrm{W}_0}+\Delta\phi^{\Lambda}_{\mathrm{W}_0}$ is a multiple of $2\pi$. Considering the case $\delta\phi_{\mathrm{W}_0}=t_{N/2}\otimes\chi_L$, we have
\begin{align} 
\Lambda^TK\delta\phi_{\mathrm{W}_0}&=\pi(\Lambda_1+\Lambda_3+\dotsb+\Lambda_{N-1})\n
&=\pi\sum_{\text{odd}\ I}\Lambda_I.
\end{align} 
On the other hand, the statistical phase factor associated with $\Lambda$ is given by Eq. (\ref{statistical_phase_1}):
\begin{align}
\Delta\phi^{\Lambda}_{\mathrm{W}_0}
&\equiv \frac{1}{2i}\sum_{I<J} \Lambda_I\Lambda_J\left([\mathcal{W}_0(iK\phi)_I\mathcal{W}_0^{-1},\mathcal{W}_0 (iK\phi)_J\mathcal{W}_0^{-1}] \right. \n
&\qquad\qquad\left.- \mathcal{W}_0[(iK\phi)_I, (iK\phi)_J]\mathcal{W}_0^{-1}\right) \n
&=\pi \sum_{I<J}\Lambda_IQ_I\Lambda_JQ_J \mod 2\pi \n
&=\pi \sum_{I<J}\Lambda_I\Lambda_J \mod 2\pi,
\end{align}
where we have used Eq. (\ref{CCR_K-matrix_2}) to arrive the second equality (and remember $\mathcal{W}_0$ is antiunitary) and the fact that $Q_I$'s are odd for a fermionic system to arrive the third equality. Then, since $\Lambda$ satisfies $\Lambda^TK\Lambda=\sum_{\text{odd}\ I}\Lambda^2_I-\sum_{\text{even}\ I}\Lambda^2_I=0$ (and thus $\sum_{\text{odd}\ I}\Lambda_I=\sum_{\text{even}\ I}\Lambda_I \mod 2$ ), we have
\begin{align} 
&\quad\frac{1}{\pi}\left(\Lambda^TK\delta\phi_{\mathrm{W}_0}+\Delta\phi^{\Lambda}_{\mathrm{W}_0}\right)\n
&=\sum_{\text{odd}\ I}\Lambda_I+\sum_{I<J}\Lambda_I\Lambda_J \mod 2 \n
&=\left(\sum_{\text{odd}\ I}\Lambda_I\right)\left(\sum_{\text{even}\ I}\Lambda_I\right)+\sum_{I<J}\Lambda_I\Lambda_J \mod 2 \n
&=\frac{1}{2}\left[\left(\sum_{\text{odd}\ I}\Lambda_I\right)^2-\left(\sum_{\text{even}\ I}\Lambda_I\right)^2 \right] \mod 2 \n
&=0 \mod 2.
\end{align} 
Therefore,  any gapping interactions ${S}^{\mathrm{int} }_{\mathrm{edge}}$ and the associated $\{v^T_a\phi\}$ [which are also Haldane's null vectors from the definition (\ref{elementary_bosonic_variables})] are invariant under $\mathcal{W}_0$. The argument also applies similarly to another choice $\delta\phi_{\mathrm{W}_0}=t_{N/2}\otimes\chi_R$.

Now, if a system possesses some symmetries $\{\mathcal{G}_i\}$, 
adding the trivial CPT operator $\mathcal{W}_0$ to the system would not change the stability condition of the 1D edge theory with symmetries $\{\mathcal{G}_i\}$, and thus the criterion for the corresponding 2D SPT phase by $\{\mathcal{G}_i\}$ is the same as  by $\{\mathcal{G}_i\}$ and $\mathcal{W}_0$. For a bosnic system, algebraic relations between these discrete symmetries are trivial and hence we do not need to specify the relations between $\{\mathcal{G}_i\}$ and $\mathcal{W}_0$ (i.e., $\mathcal{W}_0\mathcal{G}_i\mathcal{W}_0^{-1}\mathcal{G}^{-1}_i=1, \forall i$). For a fermionic system, we must have $S_{\mathrm{W}_0,\mathrm{G}_i}=1$ if $U_{\mathrm{G}_i}^TKU_{\mathrm{G}_i}=K$ and $S_{\mathrm{W}_0,\mathrm{G}_i}=-1$ if $U_{\mathrm{G}_i}^TKU_{\mathrm{G}_i}=-K$. To show this, we can look at the identity equations for $\mathcal{G}_i$ (for clarity we drop index $i$ of $\mathcal{G}_i$ in the following discussion):
\begin{align}
&\mathcal{G}^2=S_{\mathrm{G}}^{N_f}:\
U_{\mathrm{G}}^2=I_N, \  (I_N+\alpha_{\mathrm{G}} U_{\mathrm{G}})\delta\phi_{\mathrm{G}}=\eta_{\mathrm{G}}\pi t_N,  \n
&(\mathcal{W}_0\mathcal{G})^2=\left(S_{\mathrm{W}_0}S_{\mathrm{G}}S_{\mathrm{W}_0, \mathrm{G}}\right)^{N_f}:\ (U_{\mathrm{W}_0}U_{\mathrm{G}})^2=I_N, \n 
&(I_N-\alpha_{\mathrm{G}}U_{\mathrm{G}}U_{\mathrm{W}_0})
\left(\delta\phi_{\mathrm{G}}+\alpha_{\mathrm{G}}U_{\mathrm{G}}\delta\phi_{\mathrm{W}_0}\right)
=\eta_{\mathrm{W}_0,\mathrm{G}}\pi t_N,
\label{id_eqs_W_0&G_1}
\end{align} 
where $\alpha_{\mathrm{G}}=1\ (-1)$ represents a unitary (antiunitary) operator $\mathcal{G}$. Note that in the above (and the following) equations all phases are mod $2\pi$. For $\{\eta_{\mathrm{W}_0}, U_{\mathrm{W}_0}, \delta\phi_{\mathrm{W}_0}\}=\{0, -I_{N}, t_{N/2}\otimes\chi_L\}$, (\ref{id_eqs_W_0&G_1}) gives
\begin{align}
(I_N+\alpha_{\mathrm{G}}U_{\mathrm{G}})\delta\phi_{\mathrm{W}_0}
=\left(\eta_{\mathrm{G}}+\eta_{\mathrm{W}_0,\mathrm{G}}\right)\pi t_N.
\label{id_eqs_W_0&G_2}
\end{align} 
Now, since $\mathcal{G}$ satisfies $U_{\mathrm{G}}^TKU_{\mathrm{G}}=\epsilon K$ ($K=I_{N/2}\otimes\sigma_z$) with either $\epsilon=1$ or $-1$, $U_{\mathrm{G}}$ has the general form 
\begin{align}
U_{\mathrm{G}} &
=
\left\{
 \begin{array}{ll}
  V_{\mathrm{G}}\otimes I_2\ \mbox{or}\ V_{\mathrm{G}}\otimes\sigma_z&  \mbox{for $\epsilon=1$} \\
  V_{\mathrm{G}}\otimes\sigma_x& \mbox{for $\epsilon=-1$} \\
 \end{array}
 \right. , 
\label{general_form_U_G}
\end{align}
where $V_{\mathrm{G}}$ is some $N/2\times N/2$ integer matrix with the constraint $(I_{N/2}\pm V_{\mathrm{G}})t_{N/2}=0 \mod 2$ [from $(I_N\pm U_{\mathrm{G}})Q=0$]. Substitute (\ref{general_form_U_G}) to (\ref{id_eqs_W_0&G_2}), we then obtain
\begin{align}
\eta_{\mathrm{G}}+\eta_{\mathrm{W}_0,\mathrm{G}} &
=
\left\{
 \begin{array}{ll}
  0 \mod 2&  \mbox{for $\epsilon=1$} \\
  1 \mod 2& \mbox{for $\epsilon=-1$} \\
 \end{array}
 \right.,
\end{align}
which can be stated as $S_{\mathrm{W}_0, \mathrm{G}}=\epsilon$. This completes the proof.



\bibliography{reference_CPT}

\end{document}